\documentclass[11pt,a4paper]{article}

\usepackage{jheppub}
\usepackage{bbold}

\usepackage{amsmath}
\usepackage{amssymb}
\usepackage{makecell}
\usepackage{amsthm}
\usepackage{braket}
\usepackage{tabu}
\usepackage{hyperref}
\usepackage{mathtools}
\usepackage{enumerate}
\usepackage{tikz}
\usepackage{extpfeil}
\usepackage{tabularx}
\usepackage{graphicx}
\usepackage{color}
\usepackage{lscape}
\usetikzlibrary{decorations.pathmorphing}
\usepackage{pdflscape}
\usepackage{comment}
\usepackage[utf8]{inputenc}

	\usepackage{enumitem}

\newcommand{\bZ}{\mathbb{Z}}

\newcommand{\bC}{\mathbb{C}}
\newcommand{\bN}{\mathbb{N}}

\newcommand{\cM}{\mathcal{M}}

\newcommand{\Hom}{\text{Hom}}
\renewcommand{\d}{\text{d}}
\newcommand{\id}{\text{id}}
\newcommand{\ev}{\text{ev}}
\newcommand{\coev}{\text{coev}}
\newcommand{\tev}{\widetilde{\text{ev}}}
\newcommand{\tcoev}{\widetilde{\text{coev}}}
\newcommand{\Res}[1]{\text{Res}\left[#1\right]}

\newcommand{\aeq}[1]{\begin{equation}\begin{aligned}#1\end{aligned}\end{equation}}

\newcommand{\uaeq}[1]{\begin{equation*}\begin{aligned}#1\end{aligned}\end{equation*}}

\newcommand{\beq}{\begin{equation}}
\newcommand{\eeq}{\end{equation}}
\newcommand{\baeq}{\begin{equation}\begin{aligned}}
\newcommand{\eaeq}{\end{aligned}\end{equation}}

\usetikzlibrary{decorations.markings}
	\definecolor{WScolor}{RGB}{191,191,255}
	\definecolor{WScolor light}{RGB}{224,224,255}
	\definecolor{DefectColor}{RGB}{0,0,128}
	\definecolor{SymmetryDefectColor}{RGB}{0,127,0}

	\newcommand{\worldsheets}{
		\fill[WScolor] (0,0) ellipse (1.2 and 1);
	}
	\newcommand{\worldsheetLights}{
		\fill[WScolor light] (0,0) ellipse (1.2 and 1);
	}
	\newcommand{\offset}{-3pt} 

\tikzset{snake/.style={decorate, decoration=snake}}
\tikzset{
	symmetry defect/.style={color=SymmetryDefectColor, line width=1.5}, 
	defect/.style={color=DefectColor, line width=1.5}, 
	arrow position/.style={postaction={decorate,decoration={
		markings,
		mark=at position #1 with {\arrow{>}}
	}}},
	opp arrow position/.style={postaction={decorate,decoration={
		markings,
		mark=at position #1 with {\arrow{<}}
	}}},
	defect node/.style={circle,inner sep=1.5pt,fill, DefectColor}, 
	symmetry node/.style={circle,inner sep=1.5pt, fill, color=SymmetryDefectColor},
	defect unit node/.style={draw=DefectColor, circle, inner sep=0, minimum width=3, line width=1.3},
	unit node/.style={draw=SymmetryDefectColor, circle, inner sep=0, minimum width=3, line width=1.3}, 
	dot/.style={circle,inner sep=1pt,fill},
	black node/.style={circle,inner sep=1.5pt,fill},
	number node/.style={draw=SymmetryDefectColor, circle,inner sep=1.5pt, DefectColor} 
}
\usetikzlibrary{patterns,snakes,arrows}

\usetikzlibrary{decorations.pathmorphing,shapes}

\newcounter{sarrow}
\newcommand\xrsquigarrow[1]{%
\stepcounter{sarrow}
\begin{tikzpicture}[decoration=snake]
\node (\thesarrow) {\strut#1};
\draw[->,decorate] (\thesarrow.south west) -- (\thesarrow.south east);
\end{tikzpicture}%
}

\tikzset{
	slim defect/.style={color=DefectColor, line width=1}
}

\usepackage{caption}
\usepackage{multirow}

\usepackage{makecell}

\title{Realizing IR theories by projections in the UV}
\author[1]{Fabian Klos,}
\author[2]{Daniel Roggenkamp}
\affiliation[1]{Institut f\"ur Theoretische Physik, Universit\"at Heidelberg,\\
Philosophenweg 19, 69120 Heidelberg, Germany}
\affiliation[2]{Institut f\"ur Mathematik, Universit\"at Mannheim,\\
B6, 26,  68131 Mannheim, Germany}
\abstract{
We show how (topologically twisted) quantum field theories in the IR of bulk RG flows can be represented within the respective UV theories by means of codimenion-one projection defects.
Indeed, from this perspective, RG flows of bulk theories can be described in terms of RG flows of
the codimension-one identity defect in the fixed UV bulk theory. We illustrate this in the example of RG flows between supersymmetric Landau-Ginzburg orbifold models, for which the respective defects can be described in terms of matrix factorizations.}

\begin{document}
\maketitle
	
\section{Introduction}

In quantum field theories, the renormalization group (RG) flow drives theories at high ultra-violet (UV) energies to theories at low infrared (IR) energy. In this paper, we describe a method to construct IR correlators directly within the UV theory, by inserting certain codimension-one defects into UV correlators. This allows us to represent the entire IR theory in terms of the UV theory.

While this approach might also be useful in more general situations, in this paper we will only deal with quantum field theories admitting a topological twist compatible with the RG flow. The topological twist provides a good handle on defects allowing us to relate the twisted theories in the IR and UV of RG flows.
For brevity and concreteness we will restrict our discussion to two-dimensional quantum field theories, such as $2d$ $N=(2,2)$ supersymmetric QFTs, but we expect our method to be applicable in any dimension.

Starting point of the construction are RG defects as introduced in \cite{Brunner:2007ur}. These are domain walls between UV and IR theories obtained in the following way. Consider a perturbation of a scale invariant quantum field theory by a relevant local operator. The RG flow drives the theory from the original theory in the UV to some other theory in the IR. If the perturbation is restricted to a finite region of space-time, the RG flow drives the theory to the IR on the domain of the perturbation, while leaving it at the UV on the rest of space-time. Along the way, it creates a domain wall $R$ on the boundary of the perturbation domain, separating the IR theory from the UV theory:
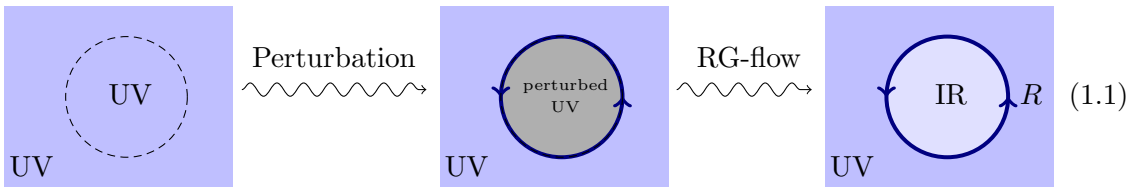
\aeq{\label{eq:Pert and flow}\tikz[anchor=base, baseline=0]{
		\fill[WScolor] (-1.6,-1.2) rectangle (1.4,1.2);
		\fill[WScolor] (0,0) ellipse (.8 and .8);
		\draw[densely dashed] (0,0) ellipse (.8 and .8);
		\node at (.05,-.1) {UV};
		\node at (-1.25,-1.05) {UV};
	}\;\xrsquigarrow{Perturbation}\;
	\tikz[anchor=base, baseline=0]{
		\fill[WScolor] (-1.6,-1.2) rectangle (1.4,1.2);
		\fill[color={rgb,255:red,175; green,175; blue,175}] (0,0) ellipse (.8 and .8);
		\draw[arrow position = .5, arrow position = 0, defect] (0,0) ellipse (.8 and .8);
		\draw[densely dashed] (0,0) ellipse (.8 and .8);
		\node at (.05,-.2) {\tiny\shortstack{perturbed\\UV}};
		\node at (-1.25,-1.05) {UV};
	}\;\xrsquigarrow{ RG-flow }\;
	\tikz[anchor=base, baseline=0]{
		\fill[WScolor] (-1.6,-1.2) rectangle (1.4,1.2);
		\fill[WScolor light] (0,0) ellipse (.8 and .8);
		\draw[arrow position = .5, arrow position = 0, defect] (0,0) ellipse (.8 and .8);
		\node at (1.1,-.1) {$R$};
		\node at (.05,-.1) {IR};
		\node at (-1.25,-1.05) {UV};
	}}
The \emph{RG defects} $R$ obtained in this way capture the entire relation between UV and IR theories. They project UV degrees of freedom onto the IR theory and embed IR degrees of freedom into the UV theory.  

In order to get a good handle on defect lines, in particular the behavior of correlation functions under changes of their position, we now pass to the topologically twisted theory. Compatibility of the RG flow with the topological twist assures that the respective RG defect descends to a defect between the topologically twisted IR and UV theories. We still refer to this defect as RG defect and to the topologically twisted theories as IR and UV theories. 
(Note that the notion of RG defects as defined in \cite{Brunner:2007ur} does not require a topological twist. In fact, examples of RG defects are known in full CFTs \cite{Gaiotto:2012np}. We expect the ideas presented below to also be applicable in this more general context, albeit in a more intricate way.)

	Fusion\footnote{In a TQFT one can move parallel defects infinitely close together resulting in a new, fused defect. We will denote fusion of defects by `$\otimes$'.} of RG defects $R$ with their downward oriented versions $R^\dagger$ gives rise to the trivial identity defect in the IR theory, $R\otimes R^\dagger\cong I_\text{IR}$, while fusion in the opposite order yields non-trivial defects $P=R^\dagger\otimes R$ in the UV theory:
	\aeq{\label{eq:rfuseprop}\tikz[baseline=-5]{
		\fill[WScolor light] (-1,-1) rectangle (1,1);
		\fill[WScolor] (-.5,-1) rectangle (.5,1);
		\draw[defect,arrow position=.5] (-.5,-1) -- (-.5,1);
		\draw[defect,opp arrow position=.5] (.5,-1) -- (.5,1);
		\node[defect] at (-.7,.8) {$R$};
		\node[defect] at (.8,.85) {$R^\dagger$};}
	= \tikz[baseline=-5]{
		\fill[WScolor light] (-.55,-1) rectangle (.55,1);
		\draw[densely dashed] (0,-1) -- (0,1);
		\node[defect] at (.3,.8) {$I_\text{IR}$};
	}
	=\tikz[baseline=-5]{
		\fill[WScolor light] (-.55,-1) rectangle (.55,1);
	}\qquad\text{and}\qquad
	\tikz[baseline=-5]{
		\fill[WScolor] (-1,-1) rectangle (1,1);
		\fill[WScolor light] (-.5,-1) rectangle (.5,1);
		\draw[defect,opp arrow position=.5] (-.5,-1) -- (-.5,1);
		\draw[defect,arrow position=.5] (.5,-1) -- (.5,1);
		\node[defect] at (-.75,.85) {$R^\dagger$};
		\node[defect] at (.7,.8) {$R$};}
	= \tikz[baseline=-5]{
		\fill[WScolor] (-.5,-1) rectangle (.5,1);
		\draw[symmetry defect] (0,-1) -- (0,1);
		\node[symmetry defect] at (.2,.8) {$P$};
	}}
	The first equation is a central property of RG defects, which ensures locality in the sense that islands of IR theories trivially connect:
	\uaeq{\tikz[scale=1.5, baseline=-5]{
		\fill[WScolor] (0,-1) rectangle (2,1);
		\fill[WScolor light] (.4,0) ellipse (.3 and .85);
		\fill[WScolor light] (1.4,.3) ellipse (.55 and .4);
			\draw[defect, arrow position=0.5] (.4,0) ellipse (.3 and .85);
			\draw[defect, arrow position=1] (1.4,.3) ellipse (.55 and .4);
			\node at (0.75,-0.6) {$R$};
			\node at (1.8,-0.15){$R$};
			\node at (1.5,-.8){UV};
			}
						=
			\tikz[scale=1.5, baseline=-5]{
		\fill[WScolor] (0,-1) rectangle (2,1);
		\fill[WScolor light] (.7,0)  arc (0:-335:.3 and .85) .. controls (0.7,0.3) .. (0.9,0.45) arc (158:-140:.55 and .4)..controls (0.7,0.2)..(0.7,0.0);		
				\draw[defect, opp arrow position=0.3] (.7,0)  arc (0:-335:.3 and .85) .. controls (0.7,0.3) .. (0.9,0.45) arc (158:-140:.55 and .4)..controls (0.7,0.2)..(0.7,0.0);
		\node at (0.75,-0.6) {$R$};
			\node at (1.5,-.8){UV};
			}}
	It also implies that the defects $P$ are projection defects, i.e. they are idempotent under fusion, $P\otimes P\cong P$. They project onto IR degrees of freedom in the UV theory.
	
Another consequence of (\ref{eq:rfuseprop}) is that 
right $R$-loops evaluate to the identity:
\aeq{\label{eq:loopcondition}\tikz[anchor=base, baseline=0]{
		\fill[WScolor light] (-1.6,-1.2) rectangle (1.4,1.2);
		\fill[WScolor] (0,0) ellipse (.8 and .8);
		\draw[opp arrow position = .5, opp arrow position = 0, defect] (0,0) ellipse (.8 and .8);
		\node at (-1.1,-.1) {$R$};
		\node at (.05,-.1) {UV};
		\node at (-1.25,-1.05) {IR};
	}=\tikz[anchor=base, baseline=0]{
		\fill[WScolor light] (-1.6,-1.2) rectangle (1.4,1.2);
		\node at (-1.25,-1.05) {IR};
	}}
	(Since the IR carries less information than the UV, the above loop-condition does not hold for left $R$-loops.)
This can be used to express correlation functions of the IR theory in terms of 
UV correlators by the following trick familiar from the discussion of dualities and generalized orbifolds
\cite{Frohlich:2006ch,Frohlich:2009gb,Carqueville:2012dk,Carqueville:2013mpa}:
Because of equation~(\ref{eq:loopcondition}), 
a given IR correlator is not changed upon insertion of right $R$ loops, c.f.~step I in the example \eqref{eq:IR to UV network} below. Since we are dealing with a topological quantum field theory, the UV islands created in this way can be expanded without changing the correlation function until they cover the entire space-time surface, see step II in \eqref{eq:IR to UV network}.
The result is a correlation function in the UV theory with a network of the projection defect $P$ inserted. For instance, a disk correlator in the IR with boundary condition $B_\text{IR}$ can be represented as UV correlator
		\aeq{\label{eq:IR to UV network}\langle{\tikz[anchor=base, baseline=0,scale=0.92]{
		\fill[color = WScolor light] (0,0) circle (1.5cm);
		\draw[black] (0,0) circle (1.5cm);
		\node[dot] at (-.4,.8) {};
		\node[dot] at (-.63,-.7){};
		\node[dot] at (.8,0) {};
		\node at (.1,1) {IR};
		\node at (-1.5,1.1) {$B_\text{IR}$};
	}}\;\;\;\rangle \stackrel{\text{I}}{=}\langle{\tikz[anchor=base, baseline=0,scale=0.92]{
		\fill[color = WScolor light] (0,0) circle (1.5cm);
		\draw[black] (0,0) circle (1.5cm);
		
		\fill[color = WScolor] (.6,.5) circle (.4cm);
		\draw[defect, opp arrow position=.5, opp arrow position=0] (.6,.5) circle (.4cm);
		\fill[color = WScolor] (-.4,0) circle (.53cm);
		\draw[defect, opp arrow position=.5, opp arrow position=0] (-.4,0) circle (.53cm);
		\fill[color = WScolor] (.6,-.6) circle (.4cm);
		\draw[defect, opp arrow position=.5, opp arrow position=0] (.6,-.6) circle (.4cm);
		\fill[color = WScolor] (-.3,-1) circle (.25cm);
		\draw[defect, opp arrow position=.5, opp arrow position=0] (-.3,-1) circle (.25cm);
		\node[dot] at (-.4,.8) {};
		\node[dot] at (-.63,-.7) {};
		\node[dot] at (.8,0.0) {};
		\node[DefectColor] at (-1.2,-.1) {$R$};
		\node at (-.4,-.1) {UV};
		\node at (.1,1) {IR};
		\node at (-1.5,1.1) {$B_\text{IR}$};
	}}\;\;\;\rangle \stackrel{\text{II}}{=} \langle{\tikz[anchor=base, baseline=0,scale=0.92]{
		\fill[color = WScolor] (0,0) circle (1.5cm);
		\draw[defect] (0,0) circle (1.5cm);
		\draw[symmetry defect] (.5,0) -- (-.86,1.22);
		\draw[symmetry defect] (.5,0) -- (1.5,0);
		\draw[symmetry defect] (.5,0) -- (-.2,-.6);
		\draw[symmetry defect] (-.2,-.6) -- (-1.21,-.86);
		\draw[symmetry defect] (-.2,-.6) -- (0,-1.5);
		\node[symmetry node] at (-.4,.8) {};
		\node[symmetry node] at (-.63,-.7) {};
		\node[symmetry node] at (.8,0.0) {};
		\node at (-.4,-.1) {UV};
		\node at (-1.5,1.1) {$B_\text{UV}$};
		\draw[densely dashed] (-.4,.8) circle (.2);
		\draw[densely dashed] (-.2,.9) -- (1.9,2);
		\begin{scope}[shift={(2,.6)},rotate=50]
			\clip (1,1) circle (.8);
			\fill[WScolor] (0,0) rectangle (2,2);
			\draw[symmetry defect] (1,0) -- (1,2);
			\fill[WScolor light] (1,1) circle (.5);
			\draw[defect, arrow position =.85] (1,1) circle (.5);
			\node[defect node] at (1,1) {};
		\end{scope}
		\draw[densely dashed] (1.9,2) circle (.8);
		\draw[densely dashed] (1.4,-.5) circle (.2);
		\draw[densely dashed] (1.5,-.67) -- (2.2,-1.7);
		\begin{scope}[shift={(2.2,-1.7)}]
			\clip (0,0) circle (.8);
			\fill[WScolor light] (-.5,-1) -- (.2,-1) -- (.5,1) -- (-.2,1) -- cycle;
			\fill[WScolor] (-.5,-1) -- (-.2,1) -- (-1,1) -- (-1,-1) -- cycle;
			\draw[black] (.2,-1) -- (.5,1);
			\draw[defect, arrow position =.6] (-.5,-1) -- (-.2,1);
		\end{scope}
		\draw[densely dashed] (2.2,-1.7) circle (.8);
	}}\!\!\!\!\!\!\!\!\!\!\!\!\!\!\!\!\!\!\rangle\phantom{*****}
	}
	Of course, steps I and II involve many choices leading to representations of one and the same IR
	 correlation function by possibly different $P$-networks in the UV.  The latter can be related by sequences of local transformations generated by 
 identities satisfied by the defects $P$ and their junctions.
	
	Carrying out this procedure on the level of correlators immediately reveals how objects of the IR theory are represented in the UV. For instance, 
	IR bulk fields appear as field insertions on the defect $P$. Right boundary conditions $B_\text{IR}$ are mapped to boundary conditions $B_\text{UV} = R^\dagger\otimes B_\text{IR}$ in the UV. Similarly, IR defects $D_\text{IR}$ are mapped to defects $D_\text{UV}=R^\dagger\otimes D_\text{IR}\otimes R$ in the UV. 
	This in particular applies to the defects associated to symmetries of the IR theory.
These symmetry defects encode the action of the symmetry group on all objects of the theory, and they fuse according to the multiplication in the symmetry group. Lifting IR symmetry defects to the UV one obtains UV defects, whose fusion is still governed by the IR symmetry group. This yields a realization of the IR symmetry group in the UV, which however is not an honest representation. After all, the identity defect in the IR corresponding to the neutral element in the IR symmetry group is lifted to the projection defect $P$ in the UV.
Thus, the lifted symmetries are only invertible on the IR degrees of freedom.
	
		In fact, given the projection defect $P$, the objects in the UV theory representing IR objects can be characterized completely within the UV theory without any reference to $R$:
	IR bulk fields are represented by defect fields on $P$, right IR boundary conditions are represented by right UV boundary conditions $B_\text{UV}$ which are invariant under fusion with $P$, $P\otimes B_\text{UV}\cong B_\text{UV}$. IR defects are represented by
defects $D_\text{UV}$ in the UV, which are invariant under fusion with $P$ from both sides, $D_\text{UV}\otimes P\cong D_\text{UV}\cong P \otimes D_\text{UV}$, etc. 
	Given the respective projection defect $P$, one can therefore completely describe the IR theory in the framework of the UV theory.
		
		Through perturbations by different relevant operators, a
				UV theory might permit many different RG flows leading to possibly different IR theories at various engery scales. All of these theories with all their symmetries etc. can be described by projection defects in one and the same UV theory. This applies in particular if the theory is asympotically free in the UV, in which case all possible IR theories can be realized by means of projection defects in a free theory.

		Remarkably,	the description of
	IR correlators in terms of UV correlators containing networks of the projection defect $P=R^\dagger\otimes R$  provides a radically new view on bulk perturbations: instead of perturbing the theory on the entire space-time, we can restrict the perturbation on a network of thin strips. 
These strips can even be made infinitely thin, effectively reducing the bulk perturbation to a (one-dimensional) perturbation of the identity defect in the UV theory.  	
RG flow leaves the bulk theory at the UV, but drives the identity defect to some projection defect $P$ in the IR.
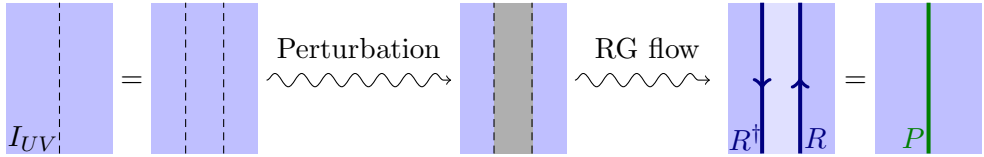
\uaeq{\tikz[baseline=25]{
		\fill[WScolor] (-.7,0) rectangle (.7,2);
		\node at (-.35,0.2) {$I_{UV}$};
		\draw[densely dashed] (0,0) -- (0,2);
	}=\tikz[baseline=25]{
		\fill[WScolor] (-.7,0) rectangle (.7,2);
		\draw[densely dashed] (-.25,0) -- (-.25,2);
		\draw[densely dashed] (.25,0) -- (.25,2);
	}\;\xrsquigarrow{Perturbation}\;\tikz[baseline=25]{
		\fill[WScolor] (-.7,0) rectangle (.7,2);
		\fill[color={rgb,255:red,175; green,175; blue,175}] (-.25,0) rectangle (.25,2);
		\draw[densely dashed] (-.25,0) -- (-.25,2);
		\draw[densely dashed] (.25,0) -- (.25,2);
	}\;\xrsquigarrow{ RG flow }\;\tikz[baseline=25]{
		\fill[WScolor] (-.7,0) rectangle (.7,2);
		\fill[WScolor light] (-.25,0) rectangle (.25,2);
		\draw[defect, opp arrow position=.5] (-.25,0) -- (-.25,2);
		\draw[defect, arrow position=.5] (.25,0) -- (.25,2);
		\node[defect] at (.45,0.2) {$R$};
		\node[defect] at (-.47,0.22) {$R^\dagger$};
	}=\tikz[baseline=25]{
		\fill[WScolor] (-.7,0) rectangle (.7,2);
		\draw[symmetry defect] (0,0) -- (0,2);
		\node[symmetry defect] at (-.2,0.2) {$P$}
	}}
	Concretely, one obtains the correlation function of the IR theory from the one at UV by first inserting an (invisible) network of the identity defect, which in particular passes through all bulk insertions and runs parallel to every boundary and also on both sides of any defect. The IR correlation function can then be obtained by a defect RG flow on this network.
	\aeq
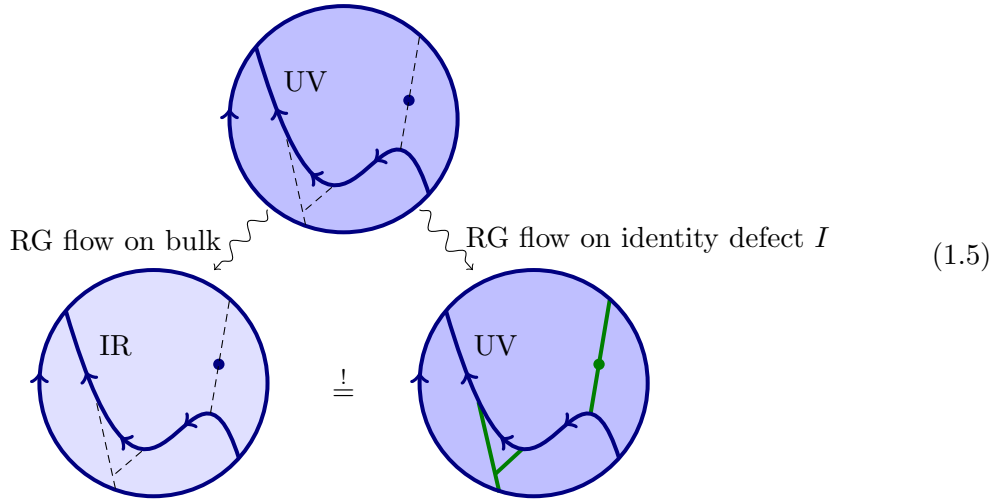
{\label{idflow}\tikz{
		\begin{scope}[scale=.5]
			\clip (-3.3,-3.3) rectangle (3.3,3.3);
			\fill[WScolor] (0,0) circle (3);
			\draw[defect, opp arrow position=.5] (0,0) circle (3);
				
			\draw[defect,opp arrow position=.75, opp arrow position=.5, opp arrow position=.25] (-2.3,1.9) .. controls (0,-6) and (1,2) .. (2.25,-2);
				
			\node at (-1,1) {UV};
			\node[defect node] at (1.72,.5) {};
			\draw[densely dashed] (1.5,-.8) -- (2,2.2);
			\draw[densely dashed] (-1,-2.8) -- (-1.5,-.5);
			\draw[densely dashed] (-1,-2.4) -- (-.3,-1.8);
		\end{scope}
		\begin{scope}[scale=.5,shift={(-5,-7)}]
			\clip (-3.3,-3.3) rectangle (3.3,3.3);
			\fill[WScolor light] (0,0) circle (3);
			\draw[defect, opp arrow position=.5] (0,0) circle (3);
			\draw[defect,opp arrow position=.75, opp arrow position=.5, opp arrow position=.25] (-2.3,1.9) .. controls (0,-6) and (1,2) .. (2.25,-2);
			\node at (-1,1) {IR};
			\node[defect node] at (1.72,.5) {};
			\draw[densely dashed] (1.5,-.8) -- (2,2.2);
			\draw[densely dashed] (-1,-2.8) -- (-1.5,-.5);
			\draw[densely dashed] (-1,-2.4) -- (-.3,-1.8);
		\end{scope}
		\begin{scope}[scale=.5,shift={(5,-7)}]
			\clip (-3.3,-3.3) rectangle (3.3,3.3);
			\fill[WScolor] (0,0) circle (3);
			\node[symmetry node] at (1.72,.5) {};
			\draw[symmetry defect] (1.5,-.8) -- (2,2.2);
			\draw[symmetry defect] (-.9,-2.9) -- (-1.45,-.5);
			\draw[symmetry defect] (-1,-2.4) -- (-.3,-1.75);
			\draw[defect, opp arrow position=.5] (0,0) circle (3);
			\draw[defect,opp arrow position=.75, opp arrow position=.5, opp arrow position=.25] (-2.3,1.9) .. controls (0,-6) and (1,2) .. (2.25,-2);
			\node at (-1,1) {UV};
		\end{scope}
		\draw [->,decorate,decoration=snake] (-1,-1.2) -- (-1.7,-2);
		\draw [->,decorate,decoration=snake] (1,-1.2) -- (1.7,-2);
		\node at (-3,-1.6) {RG flow on bulk};
		\node at (4,-1.6) {RG flow on identity defect $I$};
		\node at (0,-3.5) {$\stackrel{!}=$};
	}}
	Under the flow UV boundary conditions and defects flow to their respective fusion with $P$. In this way, bulk RG flows can be entirely studied in the fixed UV bulk theory by means of perturbations of the UV identity defect.

The fact that one can describe 		
IR theories in the UV without reference to the RG defects $R$ by using the respective projection defects $P=R^\dagger\otimes R$  suggests applying this procedure to general projection defects $P$, which do not a priori arise from RG flows. In this way, new `$P$-projected' theories can be constructed from  {\it any} projection defect $P$ in a given TQFT.

Interestingly,
it turns out that any projection defect factorizes as $P=R^\dagger\otimes R$ for 
some defect $R$ separating the $P$-projected theory from the original one. What is more, the defect $R$ satisfies the locality property of RG defects, i.e. the left equation of (\ref{eq:rfuseprop}).
Hence, in fact all projection defects factorize into RG type defects.

The construction described above is in fact closely related to the generalized orbifold procedure \cite{Frohlich:2009gb, Brunner:2013xna, Brunner:2013ota, Carqueville:2012dk}. It differs from it in that we drop  the technical assumption that left and right adjoints of defects agree $D^\dagger \cong {}^\dagger D$, which is in particular not satisfied in the  examples we present in this paper: the flows between orbifolds of Landau-Ginzburg models with a single chiral superfield  studied in \cite{Brunner:2007ur}.

The $\mathbb{Z}_d$-orbifold of the Landau-Ginzburg model with chiral superfield $X$ and superpotential $W(X)=X^d$, which we denote by $\cM_d$ admits RG flows to Landau-Ginzburg orbifolds $\cM_{d'}$ for all $d'<d$. Applying the procedure described above to these flows yields a realization
of all models $\cM_{d'}$ in terms of projection defects in $\cM_d$ for $d'<d$. In particular, taking $d\to\infty$ one obtains a representation of all models $\cM_{d'}$ in the theory of a free twisted chiral field. 

The paper is organized as follows: In section~\ref{Topological defects in Landau-Ginzburg orbifolds} we briefly introduce defect lines in 2d TQFTs, focussing on aspects which are important for our construction. The construction  is then spelled out in section~\ref{sec:rg networks in 2d tqfts}. Section~\ref{sec:rg networks in lg minimal model orbifolds} is devoted to a detailed discussion of the example of RG flows between Landau-Ginzburg orbifolds, in which the respective defects can be concretely described by means of matrix factorizations. We conclude with a discussion of open questions in section~\ref{sec:discussion}. Some more technical discussions and a brief outline of the generalized orbifold construction are relegated to various appendices.
	
\section{Defects in 2d TQFTs}
\label{Topological defects in Landau-Ginzburg orbifolds}

In this section, we set the stage by briefly introducing some aspects of defect lines in 2d topological quantum field theories (TQFTs). We do this mainly to introduce notation. For more details on defect lines, see e.g. \cite{Kapustin:2010ta,Davydov:2011kb,Carqueville:2016nqk}.

Defect lines are line operators, which (since they have codimension 1) can also separate different
2d TQFTs on the same space-time surface. Locally, a neighborhood around a point on a defect $D:T\rightarrow T'$
separating two TQFTs $T$ and $T'$ can be depicted as\footnote{Note that defect lines are oriented.}
\uaeq{ \tikz[anchor=base, baseline=\offset]{
		\worldsheetLights;
		\begin{scope}
			\clip(0,-1.1) rectangle (1.2,1);
			\worldsheets;
		\end{scope}
		\draw[defect, arrow position=.5] (0,-1) -- (0,1);
		\node at (-.3,-.2) {$D$};
		\node at (-.7,.2) {$T'$};
		\node at (.7,-.2) {$T$};
	}}
Defect lines carry local degrees of freedom, called defect fields, which can be inserted at points on defects. Defect fields can also separate different defects or glue together defects at junctions. We denote the space of defect changing fields between two defects $D,D':T\rightarrow T'$ by $\Hom(D,D')$. Every defect carries the identity field $1_D\in\Hom(D,D)$.
\uaeq
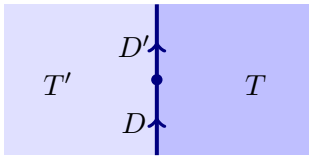
{ \tikz[anchor=base, baseline=\offset]{
		\fill[WScolor light] (-2,-1) rectangle (0,1);
		\fill[WScolor] (0,-1) rectangle (2,1);
		\draw[defect, arrow position=.25, arrow position=.75] (0,-1) -- (0,1);
		\node at (-.3,-.7) {$D$};
		\node[defect node] at (0,0) {};
		\node at (-.3,.3) {$D'$};
		\node at (-1.3,-.2) {$T'$};
		\node at (1.3,-.2) {$T$};
	}}
In every 2d TQFT $T$ there is a special invisible or identity defect $I_T$, whose insertion does not change correlation functions, and which can be connected to any other defect. The defect fields on this defect are just the bulk fields of the underlying 2d TQFT, $\Hom(I_T,I_T)\cong\mathcal{H}_T$.

Due to topological invariance, defects and field insertions can be moved on the space-time surface without changing correlation functions, as long as field insertions or defects do not cross. This in particular implies an associative operator product on defect fields $\Hom(D,D')\otimes\Hom(D',D'')\rightarrow\Hom(D,D'')$.

Similarly, parallel defect lines can be brought close together, leading to the notion of defect fusion: When brought close together, defects $D':T'\rightarrow T''$ and $D:T\rightarrow T'$ fuse to the defect $D'\otimes D:T\rightarrow T''$:
\uaeq
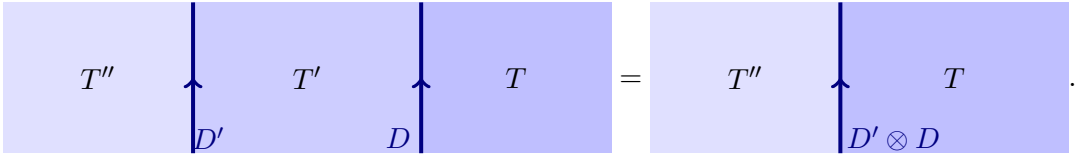
{\tikz[baseline=\offset]{
		\fill[WScolor light] (-4,-1) rectangle (-1.5,1);
		\fill[color={rgb,255:red,205; green,205; blue,255}] (-1.5,-1) rectangle (1.5,1);
		\fill[WScolor] (1.5,-1) rectangle (4,1);
		\draw[defect, arrow position=.5] (-1.5,-1) -- (-1.5,1);
		\draw[defect, arrow position=.5] (1.5,-1) -- (1.5,1);
		\node[DefectColor] at (-1.3,-.8) {$D'$};
		\node[DefectColor] at (1.2,-.8) {$D$};
		\node at (-2.75,0) {$T''$};
		\node at (0,0) {$T'$};
		\node at (2.75,0) {$T$};
	}=\tikz[baseline=\offset]{
		\fill[WScolor light] (-4,-1) rectangle (-1.5,1);
		\fill[WScolor] (-1.5,-1) rectangle (1.5,1);
		\draw[defect, arrow position=.5] (-1.5,-1) -- (-1.5,1);
		\node[DefectColor] at (-.8,-.8) {$D'\otimes D$};
		\node at (-2.75,0) {$T''$};
		\node at (0,0) {$T$};
	}.}
Topological invariance yields certain obvious compatibility relations between the operator product of defect fields and defect fusion, which we won't spell out here. We will however briefly mention one feature of topological defects, which will be of particular importance for our construction.

Due to topological invariance one can bend a defect (to the right or left) without changing correlators. This is described by the following two Zorro move identities (relations like this hold locally when inserted in any correlator):
	\aeq
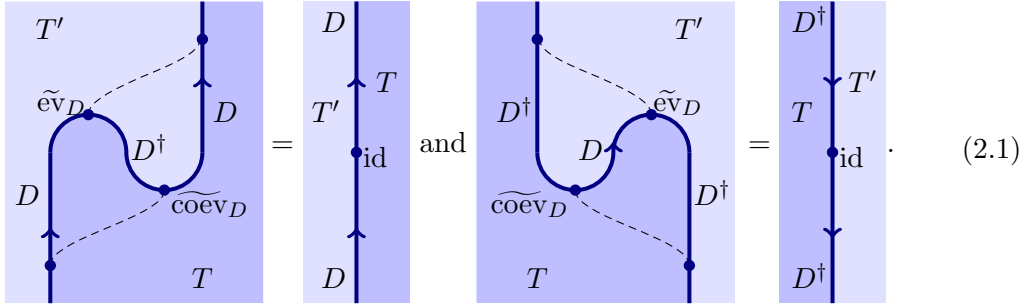
{\label{ZorroMoves2}\tikz[anchor=base, baseline=0pt]{
		\begin{scope} 
			\clip (-1,-2) -- (-1,0) arc(180:0:.5) arc(180:360:.5) -- (1,2) -- (-1.6,2) -- (-1.6,-2) -- cycle;
			\fill[WScolor light] (-1.6,-2) rectangle (1,2);
		\end{scope}
		\begin{scope} 
			\clip (-1,-2) -- (-1,0) arc(180:0:.5) arc(180:360:.5) -- (1,2) -- (1.8,2) -- (1.8,-2) -- cycle;
			\fill[WScolor] (-1,-2) rectangle (2,2);
		\end{scope}
		\draw[arrow position=.5, defect] (-1,-2) -- (-1,0);
		\draw[defect] (-1,0) arc(180:0:.5) arc(180:360:.5);
		\draw[arrow position=.5, defect] (1,0) -- (1,2);
		\draw[rotate=90, densely dashed] (.5,.5) cos (1,-.2) sin (1.5,-.97);
		\begin{scope}[shift={(-.5,-2)}]
			\draw[rotate=90, densely dashed] (.5,.5) cos (1,-.25) sin (1.5,-1);
		\end{scope}
		\node at (-1.3,-.7) {$D$};
		\node at (1.3,.4) {$D$};
		\node at (.3,-.1) {$D^\dagger$};
		\node[defect node] at (-.5,.5) {};
		\node at (-.85,.55) {$\widetilde\ev_D$};
		\node[defect node] at (-1,-1.5) {};
		\node[defect node] at (.5,-.5) {};
		\node at (1.1,-.8) {$\widetilde\coev_D$};
		\node[defect node] at (1,1.5) {};
		\node at (-1,1.5) {$T'$};
		\node at (1,-1.8) {$T$};
	} = \tikz[anchor=base, baseline=0pt]{
		\begin{scope} 
			\clip (-.5,-2) -- (-.5,2)  -- (-1.2,2) -- (-1.2,-2) -- cycle;
			\fill[WScolor light] (-1.2,-2) rectangle (-.5,2);
		\end{scope}
		\begin{scope} 
			\clip (-.5,-2) -- (-.5,2)  -- (.2,2) -- (.2,-2) -- cycle;
			\fill[WScolor] (-.5,-2) rectangle (.2,2);
		\end{scope}
		\draw[arrow position=.5, defect] (-.5,-2) -- (-.5,0);
		\draw[arrow position=.5, defect] (-.5,0) -- (-.5,2);
		\node at (-.8,-1.8) {$D$};
		\node at (-0.8,1.6) {$D$};
		\node at (-0.9,.4) {$T'$};
		\node at (-0.1,.8) {$T$};
		\node[defect node] at (-.5,0) {}; \node at (-0.25,-.2) {$\id$};
	}\text{ and }\tikz[anchor=base, baseline=0pt]{
		\begin{scope}[yscale=1,xscale=-1]
			\begin{scope} 
				\clip (-1,-2) -- (-1,0) arc(180:0:.5) arc(180:360:.5) -- (1,2) -- (-1.6,2) -- (-1.6,-2) -- cycle;
				\fill[WScolor light] (-1.6,-2) rectangle (1,2);
			\end{scope}
			\begin{scope} 
				\clip (-1,-2) -- (-1,0) arc(180:0:.5) arc(180:360:.5) -- (1,2) -- (1.8,2) -- (1.8,-2) -- cycle;
				\fill[WScolor] (-1,-2) rectangle (2,2);
			\end{scope}
			\draw[defect] (-1,-2) -- (-1,0);
			\draw[opp arrow position = .5, defect] (-1,0) arc(180:0:.5) arc(180:360:.5);
			\draw[defect] (1,0) -- (1,2);
			\draw[rotate=90, densely dashed] (.5,.5) cos (1,-.2) sin (1.5,-.97);
			\begin{scope}[shift={(-.5,-2)}]
				\draw[rotate=90, densely dashed] (.5,.5) cos (1,-.25) sin (1.5,-1);
			\end{scope}
			\node[defect node] at (-.5,.5) {};
			\node[defect node] at (-1,-1.5) {};
			\node[defect node] at (.5,-.5) {};
			\node[defect node] at (1,1.5) {};
			\node at (.3,-.1) {$D$};
			\node at (-1.3,-.7) {$D^\dagger$};
			\node at (1.3,.4) {$D^\dagger$};
			\node at (-1,1.5) {$T'$};
			\node at (1,-1.8) {$T$};
			\node at (1.1,-.8) {$\widetilde\coev_D$};
			\node at (-.85,.55) {$\widetilde\ev_D$};
		\end{scope}
	} = \tikz[anchor=base, baseline=0pt]{
		\begin{scope} 
			\clip (-.5,-2) -- (-.5,2)  -- (-1.2,2) -- (-1.2,-2) -- cycle;
			\fill[WScolor] (-1.2,-2) rectangle (-.5,2);
		\end{scope}
		\begin{scope} 
			\clip (-.5,-2) -- (-.5,2)  -- (.2,2) -- (.2,-2) -- cycle;
			\fill[WScolor light] (-.5,-2) rectangle (.2,2);
		\end{scope}
		\draw[opp arrow position=.5, defect] (-.5,-2) -- (-.5,0);
		\draw[opp arrow position=.5, defect] (-.5,0) -- (-.5,2);
		\node at (-.8,-1.8) {$D^\dagger$};
		\node at (-.8,1.6) {$D^\dagger$};
		\node at (-0.9,.4) {$T$};
		\node at (-0.1,.8) {$T'$};
		\node[defect node] at (-.5,0) {}; \node at (-0.25,-.2) {$\id$};
	}.}
These diagrams involve additional structure: First of all, bending $D$ to the right results in a downwards oriented version $D^\dagger$ of $D$, its right-adjoint.
Secondly, dotted lines depict the (invisible) identity defect $I$, which connects to the defects $D$ and $D^\dagger$ in defect (junction) fields
	\uaeq{
		\widetilde\ev_D &: D \otimes D^\dagger \rightarrow I_{T'} \\
		\widetilde\coev_D &: I_T \rightarrow D^\dagger \otimes D,
	}
	called evaluation and coevaluation maps, respectively. Of course, one can equally well bend the defect $D$ to the left
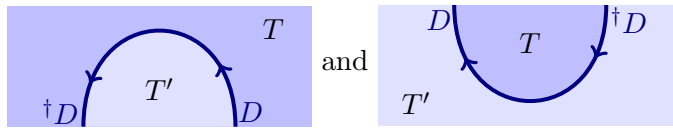
\uaeq{\tikz[baseline=-7]{
		\fill[WScolor] (-2,-1) rectangle (2,.6);
		\begin{scope}
			\clip (-1,-1) .. controls (-1,.7) and (1,.7) .. (1,-1) -- cycle;
			\fill[WScolor light] (-2,-1) rectangle (2,.6);
		\end{scope}
		\draw[defect, opp arrow position=.2, opp arrow position=.8] (-1,-1) .. controls (-1,.7) and (1,.7) .. (1,-1);
		\node[DefectColor] at (-1.3,-.8) {${}^\dagger D$};
		\node[DefectColor] at (1.2,-.8) {$D$};
		\node at (1.5,.3) {$T$};
		\node at (0,-.5) {$T'$};
	}\text{ and }\tikz[rotate = 180, baseline = 4]{
		\fill[WScolor light] (-2,-1) rectangle (2,.6);
		\begin{scope}
			\clip (-1,-1) .. controls (-1,.7) and (1,.7) .. (1,-1) -- cycle;
			\fill[WScolor] (-2,-1) rectangle (2,.6);
		\end{scope}
		\draw[defect, arrow position=.2, arrow position=.8] (-1,-1) .. controls (-1,.7) and (1,.7) .. (1,-1);
		\node[DefectColor] at (-1.3,-.8) {${}^\dagger D$};
		\node[DefectColor] at (1.2,-.8) {$D$};
		\node at (1.5,.3) {$T'$};
		\node at (0,-.5) {$T$};
	}}
	giving rise to the left-adjoint ${}^\dagger D$ of $D$. Topological invariance implies an analogous Zorro move identity involving $D$ and ${}^\dagger D$ and the respective (co-)evaluation maps
	\uaeq{
		\ev_D &: {}^\dagger D \otimes D \rightarrow I_{T} \\
		\coev_D &: I_{T'} \rightarrow D\otimes {}^\dagger D.
	}
Of course for all defects $D$, $\left({}^\dagger D\right)^\dagger \cong D \cong {}^\dagger\left(D^\dagger\right)$. For more details on adjunctions of defects, see \cite{Carqueville:2012st}.

In the main text of this paper, defect loops will play an important role. The description of loops of a defect $D$ in this framework requires a morphism $\phi\in\Hom({}^\dagger D,D^\dagger)$, by which such loops can be closed
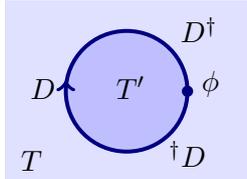
\uaeq{\tikz[anchor=base, baseline=0]{
		\fill[WScolor light] (-1.6,-1.2) rectangle (1.6,1.2);
		\fill[WScolor] (0,0) ellipse (.8 and .8);
		\draw[opp arrow position = .5, defect] (0,0) ellipse (.8 and .8);
		\node at (-1.1,-.1) {$D$};
		\node[defect node] at (.8,0) {};
		\node at (1.1,0) {$\phi$};
		\node at (.8,-1) {${}^\dagger D$};
		\node at (.95,.65) {$D^\dagger$};
		\node at (.05,-.1) {$T'$};
		\node at (-1.25,-1.05) {$T$};
	}.}
	For many
classes of 2d TQFTs (such as non-orbifold LG-models with an even number of chiral fields),
there is a canonical isomorphism $D^\dagger \cong {}^\dagger D$\footnote{or equivalently $D^{\dagger\dagger} \cong D$} for any defect $D$, which can be used for this purpose. This is not true in general, however. So we cannot resort to these canonical maps. Instead, we will
construct natural loop closing homomorphisms for the special class of defects which appear in our construction.

\section{RG-networks in 2d TQFTs}
\label{sec:rg networks in 2d tqfts}
\subsection{Projections from RG defects}
\label{sec:projfromRG}
Starting point of our construction are RG defects as defined in \cite{Brunner:2007ur}. These defects arise when 2d field theories are perturbed by local operators only on part of the space-time surface. The RG flow drives the theory to the IR on the perturbation domain, while leaving it at the UV on the rest, thus creating a defect on the boundary of the perturbation domain separating the IR from the UV theory as in \eqref{eq:Pert and flow}.
This RG defect encodes all aspects of the relationship between UV and IR theories.\footnote{For instance, in \cite{Brunner:2007ur} RG-defects are used to describe how boundary conditions behave under perturbations of the bulk theory.} 

Next, we pass to the context of 2d topological quantum field theories via the  topological twist. Indeed, we assume that the 2d QFT under consideration allows for a topological twist which moreover is compatible with the RG flow.\footnote{Examples of such theories are $2d$ $N=(2,2)$ superconformal field theories perturbed by chiral or twisted chiral fields.}
Then the RG defects descend to topological defects between topologically twisted IR and UV theories, which we will still refer to as RG defects.

Arising from local perturbations, RG defects have rather special properties. Locality postulates that
perturbations on two adjacent domains is nothing but the perturbation on the union of the domains. This implies that fusion of an RG defect $R$ with its opposite defect $R^\dagger$ in the UV theory yields the identity defect $I_\text{IR}$ in the IR:
	\uaeq{
	\tikz[baseline=\offset]{
		\fill[WScolor light] (-3,-1) rectangle (-1.5,1);
		\fill[WScolor] (-1.5,-1) rectangle (1.5,1);
		\fill[WScolor light] (1.5,-1) rectangle (3,1);
		\draw[defect, arrow position=.5] (-1.5,-1) -- (-1.5,1);
		\draw[defect, opp arrow position=.5] (1.5,-1) -- (1.5,1);
		\node[DefectColor] at (-1.3,-.8) {$R$};
		\node[DefectColor] at (1.2,-.8) {$R^\dagger$};
		\node at (-2.25,0) {IR};
		\node at (0,0) {UV};
		\node at (2.25,0) {IR};
	}=\tikz[baseline=\offset]{
		\fill[WScolor light] (-2,-1) rectangle (2,1);
		\draw[densely dashed, arrow position=.5] (0,-1) -- (0,1);
		\node at (.4,-.8) {$I_\text{IR}$};
		\node at (-1,0) {IR};
		\node at (1,0) {IR};
	}.} In other words, there is an isomorphism $R \otimes R^\dagger \xrightarrow{\cong} I_\text{IR}$ to the identity defect in the IR theory, which together with its inverse yields the following relations:
\aeq{\label{eq:RRop}\tikz[anchor=base, baseline=0]{
		\fill[WScolor light] (-1.6,-1.2) rectangle (1.4,1.2);
		\fill[WScolor] (0,0) ellipse (.8 and .8);
		\draw[densely dashed] (0,.8) -- (0,1.2);
		\node at (.2,.85) {$\cong$};
		\draw[densely dashed] (0,-.8) -- (0,-1.2);
		\node at (.2,-1.12) {$\cong$};
		\draw[opp arrow position = .5, opp arrow position = 0, defect] (0,0) ellipse (.8 and .8);
		\node at (-1.1,-.1) {$R$};
		\node at (1.15,-.1) {$R^\dagger$};
		\node at (.05,-.1) {UV};
		\node at (-1.25,-1.05) {IR};
	}=\tikz[anchor=base, baseline=0]{
		\fill[WScolor light] (-1,-1.2) rectangle (1,1.2);
		\node at (-.75,-1.05) {IR};
		\draw[densely dashed] (0,-1.2) -- (0,1.2);
	}\quad\text{and}\quad\tikz[baseline=0]{
		\fill[WScolor light] (-1.6,-1.2) rectangle (1.4,1.2);
		\node at (-1.25,-1.05) {IR};
		\draw[densely dashed] (0,-.5) -- (0,.5);
		\fill[WScolor] (-.7,1.2) arc(180:360:.7)--cycle;
		\draw[defect,opp arrow position=.3] (-.7,1.2) arc(180:360:.7);
		\fill[WScolor] (-.7,-1.2) arc(180:0:.7)--cycle;
		\draw[defect,arrow position=.3] (-.7,-1.2) arc(180:0:.7);
		\node at (-.8,-.8) {$R$};
		\node at (.8,-.6) {$R^\dagger$};
		\node at (-.8,.6) {$R$};
		\node at (.8,.6) {$R^\dagger$};
		\node at (0,.75) {$\cong$};
		\node at (0,-.8) {$\cong$};
	}=\tikz[baseline=0]{
		\fill[WScolor light] (-1.6,-1.2) rectangle (1.4,1.2);
		\node at (-1.25,-1.05) {IR};
		\fill[WScolor] (-.7,-1.2) rectangle (.7,1.2);
		\draw[defect,arrow position=.5] (.7,1.2) -- (.7,-1.2);
		\draw[defect,arrow position=.5] (-.7,-1.2) -- (-.7,1.2);
		\node at (-.9,.8) {$R$};
		\node at (1.05,.8) {$R^\dagger$};
	}}
	Since $I_\text{IR}$ is self-adjoint, $R\otimes R^\dagger\cong I_\text{IR}$ is equivalent to $R\otimes {}^\dagger R\cong I_\text{IR}$, and similar relations hold for ${}^\dagger R$.
	
	One important consequence of this is that one can close right $R$ loops in a way that makes them evaluate to the identity. (This is not true for left $R$ loops, i.e.~those enclosing the IR theory, which are not invertible in the case of non-trivial RG flows.) The loop-closing morphism $\phi:{}^\dagger R\rightarrow R^\dagger$ is obtained by moving the isomorphisms in the first relation in (\ref{eq:RRop}) along the defect to the right:
		\uaeq{\tikz[anchor=base, baseline=0]{
		\fill[WScolor light] (-1.4,-1.2) rectangle (1.4,1.2);
		\fill[WScolor] (0,0) ellipse (.8 and .8);
		\draw[densely dashed] (0,.8) -- (0,1.2);
		\node at (.2,.85) {$\cong$};
		\draw[densely dashed] (0,-.8) -- (0,-1.2);
		\node at (.2,-1.12) {$\cong$};
		\draw[opp arrow position = .5, opp arrow position = 0, defect] (0,0) ellipse (.8 and .8);
		\node at (-1.1,-.1) {$R$};
		\node at (1.15,-.1) {$R^\dagger$};
		\node at (.05,-.1) {UV};
		\node at (-1,-1.05) {IR};
	}=
		\tikz[anchor=base, baseline=0]{
		\fill[WScolor light] (-1.4,-1.2) rectangle (1.4,1.2);
		\fill[WScolor] (0,0) ellipse (.8 and .8);
		\draw[densely dashed] (0,.8) -- (0,1.2);
		\node at (.2,.85) {};
		\node at (0.7,0.7) {$R^\dagger$};
		\node at (0.5, -1.1) {${}^\dagger R$};
		\draw[densely dashed] (0,-.8) -- (0,-1.2);
		\node at (.2,-1.12) {};
		\draw[opp arrow position = .5, opp arrow position = 0, defect] (0,0) ellipse (.8 and .8);
		\node at (-1.1,-.1) {$R$};
		\node at (1.1,-.5) {};
		\node at (.05,-.1) {UV};
		\node at (-1,-1.05) {IR};
		\node[defect node] at (0.7,-0.4){};
		\node at (1,-0.5){$\phi$}
	}		\quad\text{with}\quad
		\phi:=\tikz[anchor=base,baseline=-30]{
		\fill[WScolor light] (-2,-1.3) rectangle (1.5,1.5);
		\fill[WScolor] (-2,-1.3)--(0,-1.3)--(0,.3)..controls(-.4,-1) and (-1,-1) ..(-1,1.5)--(-2,1.5)--cycle;
		\draw[defect, opp arrow position = .5, opp arrow position=.7, opp arrow position=.2] (0,-1.3)--(0,.3)..controls(-.4,-1) and (-1,-1) .. (-1,1.5);
		\node[defect node] at (0,.35) {};
		\node at (0.2,.5) {$\cong$};
		\draw[densely dashed] (0,.35) -- (-1,1);
		\node[DefectColor] at (.3,-1.1) {${}^\dagger R$};
		\node[DefectColor] at (-1.3,1) {$R^\dagger$};
		\node[DefectColor] at (-.3,0) {$R$};
		\node at (1,1) {IR};
		\node at (-.3,.7) {$I$};
		\node at (-1.5,-1) {UV};
		\begin{scope}[yscale=-1,shift={(0,2)}]
			\fill[WScolor light] (-2,-1.3) rectangle (1.5,1.5);
			\fill[WScolor] (-2,-1.3)--(0,-1.3)--(0,.3)..controls(-.4,-1) and (-1,-1) ..(-1,1.5)--(-2,1.5)--cycle;
			\draw[defect, arrow position = .5, arrow position=.7] (0,-1.3)--(0,.3)..controls(-.4,-1) and (-1,-1) .. (-1,1.5);
			\node[defect node] at (0,.35) {};
			\node at (0.2,.7) {$\cong$};
			\draw[densely dashed] (0,.35) -- (-1,1);
			\node[DefectColor] at (.4,-.7) {$R^\dagger$};
			\node[DefectColor] at (-1.3,1) {${}^\dagger R$};
			\node[DefectColor] at (-.4,0) {$R$};
		\end{scope}
	}:{}^\dagger R\longrightarrow R^\dagger}
	
Utilizing that right $R$ loops evaluate to the identity it is possible to express correlation functions of the IR theory
in terms of correlation functions in the UV by the following trick: Given a correlation function of the IR theory,
one can insert right $R$ loops without changing it. Expanding these islands of UV theory until they cover the entire surface, the IR correlation function is transformed into a correlation function of the UV theory with a network of defects as in equation~\eqref{eq:IR to UV network}.

	The network is built out of the defect $P:=R^\dagger\otimes R$ (in the following represented by green lines which we take as upwards oriented if an orientation is not specified) 
	and its junctions
	\uaeq
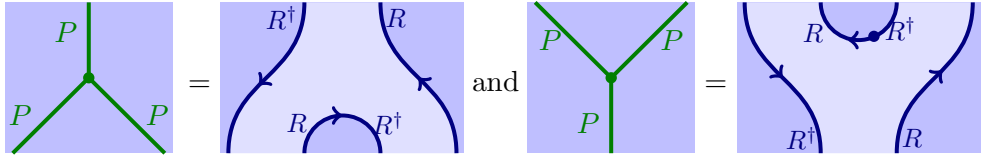
{\label{eq:(co)multiplication}\tikz[baseline=-5]{
		\fill[WScolor] (-1.1,-1) rectangle (1.1,1);
		\draw[symmetry defect] (-1,-1) -- (0,0);
		\draw[symmetry defect] (1,-1) -- (0,0);
		\draw[symmetry defect] (0,1) -- (0,0);
		\node[symmetry node] at (0,0) {};
		\node[symmetry defect] at (-.9,-.5) {$P$};
		\node[symmetry defect] at (.9,-.5) {$P$};
		\node[symmetry defect] at (-.3,.6) {$P$};
	}\text{ = }\tikz[baseline=-5]{
		\fill[WScolor] (-1.6,-1) rectangle (1.6,1);
		\fill[WScolor light] (-1.5,-1) .. controls (-1.5,0) and (-.5,0) .. (-.5,1) -- (.5,1) .. controls (.5,0) and (1.5,0) .. (1.5,-1) -- (.5,-1) arc (0:180:.5) -- cycle;
		\draw[defect, opp arrow position=.5] (-1.5,-1) .. controls (-1.5,0) and (-.5,0) .. (-.5,1);
		\draw[defect, arrow position=.5] (1.5,-1) .. controls (1.5,0) and (.5,0) .. (.5,1);
		\draw[defect, arrow position=.5] (-.5,-1) arc (180:0:.5);
		\node[defect] at (-.8,.8) {\small$R^{\dagger}$};
		\node[defect] at (.6,-.6) {\small$R^{\dagger}$};
		\node[defect] at (.7,.8) {\small$R$};
		\node[defect] at (-.6,-.6) {\small$R$};
	}\;\text{and}\;\tikz[baseline=-5,yscale=-1]{
		\fill[WScolor] (-1.1,-1) rectangle (1.1,1);
		\draw[symmetry defect] (-1,-1) -- (0,0);
		\draw[symmetry defect] (1,-1) -- (0,0);
		\draw[symmetry defect] (0,1) -- (0,0);
		\node[symmetry node] at (0,0) {};
		\node[symmetry defect] at (-.8,-.5) {$P$};
		\node[symmetry defect] at (.8,-.5) {$P$};
		\node[symmetry defect] at (-.3,.6) {$P$};
	}\text{ = }\tikz[baseline=-5,yscale=-1]{
		\fill[WScolor] (-1.6,-1) rectangle (1.6,1);
		\fill[WScolor light] (-1.5,-1) .. controls (-1.5,0) and (-.5,0) .. (-.5,1) -- (.5,1) .. controls (.5,0) and (1.5,0) .. (1.5,-1) -- (.5,-1) arc (0:180:.5) -- cycle;
		\draw[defect, arrow position=.5] (-1.5,-1) .. controls (-1.5,0) and (-.5,0) .. (-.5,1);
		\draw[defect, opp arrow position=.5] (1.5,-1) .. controls (1.5,0) and (.5,0) .. (.5,1);
		\draw[defect, opp arrow position=.5] (-.5,-1) arc (180:0:.5);
		\node[defect] at (-.75,.8) {\small$R^{\dagger}$};
		\node[defect] at (.55,-.65) {\small$R^{\dagger}$};
		\node[defect] at (.7,.8) {\small$R$};
		\node[defect] at (-.6,-.6) {\small$R$};
		\node[defect node] at (.2,-.55) {};
	}}
which we call multiplication and comultiplication, respectively.

The defect $P$ together with its junctions has some rather special features, which easily follow from the properties of $R$. In particular, $P\otimes P\cong P$, and
the following relations hold:
	\uaeq
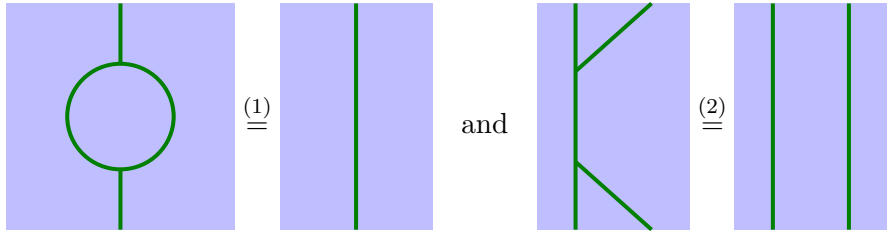
{\tikz[baseline=-6]{
		\fill[WScolor] (-1.5,-1.5) rectangle (1.5,1.5);
		\draw[symmetry defect] (0,-1.5) -- (0,-.7);
		\draw[symmetry defect] (0,-.7) arc(270:630:.7);
		\draw[symmetry defect] (0,1.5) -- (0,.7);
	} \stackrel{(1)}= \tikz[baseline=-6]{
		\fill[WScolor] (-1,-1.5) rectangle (1,1.5);
		\draw[symmetry defect] (0,-1.5) -- (0,1.5);
	} \quad\text{and}\quad \tikz[baseline=-6]{
		\fill[WScolor] (-1,-1.5) rectangle (1,1.5);
		\draw[symmetry defect] (-.5,-1.5) -- (-.5,1.5);
		\draw[symmetry defect] (-.5,.6) -- (.5,1.5);
		\draw[symmetry defect] (-.5,-.6) -- (.5,-1.5);
	} \stackrel{(2)}{=}\tikz[baseline=-6]{
		\fill[WScolor] (-1,-1.5) rectangle (1,1.5);
		\draw[symmetry defect] (-.5,-1.5) -- (-.5,1.5);
		\draw[symmetry defect] (.5,-1.5) -- (.5,1.5);
	}}
	We call the first one \emph{loop-omission property} (or \emph{separability}) and  the second one \emph{projection property}. Beyond these, $P$ also obeys the 
	 following identities:
		\uaeq{\text{associativity:}\;\;
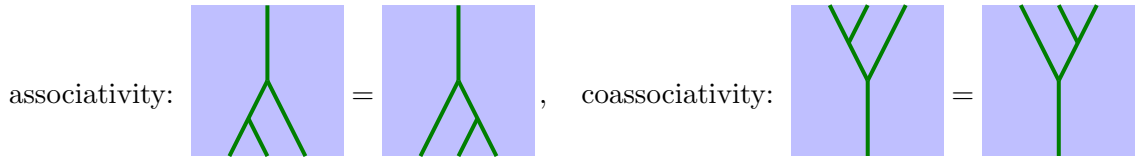
\tikz[baseline=20]{
		\fill[WScolor] (0,0) rectangle (2,2);
		\draw[symmetry defect] (.5,0) -- (1,1);
		\draw[symmetry defect] (1,1) -- (1,2);
		\draw[symmetry defect] (1,0) -- (.75,.5);
		\draw[symmetry defect] (1.5,0) -- (1,1);
	}=\tikz[baseline=20,xscale=-1]{
		\fill[WScolor] (0,0) rectangle (2,2);
		\draw[symmetry defect] (.5,0) -- (1,1);
		\draw[symmetry defect] (1,1) -- (1,2);
		\draw[symmetry defect] (1,0) -- (.75,.5);
		\draw[symmetry defect] (1.5,0) -- (1,1);
	}\,,\quad\text{coassociativity:}\;\;\tikz[baseline=-37,yscale=-1]{
		\fill[WScolor] (0,0) rectangle (2,2);
		\draw[symmetry defect] (.5,0) -- (1,1);
		\draw[symmetry defect] (1,1) -- (1,2);
		\draw[symmetry defect] (1,0) -- (.75,.5);
		\draw[symmetry defect] (1.5,0) -- (1,1);
	}=\tikz[baseline=-37,xscale=-1,yscale=-1]{
		\fill[WScolor] (0,0) rectangle (2,2);
		\draw[symmetry defect] (.5,0) -- (1,1);
		\draw[symmetry defect] (1,1) -- (1,2);
		\draw[symmetry defect] (1,0) -- (.75,.5);
		\draw[symmetry defect] (1.5,0) -- (1,1);
	}}
	\uaeq{\text{and the Frobenius identities:}\quad
		
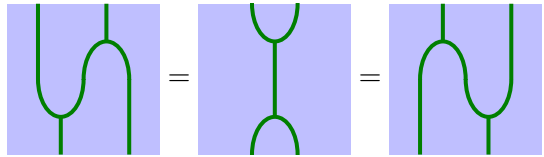
\tikz[anchor=base, baseline=12]{
			\fill[WScolor] (-1,-.5) rectangle (1,1.5);
			\begin{scope}[yscale=-1,xscale=1, shift={(-.3,-.5)}]
				\clip (-.4,0) rectangle (.4,.6);
				\draw[symmetry defect] (0,0) ellipse (.3 and .5);
			\end{scope}
			\begin{scope}[shift={(.3,.5)}]
				\clip (-.4,0) rectangle (.4,.6);
				\draw[symmetry defect] (0,0) ellipse (.3 and .5);
			\end{scope}
			\draw[symmetry defect] (.6,-.5) -- (.6,.5);
			\draw[symmetry defect] (.3,1) -- (.3,1.5);
			\draw[symmetry defect] (-.6,.5) --(-.6, 1.5);
			\draw[symmetry defect] (-.3,0) --(-.3, -.5);
		}
		=
		\tikz[anchor=base, baseline=12]{
			\fill[WScolor] (-1,-.5) rectangle (1,1.5);
			\begin{scope}[yscale=-1,xscale=1, shift={(0,-1.5)}]
				\clip (-.4,0) rectangle (.4,.6);
				\draw[symmetry defect] (0,0) ellipse (.3 and .5);
			\end{scope}
			\draw[symmetry defect] (0,0) -- (0,1);
			\begin{scope}[shift={(0,-.5)}]
				\clip (-.4,0) rectangle (.4,.6);
				\draw[symmetry defect] (0,0) ellipse (.3 and .5);
			\end{scope}
		}
		=
		\tikz[anchor=base, baseline=12]{\begin{scope}[yscale=1,xscale=-1]
			\fill[WScolor] (-1,-.5) rectangle (1,1.5);
			\begin{scope}[yscale=-1,xscale=1, shift={(-.3,-.5)}]
				\clip (-.4,0) rectangle (.4,.6);
				\draw[symmetry defect] (0,0) ellipse (.3 and .5);
			\end{scope}
			\begin{scope}[shift={(.3,.5)}]
				\clip (-.4,0) rectangle (.4,.6);
				\draw[symmetry defect] (0,0) ellipse (.3 and .5);
			\end{scope}
			\draw[symmetry defect] (.6,-.5) -- (.6,.5);
			\draw[symmetry defect] (.3,1) -- (.3,1.5);
			\draw[symmetry defect] (-.6,.5) --(-.6, 1.5);
			\draw[symmetry defect] (-.3,0) --(-.3, -.5);
		\end{scope}
		}		
			}
Moreover, $P$ comes with a unit 
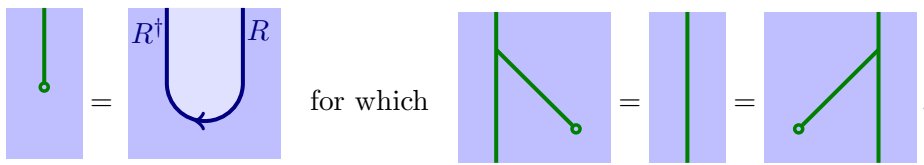
\uaeq{\tikz[baseline=-10]{
		\fill[WScolor] (-.5,-1) rectangle (.5,1);
		\draw[symmetry defect] (0,1) -- (0,0);
		\node[unit node] at (0,-.05) {};
	} = \tikz[baseline=-10]{
		\fill[WScolor] (-1,-1) rectangle (1,1);
		\fill[WScolor light] (-.5,1) -- (-.5,0) arc(180:360:.5) -- (.5,1) -- cycle;
		\draw[defect, opp arrow position=.5] (-.5,1) -- (-.5,0) arc(180:360:.5) -- (.5,1);
		\node[defect] at (-.75,.7) {$R^{\dagger}$};
		\node[defect] at (.7,.7) {$R$};
	}\quad\text{for which}\quad
		\tikz[baseline=20]{
		\fill[WScolor] (0,0) rectangle (2,2);
		\draw[symmetry defect] (.5,0) -- (.5,2);
		\draw[symmetry defect] (.5,1.5) -- (1.5,.5);
		\node[unit node] at (1.55,.45) {};
	}=\tikz[baseline=20]{
		\fill[WScolor] (0,0) rectangle (1,2);
		\draw[symmetry defect] (.5,0) -- (.5,2);
	}=\tikz[baseline=20,xscale=-1]{
		\fill[WScolor] (0,0) rectangle (2,2);
		\draw[symmetry defect] (.5,0) -- (.5,2);
		\draw[symmetry defect] (.5,1.5) -- (1.5,.5);
		\node[unit node] at (1.55,.45) {};
	}\,.}
	Indeed, instead of $P=R^\dagger\otimes R$ we could just as well have chosen $P^\prime={}^\dagger P={}^\dagger R\otimes R$ as building block of the network above. The latter defect equally satisfies the relations above with the only difference that instead of a unit, it has a counit
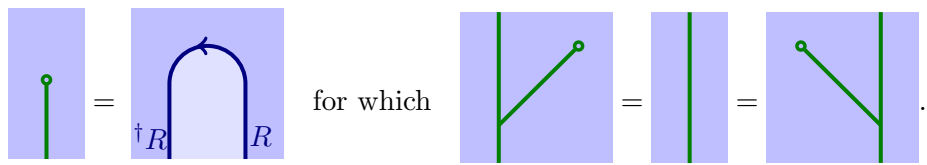
\uaeq{\tikz[baseline=-10,yscale=-1]{
		\fill[WScolor] (-.5,-1) rectangle (.5,1);
		\draw[symmetry defect] (0,1) -- (0,0);
		\node[unit node] at (0,-.05) {};
	} = \tikz[baseline=-10,yscale=-1]{
		\fill[WScolor] (-1,-1) rectangle (1,1);
		\fill[WScolor light] (-.5,1) -- (-.5,0) arc(180:360:.5) -- (.5,1) -- cycle;
		\draw[defect, opp arrow position=.5] (-.5,1) -- (-.5,0) arc(180:360:.5) -- (.5,1);
		\node[defect] at (-.75,.7) {${}^\dagger R$};
		\node[defect] at (.7,.7) {$R$};
	}\quad\text{for which}\quad
	\tikz[baseline=-37,yscale=-1]{
		\fill[WScolor] (0,0) rectangle (2,2);
		\draw[symmetry defect] (.5,0) -- (.5,2);
		\draw[symmetry defect] (.5,1.5) -- (1.5,.5);
		\node[unit node] at (1.55,.45) {};
	}=\tikz[baseline=20]{
		\fill[WScolor] (0,0) rectangle (1,2);
		\draw[symmetry defect] (.5,0) -- (.5,2);
	}=\tikz[baseline=-37,xscale=-1,yscale=-1]{
		\fill[WScolor] (0,0) rectangle (2,2);
		\draw[symmetry defect] (.5,0) -- (.5,2);
		\draw[symmetry defect] (.5,1.5) -- (1.5,.5);
		\node[unit node] at (1.55,.45) {};
	}.	
	}
	In summary, any correlation function of the IR theory  can be written as a correlator in the UV with a $P$-network inserted. The correlation function is invariant under local changes of the $P$-network generated by loop-omission and projection properties, the associativity and coassociativity relations and the Frobenius identities. 
This in particular reflects the fact that the resulting correlation function does not depend on how exactly the UV islands are inserted into the IR correlators and how they are expanded.

		\subsection{Representing the IR in the UV}
	\label{sec:Representing the IR in the UV}

	Having expressed the IR correlators in terms of UV correlators in the last section, we now discuss how
		the defining structures of IR correlators such as bulk fields, boundaries, defects and symmetries are represented in the UV theory. The results can be summarized as follows: 	If one characterizes the respective IR object by its relation to the IR identity defect $I_\text{IR}$, then its UV realization is obtained by replacing the IR identity defect by the defect $P$ of the UV theory, c.f. Table \ref{tab:8 to 6}. For simplicity we will restrict the discussion to the case of unital projection defect $P=R^\dagger\otimes R$. The results are the same for the counital case, and the argument is similar.

\begin{table}[t]
\begin{tabular}{c|c|c}
\thead{IR object} & \thead{IR realization} & \thead{UV realization} \\
\hline
Identity defect & \makecell{Separable Frobenius algebra $I_\text{IR}$ \\
	\tikz[yscale=.5]{
		\fill[WScolor light] (0,0) rectangle (2,2);
		\draw[densely dashed] (1,0) -- (1,2);
	}}
	& \makecell{(Co-)unital projection defect $P$ \\
	\tikz[yscale=.5]{
		\fill[WScolor] (0,0) rectangle (2,2);
		\draw[symmetry defect] (1,0) -- (1,2);
	}} \\
\hline
IR bulk fields & \makecell{$I_\text{IR}$-bimodule morphisms of $I_\text{IR}$\\
	\tikz{
		\fill[WScolor light] (0,0) rectangle (2,2);
		\draw[densely dashed] (1,1) -- (1,2);
		\draw[densely dashed] (.5,0) -- (1,1);
		\draw[densely dashed] (1.5,0) -- (1,1);
		\node[defect node] at (.75,.5) {};
	}}
	& \makecell{$P$-bimodule morphisms of $P$\\
	\tikz{
		\fill[WScolor] (0,0) rectangle (2,2);
		\draw[symmetry defect] (1,1) -- (1,2);
		\draw[symmetry defect] (.5,0) -- (1,1);
		\draw[symmetry defect] (1.5,0) -- (1,1);
		\node[symmetry node] at (.75,.5) {};
	}} \\
\hline
\begin{tabular}{c}
Left boundary\\ conditions\end{tabular} & \makecell{Right $I_\text{IR}$-modules \\
	\tikz[yscale=.5]{
		\fill[WScolor light] (0,0) rectangle (2,2);
		\draw[densely dashed] (1,0) -- (0,1);
		\draw[defect] (0,0) -- (0,2);
	}} & \makecell{Right $P$-modules \\
	\tikz[yscale=.5]{
		\fill[WScolor] (0,0) rectangle (2,2);
		\draw[symmetry defect] (1,0) -- (0,1);
		\draw[defect] (0,0) -- (0,2);
	}} \\
\hline
\begin{tabular}{c}
Right boundary\\
conditions\end{tabular} & \makecell{Left $I_\text{IR}$-modules \\
	\tikz[xscale=-1,yscale=.5]{
		\fill[WScolor light] (0,0) rectangle (2,2);
		\draw[densely dashed] (1,0) -- (0,1);
		\draw[defect] (0,0) -- (0,2);
	}} & \makecell{Left $P$-modules \\
	\tikz[xscale=-1,yscale=.5]{
		\fill[WScolor] (0,0) rectangle (2,2);
		\draw[symmetry defect] (1,0) -- (0,1);
		\draw[defect] (0,0) -- (0,2);
	}} \\
\hline
Defects & \makecell{$I_\text{IR}$-bimodules \\
	\tikz[xscale=-1]{
		\fill[WScolor light] (0,0) rectangle (2,2);
		\draw[densely dashed] (1.5,0) -- (1,1.5);
		\draw[densely dashed] (.5,0) -- (1,1);
		\draw[defect] (1,0) -- (1,2);
	}}
	& \makecell{$P$-bimodules \\
	\tikz[xscale=-1]{
		\fill[WScolor] (0,0) rectangle (2,2);
		\draw[symmetry defect] (1.5,0) -- (1,1.5);
		\draw[symmetry defect] (.5,0) -- (1,1);
		\draw[defect] (1,0) -- (1,2);
	}} \\
\hline
\begin{tabular}{c}
Defect changing\\ fields\end{tabular} & \makecell{$I_\text{IR}$-bimodule morphisms \\
	\tikz[xscale=-1]{
		\fill[WScolor light] (0,0) rectangle (2,2);
		\draw[densely dashed] (1.5,0) -- (1,1.5);
		\draw[densely dashed] (.5,0) -- (1,.5);
		\draw[defect] (1,0) -- (1,2);
		\node[defect node] at (1,1) {};
	}} & \makecell{$P$-bimodule morphisms \\
	\tikz[xscale=-1]{
		\fill[WScolor] (0,0) rectangle (2,2);
		\draw[symmetry defect] (1.5,0) -- (1,1.5);
		\draw[symmetry defect] (.5,0) -- (1,.5);
		\draw[defect] (1,0) -- (1,2);
		\node[defect node] at (1,1) {};
	}}
	\\
	\hline
	Defect fusion & \makecell{fusion in the IR\\$D\otimes D^\prime$} &\makecell{fusion of UV lifted defects\\$D_\text{UV}\otimes D_\text{UV}^\prime$}\\
	\hline
	Adjunction & \makecell{$D^\dagger$\\ ${}^\dagger D$}& \makecell{$D^{\dagger_P}_\text{UV}=P\otimes D^\dagger_\text{UV}\otimes P$\\ ${}^{\dagger_P}\!D_\text{UV}=P\otimes {}^\dagger D_\text{UV}\otimes P$}
		\end{tabular}
		\caption{Dictionary of IR structures lifted into the UV.}\label{tab:8 to 6}
\end{table}

		
\subsubsection*{IR bulk fields}		
Let us first discuss bulk fields of the IR theory.
Upon expanding the UV islands in the IR, bulk fields become defect fields on $P$, i.e. elements in $\Hom(P,P)$ (represented in diagrams by dots on defects). Due to topological invariance, they have to be compatible with the multiplication on $P$.
Namely,
	\uaeq
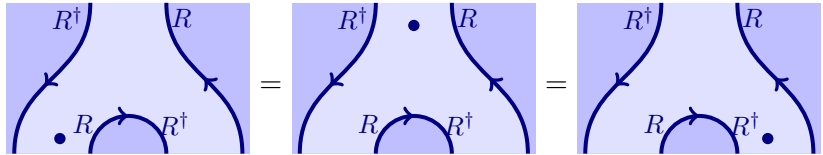
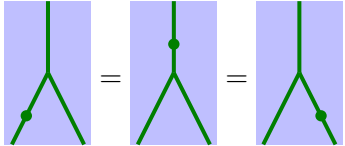
{\tikz[baseline=-5]{
		\fill[WScolor] (-1.6,-1) rectangle (1.6,1);
		\fill[WScolor light] (-1.5,-1) .. controls (-1.5,0) and (-.5,0) .. (-.5,1) -- (.5,1) .. controls (.5,0) and (1.5,0) .. (1.5,-1) -- (.5,-1) arc (0:180:.5) -- cycle;
		\draw[defect, opp arrow position=.5] (-1.5,-1) .. controls (-1.5,0) and (-.5,0) .. (-.5,1);
		\draw[defect, arrow position=.5] (1.5,-1) .. controls (1.5,0) and (.5,0) .. (.5,1);
		\draw[defect, arrow position=.5] (-.5,-1) arc (180:0:.5);
		\node[defect node] at (-.9,-.8) {};
		\node[defect] at (.7,.8) {\small$R$};
		\node[defect] at (-.6,-.6) {\small$R$};
		\node[defect] at (-.8,.8) {\small$R^{\dagger}$};
		\node[defect] at (.6,-.6) {\small$R^{\dagger}$};
	}\text{ = }\tikz[baseline=-5]{
		\fill[WScolor] (-1.6,-1) rectangle (1.6,1);
		\fill[WScolor light] (-1.5,-1) .. controls (-1.5,0) and (-.5,0) .. (-.5,1) -- (.5,1) .. controls (.5,0) and (1.5,0) .. (1.5,-1) -- (.5,-1) arc (0:180:.5) -- cycle;
		\draw[defect, opp arrow position=.5] (-1.5,-1) .. controls (-1.5,0) and (-.5,0) .. (-.5,1);
		\draw[defect, arrow position=.5] (1.5,-1) .. controls (1.5,0) and (.5,0) .. (.5,1);
		\draw[defect, arrow position=.5] (-.5,-1) arc (180:0:.5);
		\node[defect node] at (0,0.7) {};
		\node[defect] at (.7,.8) {\small$R$};
		\node[defect] at (-.6,-.6) {\small$R$};
		\node[defect] at (-.8,.8) {\small$R^{\dagger}$};
		\node[defect] at (.6,-.6) {\small$R^{\dagger}$};
	}\text{ = }\tikz[baseline=-5]{
		\fill[WScolor] (-1.6,-1) rectangle (1.6,1);
		\fill[WScolor light] (-1.5,-1) .. controls (-1.5,0) and (-.5,0) .. (-.5,1) -- (.5,1) .. controls (.5,0) and (1.5,0) .. (1.5,-1) -- (.5,-1) arc (0:180:.5) -- cycle;
		\draw[defect, opp arrow position=.5] (-1.5,-1) .. controls (-1.5,0) and (-.5,0) .. (-.5,1);
		\draw[defect, arrow position=.5] (1.5,-1) .. controls (1.5,0) and (.5,0) .. (.5,1);
		\draw[defect, arrow position=.5] (-.5,-1) arc (180:0:.5);
		\node[defect node] at (.9,-.8) {};
		\node[defect] at (.7,.8) {\small$R$};
		\node[defect] at (-.6,-.6) {\small$R$};
		\node[defect] at (-.8,.8) {\small$R^{\dagger}$};
		\node[defect] at (.6,-.6) {\small$R^{\dagger}$};
	}}
implying
\uaeq{\tikz[baseline=-5,scale=0.95]{
		\fill[WScolor] (-.6,-1) rectangle (.6,1);
		\draw[symmetry defect] (-.5,-1) -- (0,0);
		\draw[symmetry defect] (.5,-1) -- (0,0);
		\draw[symmetry defect] (0,1) -- (0,0);
		\node[symmetry node] at (-.3,-.6) {};
	}=\tikz[baseline=-5,scale=0.95]{
		\fill[WScolor] (-.6,-1) rectangle (.6,1);
		\draw[symmetry defect] (-.5,-1) -- (0,0);
		\draw[symmetry defect] (.5,-1) -- (0,0);
		\draw[symmetry defect] (0,1) -- (0,0);
		\node[symmetry node] at (0,.4) {};
	}=\tikz[baseline=-5,scale=0.95]{
		\fill[WScolor] (-.6,-1) rectangle (.6,1);
		\draw[symmetry defect] (-.5,-1) -- (0,0);
		\draw[symmetry defect] (.5,-1) -- (0,0);
		\draw[symmetry defect] (0,1) -- (0,0);
		\node[symmetry node] at (.3,-.6) {};
	}}
	Considering the algebra $P$ as $P$-bimodule, the IR bulk fields become $P$-bimodule morphisms of $P$ in the UV. By the same argument these morphisms also respect the $P$-comodule structure on $P$:
\uaeq
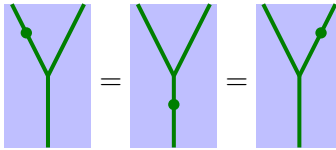
{\tikz[baseline=-5,yscale=-1,scale=0.95]{
		\fill[WScolor] (-.6,-1) rectangle (.6,1);
		\draw[symmetry defect] (-.5,-1) -- (0,0);
		\draw[symmetry defect] (.5,-1) -- (0,0);
		\draw[symmetry defect] (0,1) -- (0,0);
		\node[symmetry node] at (-.3,-.6) {};
	}=\tikz[baseline=-5,yscale=-1,scale=0.95]{
		\fill[WScolor] (-.6,-1) rectangle (.6,1);
		\draw[symmetry defect] (-.5,-1) -- (0,0);
		\draw[symmetry defect] (.5,-1) -- (0,0);
		\draw[symmetry defect] (0,1) -- (0,0);
		\node[symmetry node] at (0,.4) {};
	}=\tikz[baseline=-5,yscale=-1,scale=0.95]{
		\fill[WScolor] (-.6,-1) rectangle (.6,1);
		\draw[symmetry defect] (-.5,-1) -- (0,0);
		\draw[symmetry defect] (.5,-1) -- (0,0);
		\draw[symmetry defect] (0,1) -- (0,0);
		\node[symmetry node] at (.3,-.6) {};
	}}
Now, not only are IR bulk fields lifted to $P$-bimodule morphisms of $P$ in the UV, the Hilbert space of bulk fields of the IR theory is in fact isomorphic to the space of $P$-bimodule morphisms of $P$. More precisely,
the map
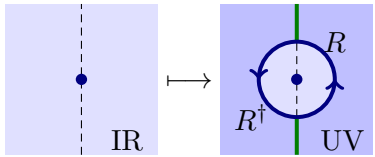
\uaeq{\tikz[baseline=25]{
		\fill[WScolor light] (0,0) rectangle (2,2);
		\draw[densely dashed] (1,0) -- (1,2);
		\node[defect node] at (1,1) {};
		\node at (1.6,0.2){IR};
	}\longmapsto\tikz[baseline=25]{
		\fill[WScolor] (0,0) rectangle (2,2);
		\draw[symmetry defect] (1,0) -- (1,2);
		\fill[WScolor light] (1,1) circle (.5);
		\draw[densely dashed] (1,.5) -- (1,1.5);
		\draw[defect, arrow position=0, arrow position=0.5] (1,1) circle (.5);
		\node[defect node] at (1,1) {};
		\node at (1.5,1.5){$R$};
		\node at (0.4,0.5){$R^\dagger$};
		\node at (1.6,0.2){UV};
	}}
sending IR bulk fields to $P$-bimodule morphisms of $P$ is an isomorphism. This is spelled out in appendix~\ref{sec:IR-UV correspondence}. 

In fact, due to the special properties of $P$, all morphisms of $P$, i.e. all defect fields on $P$ are automatically $P$-bimodule morphisms of $P$ and at the same time also $P$-bicomodule morphisms of $P$, see appendix~\ref{app:Algebra equal coalgebra morphisms}.
Thus, the IR bulk Hilbert space is isomorphic to the space of defect fields on $P$.

\subsubsection*{IR boundary conditions and defects}
Next, let us discuss left IR boundary conditions. Upon inserting and expanding UV islands in the IR theory, a left IR boundary condition $B_\text{IR}$ is lifted to the  UV boundary condition  $B_\text{UV}:=B_\text{IR}\otimes R$. The latter comes equipped with a map
	\uaeq{\begin{array}{c}B_{UV}\otimes P\rightarrow B_{UV}\\
	
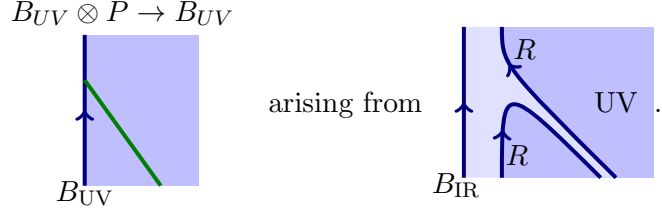
\tikz[baseline=25]{
		\fill[WScolor] (0,0) rectangle (1.5,2);
		\draw[defect, arrow position = .5] (0,0) -- (0,2);
		\node at (0,-.1) {$B_{\text{UV}}$};
		\draw[symmetry defect] (0,1.4) -- (1,0);
	}\end{array}\quad\text{arising from}
	\tikz[baseline=25]{
		\fill[WScolor] (.5,0) rectangle (2.5,2);
		\fill[WScolor light] (0,0) -- (.5,0)..controls (.5,1.3) .. (1.8,0) -- (2,0) ..controls (.5,1.5).. (.5,2) -- (0,2) -- cycle;
		\draw[defect, arrow position = .5] (0,0) -- (0,2);
		\draw[defect, arrow position = .25] (.5,0) ..controls (.5,1.3) .. (1.8,0);
		\draw[defect, opp arrow position = .25] (.5,2) ..controls (.5,1.5).. (2,0);
		\node at (-.1,-.1) {$B_{\text{IR}}$};
		\node at (.7,.3) {$R$};
		\node at (.8,1.7) {$R$};
		\node at (2,1) {UV};
	}.
	}
	It satisfies the identities
	\uaeq
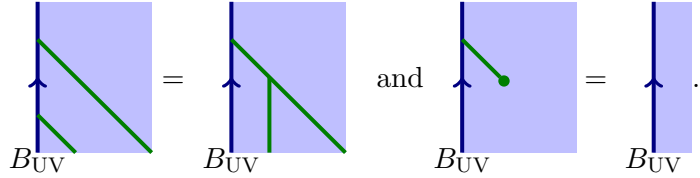
{\tikz[baseline=25]{
		\fill[WScolor] (0,0) rectangle (1.5,2);
		\draw[defect, arrow position = .5] (0,0) -- (0,2);
		\node at (0,-.1) {$B_{\text{UV}}$};
		\draw[symmetry defect] (0,1.5) -- (1.5,0);
		\draw[symmetry defect] (0,.5) -- (.5,0);
	}=\tikz[baseline=25]{
		\fill[WScolor] (0,0) rectangle (1.5,2);
		\draw[defect, arrow position = .5] (0,0) -- (0,2);
		\node at (0,-.1) {$B_{\text{UV}}$};
		\draw[symmetry defect] (0,1.5) -- (1.5,0);
		\draw[symmetry defect] (.5,1) -- (.5,0);
	}\quad\text{and}\tikz[baseline=25]{
		\fill[WScolor] (0,0) rectangle (1.5,2);
		\draw[defect, arrow position = .5] (0,0) -- (0,2);
		\node at (0,-.1) {$B_{\text{UV}}$};
		\draw[symmetry defect] (0,1.5) -- (.5,1);
		\node[symmetry node] at (.55,.95) {};
	}=\tikz[baseline=25]{
		\fill[WScolor] (0,0) rectangle (.5,2);
		\draw[defect, arrow position = .5] (0,0) -- (0,2);
		\node at (0,-.1) {$B_{\text{UV}}$};
	}.}
	In other words, $B_\text{UV}$ is a right $P$-module. In fact, the unit of $P$ induces a $P$-comodule structure on any $P$-module, hence also on $B_\text{UV}$:
	\uaeq
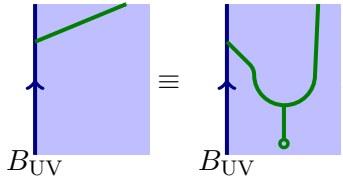
{\tikz[baseline=25]{
		\fill[WScolor] (0,0) rectangle (1.5,2);
		\draw[defect, arrow position = .5] (0,0) -- (0,2);
		\node at (0,-.1) {$B_{\text{UV}}$};
		\draw[symmetry defect] (0,1.5) -- (1.2,2);
	}\equiv\tikz[baseline=25]{
		\fill[WScolor] (0,0) rectangle (1.5,2);
		\draw[defect, arrow position = .5] (0,0) -- (0,2);
		\node at (0,-.1) {$B_{\text{UV}}$};
		\draw[symmetry defect] (0,1.5) -- (.3,1.2) arc(45:0:.2) arc(180:360:.4) -- (1.2,2);
		\draw[symmetry defect] (.75,.65) -- (.75,.2);
		\node[unit node] at (.75,.15) {};
	}.}
Therefore, left IR boundary conditions lift to right $P$-modules in the UV, which automatically are also $P$-comodules.

Conversely, all right $P$-modules arise in this way from IR boundary conditions. To see this, note that due to the special properties of the defect $P$, a left UV boundary condition $B$ is a right $P$-module iff $B\cong B\otimes P$ as shown in appendix \ref{sec:rephrasing P modules}. Hence, given a right $P$-module $B$, the IR boundary condition $B_\text{IR}=B\otimes R^\dagger$ satisfies $B_\text{IR}\otimes R=B\otimes R^\dagger\otimes R=B\otimes P\cong B$. Thus, left IR boundary conditions are in one-to-one correspondence with right $P$-modules in the UV.\footnote{Indeed, this also holds if one chooses to construct the network using $P^\prime={}^\dagger R\otimes R$ instead of $P=R^\dagger\otimes R$. In that case however $B_{UV}=B_{IR}\otimes{}^\dagger R$ inherits a natural $P^\prime$-comodule structure, which by means of the counit on $P^\prime$ also induces a $P^\prime$-module structure on $B_{UV}$.}

		Analogously one finds that right IR boundary conditions $B_\text{IR}$ lift to left $P$-modules $B_\text{UV}=R^\dagger\otimes B_\text{IR}$ in the UV,
		and defects $D_\text{IR}$ of the IR theory lift to $P$-bimodules $D_{UV}=R^\dagger\otimes D_\text{IR}\otimes R$.
	Importantly, $P$ itself is the UV lift of the IR identity defect:
	\uaeq
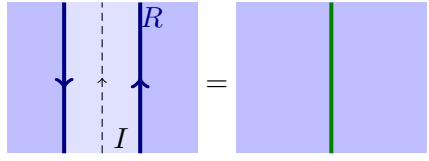
{\tikz[baseline=-5]{
		\fill[WScolor] (-2.5,-1) rectangle (0,1);
		\fill[WScolor light] (-1.75,-1) rectangle (-.75,1);
		\draw[defect,opp arrow position=.5] (-1.75,-1) -- (-1.75,1);
		\draw[defect,arrow position=.5] (-.75,-1) -- (-.75,1);
		\node[defect] at (-.6,.8) {$R$};
		\draw[densely dashed, arrow position = .5] (-1.25,-1) -- (-1.25,1);
		\node at (-1,-.8) {$I$};
	}=\tikz[baseline=-5]{
		\fill[WScolor] (-2.5,-1) rectangle (0,1);
		\draw[symmetry defect] (-1.25,-1) -- (-1.25,1);
	}}
	
	A straight-forward generalization of the discussion of IR bulk fields shows that IR defect fields are lifted to bimodule morphisms of the respective UV lifted defects, which again due to the special properties of $P$  are nothing but the defect fields of the UV lifts. 
	
	\subsubsection*{Fusion of IR defects}
	
Because of $R\otimes R^\dagger\cong I_\text{IR}$, the lift of fused IR defects is the fusion of the lifted defects:
	\uaeq
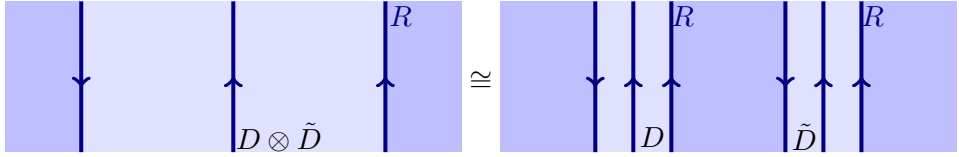
{\tikz[baseline=-5]{
		\fill[WScolor] (-3,-1) rectangle (3,1);
		\fill[WScolor light] (-2,-1) rectangle (2,1);
		\draw[defect,opp arrow position=.5] (-2,-1) -- (-2,1);
		\draw[defect,arrow position=.5] (2,-1) -- (2,1);
		\node[defect] at (2.2,.8) {$R$};
		\draw[defect, arrow position=.5] (0,-1) -- (0,1);
		\node at (.6,-.8) {$D\otimes\tilde D$};
	} \cong \tikz[baseline=-5]{
		\fill[WScolor] (-3,-1) rectangle (3,1);
		\fill[WScolor light] (-1.75,-1) rectangle (-.75,1);
		\fill[WScolor light] (.75,-1) rectangle (1.75,1);
		\draw[defect,opp arrow position=.5] (-1.75,-1) -- (-1.75,1);
		\draw[defect,arrow position=.5] (-.75,-1) -- (-.75,1);
		\draw[defect,opp arrow position=.5] (.75,-1) -- (.75,1);
		\draw[defect,arrow position=.5] (1.75,-1) -- (1.75,1);
		\node[defect] at (-.6,.8) {$R$};
		\node[defect] at (1.9,.8) {$R$};
		\draw[defect, arrow position = .5] (-1.25,-1) -- (-1.25,1);
		\draw[defect, arrow position = .5] (1.25,-1) -- (1.25,1);
		\node at (-1,-.8) {$D$};
		\node at (1,-.8) {$\tilde D$};
	}}
	This is a rather special property closely tied to the projection property of $P$.
	
	\subsubsection*{Adjunction of IR defects}	
	While fusion of defects in the IR lifts to fusion in the UV, adjunction is not compatible with the lift from IR to UV.
		If for instance, an IR defect $D_\text{IR}$ is lifted to a defect $D_\text{UV}=R^\dagger\otimes D_\text{IR}\otimes R$ in the IR, then the right adjoint of the latter in the UV theory is given by
	$D_\text{UV}^\dagger=R^\dagger\otimes D_\text{IR}^\dagger\otimes R^{\dagger\dagger}$, which in general does not coincide with the lift $R^\dagger\otimes D_\text{IR}^\dagger\otimes R$ of the right adjoint of $D_\text{IR}$ to the UV theory. However, the two are related: Selfadjointness of the IR identity defect yields $R^{\dagger\dagger}\otimes R^\dagger\cong I_\text{IR}$, and hence 
	the UV lift of the adjoint can be expressed as  $R^\dagger\otimes D_\text{IR}^\dagger\otimes R\cong D_\text{UV}^\dagger\otimes P\cong P\otimes D_\text{UV}^\dagger\otimes P$, leading to the notion of IR adjunction in the UV theory, which we denote by
$$
D_\text{UV}^{\dagger_P}:=P\otimes D_\text{UV}^\dagger\otimes P\,.
$$
Similarly, the UV lift of a left adjoint defect is given by\footnote{The same formulas for left and right IR adjunction hold if one chooses to construct the network using the counital $P^\prime={}^\dagger R\otimes R$ instead of $P=R^\dagger\otimes R$.}
$$
{}^{\dagger_P}\!D_\text{UV}=P\otimes {}^\dagger D_\text{UV}\otimes P\,.
$$
These formulas are very natural. After all, the defining relation of adjoints are the Zorro move identities (\ref{ZorroMoves2}), which involve the identity defect. Lifting these identities from the IR theory to the UV replaces the identity defect with the defect $P$:
\uaeq
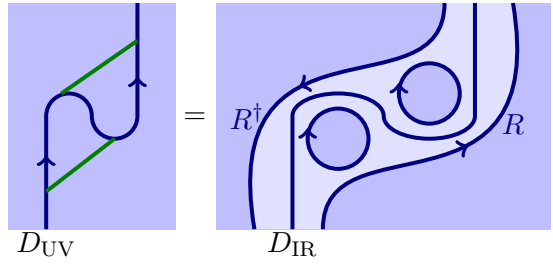
{\label{eq:adj of orb}\tikz[baseline=40]{
		\fill[WScolor] (0,0) rectangle (2.2,3);
		\draw[defect, arrow position=.2, arrow position = .8] (.5,0) -- (.5,1.5) arc(180:0:.3) arc(180:360:.3) -- (1.7,3);
		\draw[symmetry defect] (.7,1.8) -- (1.7,2.5);
		\draw[symmetry defect] (.5,.5) -- (1.4,1.2);
		\node at (.5,-.2) {$D_\text{UV}$};
	}=\tikz[xscale=2,baseline=40]{
		\fill[WScolor] (0,0) rectangle (2.2,3);
		\fill[WScolor light] (.25,0) .. controls (0,3) and (1.5,1.5) .. (1.5,3) -- (1.95,3) .. controls (2.2,0) and (.7,1.5) .. (.7,0) -- cycle;
		\draw[defect] (.5,0) -- (.5,1.5) arc(180:0:.3) arc(180:360:.3) -- (1.7,3);
		\fill[WScolor] (.8,1.2) ellipse (.2 and .4);
		\fill[WScolor] (1.4,1.8) ellipse (.2 and .4);
		\draw[defect, opp arrow position =.5] (.8,1.2) ellipse (.2 and .4);
		\draw[defect, opp arrow position =.5] (1.4,1.8) ellipse (.2 and .4);
		\draw[defect,opp arrow position =.5] (.25,0) .. controls (0,3) and (1.5,1.5) .. (1.5,3);
		\draw[defect,opp arrow position =.5] (1.95,3) .. controls (2.2,0) and (.7,1.5) .. (.7,0);
		\node at (.5,-.2) {$D_\text{IR}$};
		\node[defect] at (.2,1.5) {$R^\dagger$};
		\node[defect] at (1.95,1.4) {$R$};
	}}
For instance, lifting the IR Zorro move identities for the right adjoint to the UV results in the relations
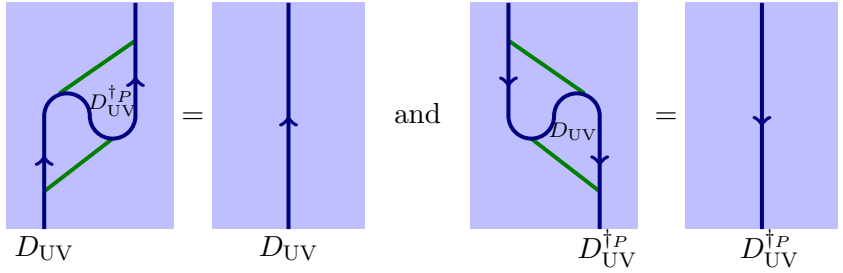
\uaeq{\tikz[baseline=40]{
		\fill[WScolor] (0,0) rectangle (2.2,3);
		\draw[symmetry defect] (.7,1.8) -- (1.7,2.5);
		\draw[symmetry defect] (.5,.5) -- (1.4,1.2);
		\draw[defect, arrow position=.2, arrow position = .8] (.5,0) -- (.5,1.5) arc(180:0:.3) arc(180:360:.3) -- (1.7,3);
		\node at (.5,-.25) {$D_\text{UV}$};
		\node at (1.4,1.7){${\scriptstyle D_\text{UV}^{\dagger_P}}$};
	}=\tikz[baseline=40]{
		\fill[WScolor] (-.5,0) rectangle (1.5,3);
		\draw[defect, arrow position=.5] (.5,0) -- (.5,3);
		\node at (.5,-.25) {$D_\text{UV}$};
	}\quad\text{and}\quad\tikz[baseline=40,xscale=-1]{
		\fill[WScolor] (0,0) rectangle (2.2,3);
		\draw[symmetry defect] (.7,1.8) -- (1.7,2.5);
		\draw[symmetry defect] (.5,.5) -- (1.4,1.2);
		\draw[defect, opp arrow position=.2, opp arrow position = .8] (.5,0) -- (.5,1.5) arc(180:0:.3) arc(180:360:.3) -- (1.7,3);
		\node at (.4,-.25) {$D_\text{UV}^{\dagger_P}$};
		\node at (0.85,1.3){${\scriptstyle D_\text{UV}}$};
	}=\tikz[baseline=40,xscale=-1]{
		\fill[WScolor] (-.5,0) rectangle (1.5,3);
		\draw[defect, opp arrow position=.5] (.5,0) -- (.5,3);
		\node at (.4,-.25) {$D_\text{UV}^{\dagger_P}$};
	}\,.}
It is easy to see that fusing the UV adjoint from both sides with $P$  yields a defect which satisfies the $P$-Zorro move identities, c.f.~appendix~\ref{app:P-adjoints}.	

	A special case is $P$ itself: Since it is the UV lift of the IR identity defect, which is selfadjoint, $P$ is selfadjoint with respect to $P$-adjunction: $P\cong P^{\dagger_P}=P\otimes P^\dagger\otimes P$.
		
		\subsubsection*{IR symmetries}

	Also symmetries of the IR theory can be easily described in the UV. As noted in \cite{Frohlich:2006ch} (see also \cite{Brunner:2007qu}), symmetries of 2d field theories can be described by symmetry defects ${}_gI$ which describe the action of an element $g$ of the symmetry group on any object in field theory. The symmetry defects fuse according to the multiplication in the symmetry group: ${}_gI\otimes {}_hI={}_{g\cdot h}I$ ($g\cdot h$ denotes
the product in the symmetry group).	
	Now, IR symmetry defects lift to the UV as any other defect: ${}_gI\mapsto {}_gI_\text{UV}=R^\dagger\otimes {}_gI\otimes R$. Since IR fusion lifts to UV fusion, the fusion of the lifted symmetry defects still respects the multiplication in the symmetry group, ${}_gI_\text{UV}\otimes {}_hI_\text{UV}={}_{g\cdot h}I_\text{UV}$. In that sense, the IR symmetry group is already present in the UV theory. 
However it is not realized as a symmetry group in the UV, since the lift of the IR identity defect, which is the symmetry defect associated to the neutral element of the symmetry group, does not lift to the identity defect, but rather to $P$. So the lifted symmetry defects are in general not invertible defects in the UV, but instead satisfy $
{}_gI_\text{UV}\otimes {}_{g^{-1}}I_\text{UV}=P$.

\subsubsection*{IR projectors and subsequent flows}

Projection defects $P_2=(R_2)^\dagger \otimes R_2$ in the IR theory associated to some RG flow from the IR theory to some theory $\text{IR}_2$ can also be lifted to the UV. The corresponding defects in the UV theory are given by
\uaeq{\widetilde{P}=R^\dagger \otimes P_2\otimes R=R^\dagger\otimes R_2^\dagger\otimes R_2\otimes R
=(R_2\otimes R)^\dagger\otimes (R_2\otimes R)\,.}
These are precisely the projection defects built out of the RG defect $R_2\otimes R$ associated to the concatenation of RG flows from the UV via IR to theory $\text{IR}_2$.

\subsubsection*{IR correlation functions}

Having described how to realize IR objects inside the UV theory, it is straight-forward to represent IR correlators in the UV theory:
First, prepare the IR correlator by placing identity defects  through all field insertions, in particular bulk fields and at defect cusps. Then replace all IR objects by the respective UV objects as described above. Importantly, this includes the IR identity defect, which has to be replaced by the UV projection defect $P$. The resulting UV correlator coincides with the original IR correlator.
	\uaeq{\braket
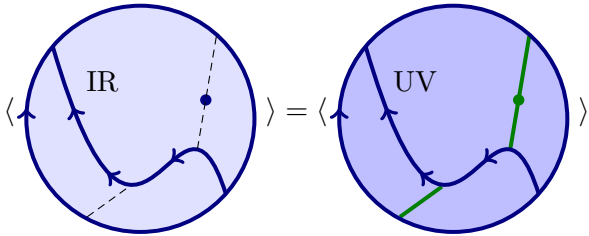
{\tikz[baseline=0]{\begin{scope}[scale=.5]
		\clip (-3.3,-3.3) rectangle (3.3,3.3);
		\fill[WScolor light] (0,0) circle (3);
		\draw[defect, opp arrow position=.5] (0,0) circle (3);
		\draw[defect,opp arrow position=.75, opp arrow position=.5, opp arrow position=.25] (-2.3,1.9) .. controls (0,-6) and (1,2) .. (2.25,-2);
		\node at (-1,1) {IR};
		\node[defect node] at (1.72,.5) {};
		\draw[densely dashed] (1.5,-.8) -- (2,2.2);
		\draw[densely dashed] (-1.4,-2.6) -- (-.3,-1.8);
		\end{scope}
	}}=\braket{\tikz[baseline=0]{\begin{scope}[scale=.5]
		\clip (-3.3,-3.3) rectangle (3.3,3.3);
		\fill[WScolor] (0,0) circle (3);
		\node[symmetry node] at (1.72,.5) {};
		\draw[symmetry defect] (1.5,-.8) -- (2,2.2);
		\draw[symmetry defect] (-1.4,-2.6) -- (-.3,-1.8);
		\draw[defect, opp arrow position=.5] (0,0) circle (3);
		\draw[defect,opp arrow position=.75, opp arrow position=.5, opp arrow position=.25] (-2.3,1.9) .. controls (0,-6) and (1,2) .. (2.25,-2);
		\node at (-1,1) {UV};
	\end{scope}}}
	}

	\subsection{Bulk RG flow as defect flow}
	
	The previous discussion suggests a radically new view on bulk RG flow. Namely, that bulk perturbations of a 2d theory can be understood as a perturbation of a defect network in the fixed UV bulk theory. More precisely, insertion and expansion of UV islands in the perturbed theory confines the perturbation on ever smaller domains, which eventually become one-dimensional. Hence,  perturbed correlation functions are nothing but UV correlation functions with networks of perturbed identity defects inserted. RG flow then does not change the bulk UV theory, but only drives the identity defect in the UV to some projection defect $P$, c.f.~(\ref{idflow}).
		
			The two-dimensional RG flow in the bulk can hence be reduced to a one-dimensional RG flow on the identity defect $I_\text{UV}$. Such defect flows are of course much easier to handle, because the underlying bulk theory does not change. For instance, UV bulk fields ($I_\text{UV}$-endomorphisms) and boundaries ($I_\text{UV}$-modules) flow to bulk fields and boundaries in the UV theory, which are compatible with $P$, i.e. to $P$-bimodule morphisms of $P$ and $P$-modules, respectively.
	
		Thus, if one can get a handle on perturbations of the identity defect in a given TQFT, the structures (bulk space, boundaries, correlators, etc.) associated to the corresponding perturbed bulk theory 
		can be easily extracted. 
			
	\subsection{IR theories from projections}
	\label{sec:ir theories from projections}
	
	In the previous discussion, we represented correlation functions of a perturbed 2d TQFT as correlation functions of the unperturbed UV theory with a defect network inserted. While the starting point of the construction were RG defects $R$, the correlation functions of the perturbed theory only depended on the projection defect $P=R^\dagger\otimes R$. This suggests applying this method to arbitrary unital or counital projection defects $P$, which 
have the same properties as the defects associated to RG flows discussed in section~\ref{sec:projfromRG}: The projection property, $P\otimes P\cong P$ means that there are two junctions  
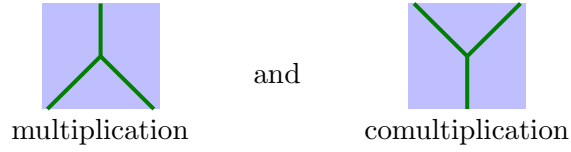
\footnote{As before, $P$ is depicted in green, oriented from bottom to top.}
	\uaeq{\begin{array}{c}{\tikz[baseline=-5, scale=0.7]{
		\fill[WScolor] (-1.1,-1) rectangle (1.1,1);
		\draw[symmetry defect] (-1,-1) -- (0,0);
		\draw[symmetry defect] (1,-1) -- (0,0);
		\draw[symmetry defect] (0,1) -- (0,0);
	}}\\{\text{multiplication}}\end{array}\qquad\text{and}\qquad
\begin{array}{c}{\tikz[baseline=-5,yscale=-1, scale=0.7]{
		\fill[WScolor] (-1.1,-1) rectangle (1.1,1);
		\draw[symmetry defect] (-1,-1) -- (0,0);
		\draw[symmetry defect] (1,-1) -- (0,0);
		\draw[symmetry defect] (0,1) -- (0,0);
	}}\\{\text{comultiplication}}\end{array}}
	satisfying the loop-omission (separability) and projection properties:
	\uaeq
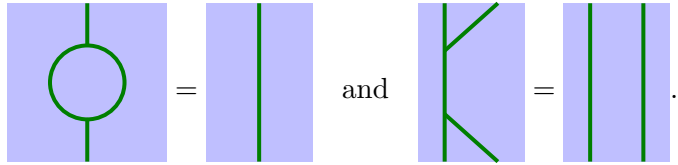
{\tikz[baseline=-6, scale=0.7]{
		\fill[WScolor] (-1.5,-1.5) rectangle (1.5,1.5);
		\draw[symmetry defect] (0,-1.5) -- (0,-.7);
		\draw[symmetry defect] (0,-.7) arc(270:630:.7);
		\draw[symmetry defect] (0,1.5) -- (0,.7);
	} = \tikz[baseline=-6, scale=0.7]{
		\fill[WScolor] (-1,-1.5) rectangle (1,1.5);
		\draw[symmetry defect] (0,-1.5) -- (0,1.5);
	} \quad\text{and}\quad \tikz[baseline=-6, scale=0.7]{
		\fill[WScolor] (-1,-1.5) rectangle (1,1.5);
		\draw[symmetry defect] (-.5,-1.5) -- (-.5,1.5);
		\draw[symmetry defect] (-.5,.6) -- (.5,1.5);
		\draw[symmetry defect] (-.5,-.6) -- (.5,-1.5);
	} =\tikz[baseline=-6, scale=0.7]{
		\fill[WScolor] (-1,-1.5) rectangle (1,1.5);
		\draw[symmetry defect] (-.5,-1.5) -- (-.5,1.5);
		\draw[symmetry defect] (.5,-1.5) -- (.5,1.5);
	}.}
	The junctions turn $P$ into an algebra as well as a coalgebra. We require $P$ to 
	either have\footnote{The special case in which $P$ has a unit as well as a counit is discussed in appendix \ref{app:Projections with unit and counit}.}
	\uaeq{\text{a unit}\;
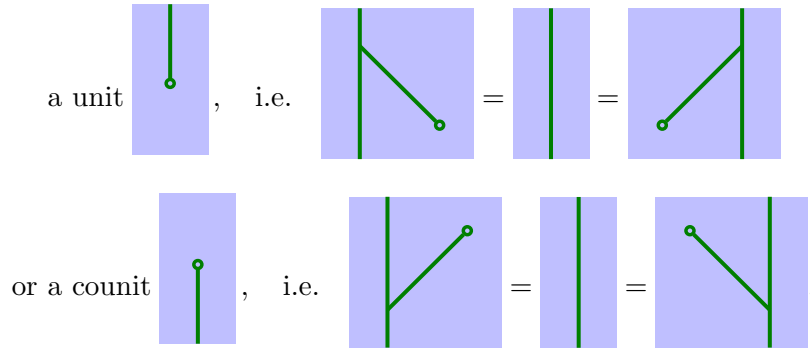
\tikz[baseline=-10]{
		\fill[WScolor] (-.5,-1) rectangle (.5,1);
		\draw[symmetry defect] (0,1) -- (0,0);
		\node[unit node] at (0,-.05) {};
	}\,,\quad\text{i.e.}\quad\tikz[baseline=20]{
		\fill[WScolor] (0,0) rectangle (2,2);
		\draw[symmetry defect] (.5,0) -- (.5,2);
		\draw[symmetry defect] (.5,1.5) -- (1.5,.5);
		\node[unit node] at (1.55,.45) {};
	}=\tikz[baseline=20]{
		\fill[WScolor] (0,0) rectangle (1,2);
		\draw[symmetry defect] (.5,0) -- (.5,2);
	}=\tikz[baseline=20,xscale=-1]{
		\fill[WScolor] (0,0) rectangle (2,2);
		\draw[symmetry defect] (.5,0) -- (.5,2);
		\draw[symmetry defect] (.5,1.5) -- (1.5,.5);
		\node[unit node] at (1.55,.45) {};
	}}
	\uaeq{\text{or a counit}\;\tikz[baseline=-10,yscale=-1]{
		\fill[WScolor] (-.5,-1) rectangle (.5,1);
		\draw[symmetry defect] (0,1) -- (0,0);
		\node[unit node] at (0,-.05) {};
	}\,,\quad\text{i.e.}\quad\tikz[baseline=-37,yscale=-1]{
		\fill[WScolor] (0,0) rectangle (2,2);
		\draw[symmetry defect] (.5,0) -- (.5,2);
		\draw[symmetry defect] (.5,1.5) -- (1.5,.5);
		\node[unit node] at (1.55,.45) {};
	}=\tikz[baseline=20]{
		\fill[WScolor] (0,0) rectangle (1,2);
		\draw[symmetry defect] (.5,0) -- (.5,2);
	}=\tikz[baseline=-37,xscale=-1,yscale=-1]{
		\fill[WScolor] (0,0) rectangle (2,2);
		\draw[symmetry defect] (.5,0) -- (.5,2);
		\draw[symmetry defect] (.5,1.5) -- (1.5,.5);
		\node[unit node] at (1.55,.45) {};
	}.}
	As is shown in appendix~\ref{app:Equivalence of (co)multiplication and Frobenius properties for projections}, the existence of a unit for a projection defect implies coassociativity, while the existence of a counit implies associativity. 
In fact, for projection defects, associativity,  coassociativity and the Frobenius identities	\uaeq
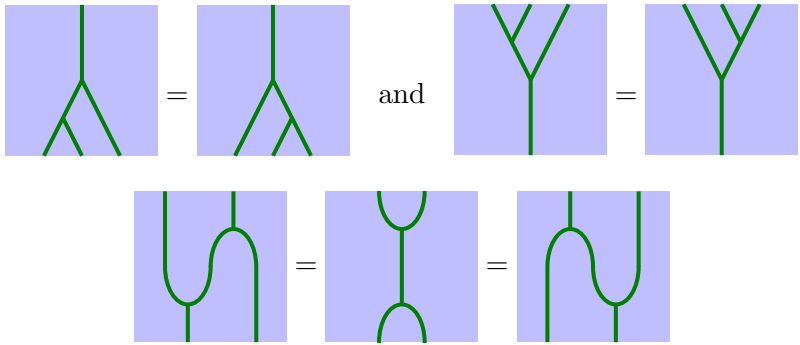
{\tikz[baseline=20]{
		\fill[WScolor] (0,0) rectangle (2,2);
		\draw[symmetry defect] (.5,0) -- (1,1);
		\draw[symmetry defect] (1,1) -- (1,2);
		\draw[symmetry defect] (1,0) -- (.75,.5);
		\draw[symmetry defect] (1.5,0) -- (1,1);
	}=\tikz[baseline=20,xscale=-1]{
		\fill[WScolor] (0,0) rectangle (2,2);
		\draw[symmetry defect] (.5,0) -- (1,1);
		\draw[symmetry defect] (1,1) -- (1,2);
		\draw[symmetry defect] (1,0) -- (.75,.5);
		\draw[symmetry defect] (1.5,0) -- (1,1);
	}\quad\text{and}\quad\tikz[baseline=-37,yscale=-1]{
		\fill[WScolor] (0,0) rectangle (2,2);
		\draw[symmetry defect] (.5,0) -- (1,1);
		\draw[symmetry defect] (1,1) -- (1,2);
		\draw[symmetry defect] (1,0) -- (.75,.5);
		\draw[symmetry defect] (1.5,0) -- (1,1);
	}=\tikz[baseline=-37,xscale=-1,yscale=-1]{
		\fill[WScolor] (0,0) rectangle (2,2);
		\draw[symmetry defect] (.5,0) -- (1,1);
		\draw[symmetry defect] (1,1) -- (1,2);
		\draw[symmetry defect] (1,0) -- (.75,.5);
		\draw[symmetry defect] (1.5,0) -- (1,1);
	}}
	\uaeq{\tikz[anchor=base, baseline=12]{
			\fill[WScolor] (-1,-.5) rectangle (1,1.5);
			\begin{scope}[yscale=-1,xscale=1, shift={(-.3,-.5)}]
				\clip (-.4,0) rectangle (.4,.6);
				\draw[symmetry defect] (0,0) ellipse (.3 and .5);
			\end{scope}
			\begin{scope}[shift={(.3,.5)}]
				\clip (-.4,0) rectangle (.4,.6);
				\draw[symmetry defect] (0,0) ellipse (.3 and .5);
			\end{scope}
			\draw[symmetry defect] (.6,-.5) -- (.6,.5);
			\draw[symmetry defect] (.3,1) -- (.3,1.5);
			\draw[symmetry defect] (-.6,.5) --(-.6, 1.5);
			\draw[symmetry defect] (-.3,0) --(-.3, -.5);
		}
		=
		\tikz[anchor=base, baseline=12]{
			\fill[WScolor] (-1,-.5) rectangle (1,1.5);
			\begin{scope}[yscale=-1,xscale=1, shift={(0,-1.5)}]
				\clip (-.4,0) rectangle (.4,.6);
				\draw[symmetry defect] (0,0) ellipse (.3 and .5);
			\end{scope}
			\draw[symmetry defect] (0,0) -- (0,1);
			\begin{scope}[shift={(0,-.5)}]
				\clip (-.4,0) rectangle (.4,.6);
				\draw[symmetry defect] (0,0) ellipse (.3 and .5);
			\end{scope}
		}
		=
		\tikz[anchor=base, baseline=12]{\begin{scope}[yscale=1,xscale=-1]
			\fill[WScolor] (-1,-.5) rectangle (1,1.5);
			\begin{scope}[yscale=-1,xscale=1, shift={(-.3,-.5)}]
				\clip (-.4,0) rectangle (.4,.6);
				\draw[symmetry defect] (0,0) ellipse (.3 and .5);
			\end{scope}
			\begin{scope}[shift={(.3,.5)}]
				\clip (-.4,0) rectangle (.4,.6);
				\draw[symmetry defect] (0,0) ellipse (.3 and .5);
			\end{scope}
			\draw[symmetry defect] (.6,-.5) -- (.6,.5);
			\draw[symmetry defect] (.3,1) -- (.3,1.5);
			\draw[symmetry defect] (-.6,.5) --(-.6, 1.5);
			\draw[symmetry defect] (-.3,0) --(-.3, -.5);
		\end{scope}
		}
	}
	 are all equivalent to one another, c.f.~appendix~\ref{app:Equivalence of (co)multiplication and Frobenius properties for projections}. Thus, unital or counital projection defects satsify all of them.

	 As in the context of RG defects discussed in section~\ref{sec:Representing the IR in the UV}, replacing the identity defect $I$ in a 2d TQFT by an arbitrary projection defect $P$, and inserting $P$ networks into the correlation functions one obtains correlation functions of a new, $P$-projected 2d TQFT. The relation between the projected and unprojected theories is exactly the same as the relation between IR and UV theories discussed in section~\ref{sec:Representing the IR in the UV}.
	 
	 While apriori, projection defects $P$ do not arise from a bulk perturbations, we will show in the next section that in fact they always factorize into RG type defects.

		
	\subsection{Factorization of projection defects}
	\label{sec:rg defects from projections}
	
	We now come full circle by showing that any (unital or counital) projection defect $P$ factorizes as
\begin{eqnarray*}
P&=&R^\dagger\otimes R\qquad\text{in case}\,P\;\text{is unital}\nonumber\\
P&=&{}^\dagger R\otimes R\qquad\text{in case}\,P\;\text{is counital}\nonumber\,,
\end{eqnarray*}
	where $R$ is an RG type\footnote{$R\otimes R^\dagger$ is isomorphic to the identity defect} defect between the $P$-projected theory on one side and the original unprojected theory on the other. By analogy to the case of RG flows, we call the original, unprojected theory UV and the $P$-projected theory IR.
	
		The basic idea is simple: as $P$ is a (co)algebra, it can be viewed as a left and/or right (co)module over itself. Thus, the defect $P$ can be regarded as a defect in the original (UV) theory (the defect $P$ itself), a defect in the $P$-projected (IR) theory (the identity defect), or a defect separating one of those from the other. To indicate which of the interpretations we are referring to, we denote the respective defects as $P_{\text{UV}|\text{UV}}$, $P_{\text{IR}|\text{IR}}$, $P_{\text{IR}|\text{UV}}$ or $P_{\text{UV}|\text{IR}}$, respectively. For instance, viewed as a left $P$-(co)module and a right $I_\text{UV}$-(co)module $P$ represents the defect $P_{\text{IR}|\text{UV}}$ between the $P$-projected (IR) theory and the original (UV) theory
	\uaeq
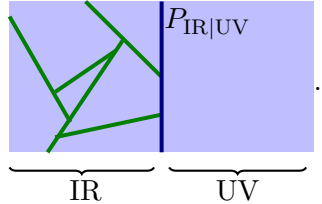
{\tikz[baseline=-5,xscale=-1]{
		\fill[WScolor] (-2,-1) rectangle (2,1);
		\draw[symmetry defect] (0,0) -- (1,1);
		\draw[symmetry defect] (.5,.5) -- (1.5,-1);
		\draw[symmetry defect] (0,-.5) -- (1.4,-.8);
		\draw[symmetry defect] (1.2,-.6) -- (2,.8);
		\draw[symmetry defect] (1.4,-.2) -- (.6,.35);
		\draw[thick,decoration={brace},decorate] (-1.9,-1.2) -- (-.1,-1.2);
		\draw[thick,decoration={brace},decorate] (.1,-1.2) -- (2,-1.2);
		\node at (-1,-1.5) {UV};
		\node at (1,-1.5) {IR};
		\draw[defect] (0,-1) -- (0,1);
		\node at (-.6,.7) {$P_{\text{IR}|\text{UV}}$};
	}.}
This defect plays the role of the RG defect $R$.

To show that it is indeed of RG type, we first need to determine its adjoints. We will restrict our discussion to the case that $P$ is unital. (There is an analogous argument for the case of counital $P$.) Since $P_{\text{IR}|\text{UV}}$ is a defect between IR and UV theory, the adjoints have to satisfy mixed Zorro identities:
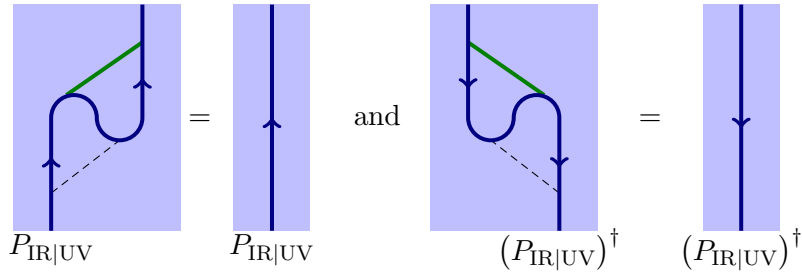
\uaeq{\tikz[baseline=40]{
		\fill[WScolor] (0,0) rectangle (2.2,3);
		\draw[symmetry defect] (.7,1.8) -- (1.7,2.5);
		\draw[densely dashed] (.5,.5) -- (1.4,1.2);
		\draw[defect, arrow position=.2, arrow position = .8] (.5,0) -- (.5,1.5) arc(180:0:.3) arc(180:360:.3) -- (1.7,3);
		\node at (.5,-.25) {$P_{\text{IR}|\text{UV}}$};
	}=\tikz[baseline=40]{
		\fill[WScolor] (0,0) rectangle (1,3);
		\draw[defect, arrow position=.5] (.5,0) -- (.5,3);
		\node at (.5,-.25) {$P_{\text{IR}|\text{UV}}$};
	}\quad\text{and}\quad\tikz[baseline=40,xscale=-1]{
		\fill[WScolor] (0,0) rectangle (2.2,3);
		\draw[symmetry defect] (.7,1.8) -- (1.7,2.5);
		\draw[densely dashed] (.5,.5) -- (1.4,1.2);
		\draw[defect, opp arrow position=.2, opp arrow position = .8] (.5,0) -- (.5,1.5) arc(180:0:.3) arc(180:360:.3) -- (1.7,3);
		\node at (.5,-.25) {$\left(P_{\text{IR}|\text{UV}}\right)^{\dagger}$};
	}=\tikz[baseline=40,xscale=-1]{
		\fill[WScolor] (0,0) rectangle (1,3);
		\draw[defect, opp arrow position=.5] (.5,0) -- (.5,3);
		\node at (.5,-.25) {$\left(P_{\text{IR}|\text{UV}}\right)^{\dagger}$};
	}}
for the right adjoint and
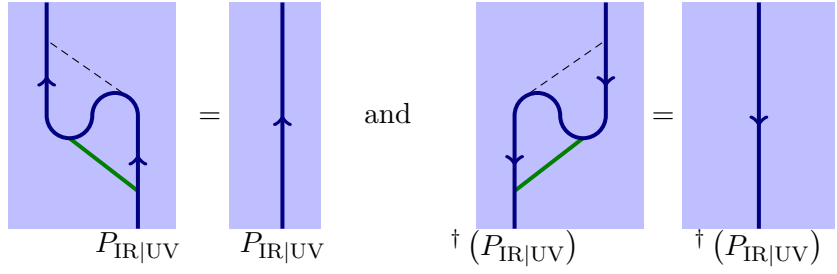
\uaeq{\tikz[baseline=40,xscale=-1]{
		\fill[WScolor] (0,0) rectangle (2.2,3);
		\draw[densely dashed] (.7,1.8) -- (1.7,2.5);
		\draw[symmetry defect] (.5,.5) -- (1.4,1.2);
		\draw[defect, arrow position=.2, arrow position = .8] (.5,0) -- (.5,1.5) arc(180:0:.3) arc(180:360:.3) -- (1.7,3);
		\node at (.5,-.25) {$P_{\text{IR}|\text{UV}}$};
	}=\tikz[baseline=40,xscale=-1]{
		\fill[WScolor] (0,0) rectangle (1.2,3);
		\draw[defect, arrow position=.5] (.5,0) -- (.5,3);
		\node at (.5,-.25) {$P_{\text{IR}|\text{UV}}$};
	}\quad\text{and}\quad\tikz[baseline=40]{
		\fill[WScolor] (0,0) rectangle (2.2,3);
		\draw[densely dashed] (.7,1.8) -- (1.7,2.5);
		\draw[symmetry defect] (.5,.5) -- (1.4,1.2);
		\draw[defect, opp arrow position=.2, opp arrow position = .8] (.5,0) -- (.5,1.5) arc(180:0:.3) arc(180:360:.3) -- (1.7,3);
		\node at (.5,-.25) {${}^{\dagger}\left(P_{\text{IR}|\text{UV}}\right)$};
	}=\tikz[baseline=40]{
		\fill[WScolor] (-.5,0) rectangle (1.5,3);
		\draw[defect, opp arrow position=.5] (.5,0) -- (.5,3);
		\node at (.5,-.25) {${}^{\dagger}\left(P_{\text{IR}|\text{UV}}\right)$};
	}}
for the left adjoint. Here, the defect $P$ plays the role of the identity defect on the IR side of the defect. 
 For unital $P$, comultiplication induces a coevaluation map $I_\text{UV}\rightarrow P\rightarrow P\otimes P$, and, as is shown in appendix~\ref{app:Adjoints of induced RG defects}
	\uaeq{
		\left(P_{\text{IR}|\text{UV}}\right)^{\dagger} &= P_{\text{UV}|\text{IR}} \\
		{}^{\dagger}\!\left(P_{\text{IR}|\text{UV}}\right) &= \left({}^\dagger P\right)_{\text{UV}|\text{IR}}\,.
	}
	(${}^\dagger P$ denotes the left adjoint of $P$ in the UV theory.) Now, since fusion over the IR theory is the same as fusion in the UV, it follows from the projection property of $P$ that
	\uaeq{
	P_{\text{UV}|\text{UV}}=P\cong P\otimes P=P_{\text{UV}|\text{IR}}\otimes P_{\text{IR}|\text{UV}}=(P_{\text{IR}|\text{UV}})^{\dagger}\otimes P_{\text{IR}|\text{UV}}\,.
	}
	Moreover, the identity defect in the IR theory is represented by $P$ in the UV theory, and hence
	\uaeq{
	I_{\text{IR}}=P_{\text{IR}|\text{IR}}=P\cong P\otimes P=P_{\text{IR}|\text{UV}}\otimes P_{\text{UV}|\text{IR}}=P_{\text{IR}|\text{UV}}\otimes (P_{\text{IR}|\text{UV}})^{\dagger}\,.\
	}
Thus, any unital projection defect $P$ factorizes as $P=R^\dagger\otimes R$, where $R=P_{\text{IR}|\text{UV}}$ has the property that $R\otimes R^\dagger\cong I_{\text{IR}}$. Note that all the defects $R$, $R^\dagger$ and $I_{\text{IR}}$ are represented by $P$ in the UV theory, and the isomorphism $R\otimes R^\dagger\rightarrow I_{\text{IR}}$ and its inverse are just given by the multiplication and comultiplication of $P$, respectively. The loop-omission and projection property of $P$ then imply
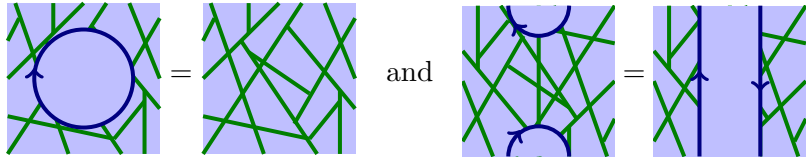
\uaeq{\tikz[baseline=-1]{
		\fill[WScolor] (0,-1) rectangle (2,1);
		\draw[symmetry defect] (0,0) -- (1,1);
		\draw[symmetry defect] (.5,.5) -- (1.5,-1);
		\draw[symmetry defect] (0,-.5) -- (1.4,-.8);
		\draw[symmetry defect] (1.2,-.6) -- (2,.8);
		\draw[symmetry defect] (1.4,-.2) -- (.6,.35);
		\draw[symmetry defect] (0,-1) -- (1.3,1);
		\draw[symmetry defect] (.8,-1) -- (.1,1);
		\draw[symmetry defect] (2,-.4) -- (1.1,.7);
		\draw[symmetry defect] (1.8,-1) -- (1.8,-.15);
		\draw[symmetry defect] (1.4,-.8) -- (1.8,-.3);
		\draw[symmetry defect] (2,0) -- (1.6,1);
		\draw[symmetry defect] (0,.5) -- (.15,.8);
		\draw[symmetry defect] (.6,.6) -- (.6,1);
		\fill[WScolor] (1,0) circle (.65);
		\draw[defect,opp arrow position=.5] (1,0) circle (.65);
		}
	=\tikz[baseline=-1]{
		\fill[WScolor] (0,-1) rectangle (2,1);
		\draw[symmetry defect] (0,0) -- (1,1);
		\draw[symmetry defect] (.5,.5) -- (1.5,-1);
		\draw[symmetry defect] (0,-.5) -- (1.4,-.8);
		\draw[symmetry defect] (1.2,-.6) -- (2,.8);
		\draw[symmetry defect] (1.4,-.2) -- (.6,.35);
		\draw[symmetry defect] (0,-1) -- (1.3,1);
		\draw[symmetry defect] (.8,-1) -- (.1,1);
		\draw[symmetry defect] (2,-.4) -- (1.1,.7);
		\draw[symmetry defect] (1.8,-1) -- (1.8,-.15);
		\draw[symmetry defect] (1.4,-.8) -- (1.8,-.3);
		\draw[symmetry defect] (2,0) -- (1.6,1);
		\draw[symmetry defect] (0,.5) -- (.15,.8);
		\draw[symmetry defect] (.6,.6) -- (.6,1);
	}
	\quad\text{and}\quad
	\tikz[scale=1,yscale=-1,xscale=-1,baseline=0]{
		\clip (0,-1) rectangle (2,1);
		\fill[WScolor] (0,-1) rectangle (2,1);
		\draw[symmetry defect] (0,0) -- (1,1);
		\draw[symmetry defect] (.5,.5) -- (1.5,-1);
		\draw[symmetry defect] (0,-.5) -- (1.4,-.8);
		\draw[symmetry defect] (1.2,-.6) -- (2,.8);
		\draw[symmetry defect] (1.4,-.2) -- (.6,.35);
		\draw[symmetry defect] (0,-1) -- (1.3,1);
		\draw[symmetry defect] (.8,-1) -- (.1,1);
		\draw[symmetry defect] (2,-.4) -- (1.1,.7);
		\draw[symmetry defect] (1.8,-1) -- (1.8,-.15);
		\draw[symmetry defect] (1.4,-.8) -- (1.8,-.3);
		\draw[symmetry defect] (2,0) -- (1.6,1);
		\draw[symmetry defect] (0,.5) -- (.15,.8);
		\draw[symmetry defect] (.6,.6) -- (.6,1);
		\draw[symmetry defect] (1,1) -- (1,-1);
		\fill[WScolor] (.6,1) arc(180:360:.4) -- cycle;
		\draw[defect, opp arrow position = .8] (.6,1) arc(180:360:.4);
		\fill[WScolor] (.6,-1) arc(180:0:.4) -- cycle;
		\draw[defect, arrow position = .8] (.6,-1) arc(180:0:.4);
		}
	= \tikz[scale=1,yscale=-1,xscale=-1,baseline=0]{
		\clip (0,-1) rectangle (2,1);
		\fill[WScolor] (0,-1) rectangle (2,1);
		\draw[symmetry defect] (0,0) -- (1,1);
		\draw[symmetry defect] (.5,.5) -- (1.5,-1);
		\draw[symmetry defect] (0,-.5) -- (1.4,-.8);
		\draw[symmetry defect] (1.2,-.6) -- (2,.8);
		\draw[symmetry defect] (1.4,-.2) -- (.6,.35);
		\draw[symmetry defect] (0,-1) -- (1.3,1);
		\draw[symmetry defect] (.8,-1) -- (.1,1);
		\draw[symmetry defect] (2,-.4) -- (1.1,.7);
		\draw[symmetry defect] (1.8,-1) -- (1.8,-.15);
		\draw[symmetry defect] (1.4,-.8) -- (1.8,-.3);
		\draw[symmetry defect] (2,0) -- (1.6,1);
		\draw[symmetry defect] (0,.5) -- (.15,.8);
		\draw[symmetry defect] (.6,.6) -- (.6,1);
		\draw[symmetry defect] (1,1) -- (1,-1);
		\fill[WScolor] (.6,1) -- (.6,-1) -- (1.4,-1) -- (1.4,1) -- cycle;
		\draw[defect, opp arrow position = .5] (.6,1) -- (.6,-1);
		\draw[defect, opp arrow position = .5] (1.4,-1) -- (1.4,1);
	}}
         Similar considerations lead to an analogous factorization of counital projection defects $P$. The role of the RG defect is again played by $R=P_{\text{IR}|\text{UV}}$. But the adjoints differ from the unital case:
           \uaeq{
		\left(P_{\text{IR}|\text{UV}}\right)^{\dagger} &= \left(P^\dagger\right)_{\text{UV}|\text{IR}} \\
		{}^{\dagger}\left(P_{\text{IR}|\text{UV}}\right) &= P_{\text{UV}|\text{IR}},
	}
which leads to slightly different factorizations
\uaeq{
	P_{\text{UV}|\text{UV}}=P\cong P\otimes P=P_{\text{UV}|\text{IR}}\otimes P_{\text{IR}|\text{UV}}={}^{\dagger}(P_{\text{IR}|\text{UV}})\otimes P_{\text{IR}|\text{UV}}\,,
	}
and
	\uaeq{
	I_{\text{IR}}=P_{\text{IR}|\text{IR}}=P\cong P\otimes P=P_{\text{IR}|\text{UV}}\otimes P_{\text{UV}|\text{IR}}=P_{\text{IR}|\text{UV}}\otimes {}^{\dagger}(P_{\text{IR}|\text{UV}})\,.\
	}
	
	If $P$ comes with both, a unit and a counit, it is self-adjoint ($P^\dagger \cong P \cong {}^\dagger P$, see appendix~\ref{app:Projections with unit and counit}), and the left and right adjoint of the induced  RG defect $R$ are isomorphic, $R^\dagger \cong {}^\dagger R$.
	
	\subsection{Relation to the generalized orbifold procedure}
	\label{sec:Relation to the generalized orbifold procedure}
	
	The method described in section~\ref{sec:rg defects from projections} above to construct a new 2d TQFT by replacing the identity defect by a projection defect $P$ is very close to and in fact inspired by the generalized orbifold procedure \cite{Frohlich:2009gb, Brunner:2013xna, Brunner:2013ota,Brunner:2014lua, Carqueville:2012dk,Fuchs:2002cm} (see appendix~\ref{sec:generalized orbifold theories} for a quick summary). In that procedure  a new 2d theory is defined from an original one by inserting networks of a defect $A$ into the correlation functions of the original theory. The difference to our construction is the requirements imposed on $A$.
	
	In the generalized orbifold construction the defect $A$ has to be a separable Frobenius algebra\footnote{a unital, counital, associative, coassociative algebra and coalgebra satisfying loop-omission and Frobenius properties}, c.f. appendix~\ref{sec:generalized orbifold theories}. This condition is very similar to the properties of projection defects with two differences: On the one hand the defect $A$ does not have to satisfy the projection property, but is on the other hand required to have both, a unit and a counit, which we do not demand of projection defects. 
	Moreover, it is often assumed in the generalized orbifold procedure that left and right adjoints of any defect $D$ are isomorphic, i.e. $D^\dagger \cong {}^\dagger D$, so that further conditions such as pivotality and symmetry can be demanded (see e.g. \cite{Carqueville:2012dk}). We do not require such a condition, and in fact it is not met in our examples discussed in section~\ref{sec:rg networks in lg minimal model orbifolds}
.

	A projection defect $P$ has both a unit and counit if and only if left and right adjoints of the respective RG defects are isomorphic
	\uaeq{
		\left(P_{\text{IR}|\text{UV}}\right)^{\dagger} \cong P_{\text{UV}|\text{IR}} \cong
		{}^{\dagger}\!\left(P_{\text{IR}|\text{UV}}\right)\,,
}
c.f.~appendix~\ref{app:Projections with unit and counit}. 
In that case $P$ is a separable Frobenius algebra, and the construction described in section~\ref{sec:rg defects from projections} is a special case of the generalized orbifold construction. Indeed, the projection property of $P$ brings about 
interesting simplifications in the generalized orbifold construction, which we will spell out in the remainder of this section.
	
	Let $A$ be a separable Frobenius algebra in a given 2d theory. We will represent it by green line segments in diagrams. Defects in the generalized orbifold theory defined by $A$ are given by defects in the underlying 2d theory, which are $A$-(bi)modules. Let $D$ and $\tilde D$ be two such (bi)modules. Their fusion in the generalized orbifold theory is given by their tensor product $D\otimes_A \tilde D$ over the algebra $A$, pictorially
	\uaeq
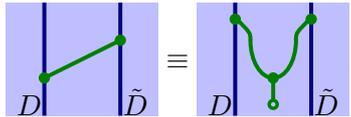
{
		\tikz[anchor=base, baseline=18]{
			\fill[WScolor] (-1.5,0) rectangle (.5,1.5);
			\draw[defect] (0,0) -- (0,1.5);
			\draw[defect] (-1,0) -- (-1,1.5);
			\node[symmetry node] at (-1,.5) {};
			\node[symmetry node] at (0,1) {};
			\draw[symmetry defect] (-1,.5) -- (0,1);
			\node at (-1.2,0) {$D$};
			\node at (.2,0) {$\tilde D$};
		}\equiv
		\tikz[anchor=base, baseline=18]{
			\fill[WScolor] (-1.5,0) rectangle (.5,1.5);
			\draw[defect] (0,0) -- (0,1.5);
			\node[symmetry node] at (0,1.28) {};
			\draw[symmetry defect] (-.2,1) .. controls (-.2,1.1) .. (0,1.3);
			\begin{scope}[yscale=-1,shift={(-.5,-1)}]
				\clip (-.4,0) rectangle (.4,.6);
				\draw[symmetry defect] (0,0) ellipse (.3 and .5);
			\end{scope}
			\node[symmetry node] at (-.5,.5) {};
			\draw[symmetry defect] (-.5,.2) -- (-.5,.5);
			\node[unit node] at (-.5,.15) {};
			\draw[symmetry defect] (-.8,1) .. controls (-.8,1.1) .. (-1,1.3);
			\draw[defect] (-1,0) -- (-1,1.5);
			\node[symmetry node] at (-1,1.28) {};
			\node at (-1.2,0) {$D$};
			\node at (.2,0) {$\tilde D$};
		}.
	}
	In general, it is different from the fusion $D\otimes\tilde D$ in the underlying unorbifolded theory.
	
	Indeed, similarly to projection defects, also separable Frobenius algebras always factorize into defects between the orbifold and the underlying unorbifolded theory and their adjoints. Namely, considered as a left $A$- and right $I$-module\footnote{$I$ is the identity defect of the underlying unorbifolded theory.}, $A$ represents a defect $R$ between orbifolded and unorbifolded theory. Considered as right $A$- and left $I$-module it represents the adjoint defect $R^\dagger\cong {}^\dagger R$. Now, for any separable Frobenius algebra we have $A\otimes_A A\cong A$, or pictorially
	\uaeq{\tikz[baseline=20]{
		\fill[WScolor] (0,0) rectangle (2,2);
		\draw[symmetry defect] (1,0) -- (1,2);
	}=\tikz[baseline=20]{
		\fill[WScolor] (0,0) rectangle (2,2);
		\draw[symmetry defect] (.5,0) -- (.5,2);
		\draw[symmetry defect] (1.5,0) -- (1.5,2);
		\begin{scope}[shift={(1.5,.25)}]
			\draw[symmetry defect] (-.2,1) .. controls (-.2,1.1) .. (0,1.3);
			\begin{scope}[yscale=-1,shift={(-.5,-1)}]
				\clip (-.4,0) rectangle (.4,.6);
				\draw[symmetry defect] (0,0) ellipse (.3 and .5);
			\end{scope}
			\draw[symmetry defect] (-.5,.2) -- (-.5,.5);
			\node[unit node] at (-.5,.15) {};
			\draw[symmetry defect] (-.8,1) .. controls (-.8,1.1) .. (-1,1.3);
		\end{scope}
	}=\tikz[baseline=20]{
		\fill[WScolor] (0,0) rectangle (2,2);
		\draw[symmetry defect] (.5,0) -- (.5,2);
		\draw[symmetry defect] (1.5,0) -- (1.5,2);
			\draw[symmetry defect] (.5,.5) -- (1.5,1);
			\draw[symmetry defect] (1,.75) -- (1.5,.25);
			\draw[symmetry defect] (1.25,.875) -- (1.5,1.5);
			\draw[symmetry defect] (1.375,1.2) -- (.5,1.5);
	}.
	}
Hence, $A$ as a defect in the unorbifolded theory factorizes as $A\cong R^\dagger\otimes_A R$. 
However, for generic $A$ the defect $R$ is not of RG type, i.e. $R\otimes R^\dagger$ is not isomorphic to the identity defect in the orbifold theory. Hence, bubbles of a generalized orbifold theory inserted in the unorbifolded theory do not in general connect trivially:
\uaeq{\tikz[scale=1.5, baseline=-5]{
		\fill[WScolor] (0,-1) rectangle (2,1);
		\draw[symmetry defect] (.2,.2) .. controls (.4,1) and (.8,1) .. (1,.5) .. controls (1.2,.2) .. (1.6,.2) .. controls (2,.2) and (2,-.2) .. (1.4,-.2) .. controls (.7,-.4) and (2,-.8) .. (1.2,-.9) .. controls (0,-1) .. cycle;
		\clip (.2,.2) .. controls (.4,1) and (.8,1) .. (1,.5) .. controls (1.2,.2) .. (1.6,.2) .. controls (2,.2) and (2,-.2) .. (1.4,-.2) .. controls (.7,-.4) and (2,-.8) .. (1.2,-.9) .. controls (0,-1) .. cycle;
		\draw[symmetry defect] (0,0) -- (1,1);
		\draw[symmetry defect] (.5,.5) -- (1.5,-1);
		\draw[symmetry defect] (0,-.5) -- (1.4,-.8);
		\draw[symmetry defect] (1.2,-.6) -- (2,.8);
		\draw[symmetry defect] (1.4,-.2) -- (.6,.35);
		\draw[symmetry defect] (0,-1) -- (1.3,1);
		\draw[symmetry defect] (.8,-1) -- (.1,1);
		\draw[symmetry defect] (2,-.4) -- (1.1,.7);
		\draw[symmetry defect] (1.8,-1) -- (1.8,-.15);
		\draw[symmetry defect] (1.4,-.8) -- (1.8,-.3);
		\draw[symmetry defect] (2,0) -- (1.6,1);
		\draw[symmetry defect] (0,.5) -- (.15,.8);
		\draw[symmetry defect] (.6,.6) -- (.6,1);
	}\neq\tikz[scale=1.5,baseline=-5]{
		\fill[WScolor] (0,-1) rectangle (2,1);
		\begin{scope}
			\draw[symmetry defect] (.4,0) ellipse (.3 and .85);
			\clip (.4,0) ellipse (.3 and .85);
			\draw[symmetry defect] (0,0) -- (1,1);
			\draw[symmetry defect] (.5,.5) -- (1.5,-1);
			\draw[symmetry defect] (0,-.5) -- (1.4,-.8);
			\draw[symmetry defect] (1.2,-.6) -- (2,.8);
			\draw[symmetry defect] (1.4,-.2) -- (.6,.35);
			\draw[symmetry defect] (0,-1) -- (1.3,1);
			\draw[symmetry defect] (.8,-1) -- (.1,1);
			\draw[symmetry defect] (2,-.4) -- (1.1,.7);
			\draw[symmetry defect] (1.8,-1) -- (1.8,-.15);
			\draw[symmetry defect] (1.4,-.8) -- (1.8,-.3);
			\draw[symmetry defect] (2,0) -- (1.6,1);
			\draw[symmetry defect] (0,.5) -- (.15,.8);
			\draw[symmetry defect] (.6,.6) -- (.6,1);
		\end{scope}
		\begin{scope}
			\draw[symmetry defect] (1.4,.3) ellipse (.55 and .4);
			\clip (1.4,.3) ellipse (.55 and .4);
			\draw[symmetry defect] (0,0) -- (1,1);
			\draw[symmetry defect] (.5,.5) -- (1.5,-1);
			\draw[symmetry defect] (0,-.5) -- (1.4,-.8);
			\draw[symmetry defect] (1.2,-.6) -- (2,.8);
			\draw[symmetry defect] (1.4,-.2) -- (.6,.35);
			\draw[symmetry defect] (0,-1) -- (1.3,1);
			\draw[symmetry defect] (.8,-1) -- (.1,1);
			\draw[symmetry defect] (2,-.4) -- (1.1,.7);
			\draw[symmetry defect] (1.8,-1) -- (1.8,-.15);
			\draw[symmetry defect] (1.4,-.8) -- (1.8,-.3);
			\draw[symmetry defect] (2,0) -- (1.6,1);
			\draw[symmetry defect] (0,.5) -- (.15,.8);
			\draw[symmetry defect] (.6,.6) -- (.6,1);
		\end{scope}
		\begin{scope}
			\draw[symmetry defect] (1.3,-.6) ellipse (.5 and .3);
			\clip (1.3,-.6) ellipse (.5 and .3);
			\draw[symmetry defect] (0,0) -- (1,1);
			\draw[symmetry defect] (.5,.5) -- (1.5,-1);
			\draw[symmetry defect] (0,-.5) -- (1.4,-.8);
			\draw[symmetry defect] (1.2,-.6) -- (2,.8);
			\draw[symmetry defect] (1.4,-.2) -- (.6,.35);
			\draw[symmetry defect] (0,-1) -- (1.3,1);
			\draw[symmetry defect] (.8,-1) -- (.1,1);
			\draw[symmetry defect] (2,-.4) -- (1.1,.7);
			\draw[symmetry defect] (1.8,-1) -- (1.8,-.15);
			\draw[symmetry defect] (1.4,-.8) -- (1.8,-.3);
			\draw[symmetry defect] (2,0) -- (1.6,1);
			\draw[symmetry defect] (0,.5) -- (.15,.8);
			\draw[symmetry defect] (.6,.6) -- (.6,1);
		\end{scope}
	}}
	Instead, pushing two bubbles of the generalized orbifold against each other creates a non-trivial defect at the interface of the two bubbles. Thus, the generalized orbifold cannot be obtained by a local perturbation of the original theory. This is only true if $A$ additionally satisfies the projection property. 
	
	In that case, fusion in the generalized orbifold simplifies dramatically -- it reduces to fusion in the unorbifolded theory.
Namely, for a separable Frobenius algebra the projection property can be rephrased as 
\uaeq{\tikz[baseline=20]{
		\fill[WScolor] (0,0) rectangle (2,2);
		\draw[symmetry defect] (.5,0) -- (1,.5) -- (1,1.5) -- (.5,2);
		\draw[symmetry defect] (1.5,0) -- (1,.5);
		\draw[symmetry defect] (1.5,2) -- (1,1.5);
	}=\tikz[baseline=20]{
		\fill[WScolor] (0,0) rectangle (2,2);
		\draw[symmetry defect] (.5,0) -- (.5,2);
		\draw[symmetry defect] (1.5,0) -- (1.5,2);
	}\Leftrightarrow\tikz[baseline=20]{
		\fill[WScolor] (0,0) rectangle (1,2);
		\begin{scope}[shift={(1,.5)}]
			\draw[symmetry defect] (-.2,1) .. controls (-.2,1.1) .. (0,1.3);
			\begin{scope}[yscale=-1,shift={(-.5,-1)}]
				\clip (-.4,0) rectangle (.4,.6);
				\draw[symmetry defect] (0,0) ellipse (.3 and .5);
			\end{scope}
			\node[symmetry node] at (-.5,.5) {};
			\draw[symmetry defect] (-.5,.2) -- (-.5,.5);
			\node[unit node] at (-.5,.15) {};
			\draw[symmetry defect] (-.8,1) .. controls (-.8,1.1) .. (-1,1.3);
		\end{scope}
	}=\tikz[baseline=20]{
		\fill[WScolor] (0,0) rectangle (1,2);
		\begin{scope}[shift={(1,.5)}]
			\draw[symmetry defect] (-.2,1) .. controls (-.2,1.1) .. (0,1.3);
			\draw[symmetry defect] (-.8,1) .. controls (-.8,1.1) .. (-1,1.3);
			\node[unit node] at (-.8,.95) {};
			\node[unit node] at (-.2,.95) {};
		\end{scope}
	}}
	leading to the following simplification for defect fusion in the orbifold theory:
		\uaeq{
		\tikz[anchor=base, baseline=18]{
			\fill[WScolor] (-1.5,0) rectangle (.5,1.5);
			\draw[defect] (0,0) -- (0,1.5);
			\node[symmetry node] at (0,1.28) {};
			\draw[symmetry defect] (-.2,1) .. controls (-.2,1.1) .. (0,1.3);
			\begin{scope}[yscale=-1,shift={(-.5,-1)}]
				\clip (-.4,0) rectangle (.4,.6);
				\draw[symmetry defect] (0,0) ellipse (.3 and .5);
			\end{scope}
			\node[symmetry node] at (-.5,.5) {};
			\draw[symmetry defect] (-.5,.2) -- (-.5,.5);
			\node[unit node] at (-.5,.15) {};
			\draw[symmetry defect] (-.8,1) .. controls (-.8,1.1) .. (-1,1.3);
			\draw[defect] (-1,0) -- (-1,1.5);
			\node[symmetry node] at (-1,1.28) {};
			\node at (-1.2,0.1) {$D$};
			\node at (.2,0.1) {$\tilde D$};
		}\stackrel{\text{unit}}{=}
		\tikz[anchor=base, baseline=18]{
			\fill[WScolor] (-1.5,0) rectangle (.5,1.5);
			\draw[defect] (0,0) -- (0,1.5);
			\node[symmetry node] at (0,1.28) {};
			\draw[symmetry defect] (-.2,1) .. controls (-.2,1.1) .. (0,1.3);
			\begin{scope}[yscale=-1,shift={(-.5,-1)}]
				\clip (-.4,0) rectangle (.4,.6);
				\draw[symmetry defect] (0,0) ellipse (.3 and .5);
			\end{scope}
			\node[symmetry node] at (-.5,.5) {};
			\draw[symmetry defect] (-.5,.2) -- (-.5,.5);
			\node[unit node] at (-.5,.15) {};
			\draw[symmetry defect] (-.8,1) .. controls (-.8,1.1) .. (-1,1.3);
			\draw[defect] (-1,0) -- (-1,1.5);
			\node[symmetry node] at (-1,1.28) {};
			\node at (-1.2,0.1) {$D$};
			\node at (.2,0.1) {$\tilde D$};
			\draw[symmetry defect] (-.5,.4) -- (-.2,.2);
			\node[unit node] at (-.16,.16) {};
		}\stackrel{\text{proj.}}=
		\tikz[anchor=base, baseline=18]{
			\fill[WScolor] (-1.5,0) rectangle (.5,1.5);
			\draw[defect] (0,0) -- (0,1.5);
			\draw[symmetry defect] (-.2,1) .. controls (-.2,1.1) .. (0,1.3);
			\node[unit node] at (-.2,.95) {};
			\begin{scope}[yscale=-1,shift={(-.5,-1)}]
				\clip (-.4,0) rectangle (.4,.6);
				\draw[densely dashed] (0,0) ellipse (.3 and .5);
			\end{scope}
			\draw[densely dashed] (-.5,.2) -- (-.5,.5);
			\node[unit node, densely dashed,black,thin] at (-.5,.15) {};
			\draw[symmetry defect] (-.8,1) .. controls (-.8,1.1) .. (-1,1.3);
			\node[unit node] at (-.8,.95) {};
			\draw[defect] (-1,0) -- (-1,1.5);
			\node at (-1.2,0.1) {$D$};
			\node at (.2,0.1) {$\tilde D$};
		}=
		\tikz[anchor=base, baseline=18]{
			\fill[WScolor] (-1.5,0) rectangle (.5,1.5);
			\draw[defect] (0,0) -- (0,1.5);
			\draw[densely dashed] (-.2,1) .. controls (-.2,1.1) .. (0,1.3);
			\begin{scope}[yscale=-1,shift={(-.5,-1)}]
				\clip (-.4,0) rectangle (.4,.6);
				\draw[densely dashed] (0,0) ellipse (.3 and .5);
			\end{scope}
			\draw[densely dashed] (-.5,.2) -- (-.5,.5);
			\node[unit node, densely dashed,black,thin] at (-.5,.15) {};
			\draw[densely dashed] (-.8,1) .. controls (-.8,1.1) .. (-1,1.3);
			\draw[defect] (-1,0) -- (-1,1.5);
			\node at (-1.2,0.1) {$D$};
			\node at (.2,0.1) {$\tilde D$};
		}.
	}

\section{RG-networks in Landau-Ginzburg orbifolds}
\label{sec:rg networks in lg minimal model orbifolds}

In this section we will apply our construction in the context of topologically twisted Landau-Ginzburg models. More precisely, we will consider Landau-Ginzburg models $\cM_d$ with a single chiral superfield $X$ and superpotential $W(X)=X^d$. These models admit relevant perturbations generated by deformations of the superpotential by lower degree polynomials. For instance, the model $\cM_d$ can be perturbed by adding a term $\lambda\,X^{d'}$ to the superpotential. For $d'<d$ this perturbation is relevant and the renormalization group flow drives the theory from the model $\cM_d$ in the UV ($\lambda=0$) to the model $\cM_{d'}$ ($\lambda=\infty$) in the IR. These perturbations are chiral and hence preserve A-type supersymmetry. The corresponding RG defects are therefore A-type defects. We prefer to work with B-type defects in Landau-Ginzburg models, because they are much better understood.
For that reason, we will consider the mirror dual situation instead: RG flows between Landau-Ginzburg orbifolds $\cM_d/\mathbb{Z}_d$ and $\cM_{d'}/\mathbb{Z}_{d'}$ generated by twisted chiral perturbations. The respective RG defects have been constructed in \cite{Brunner:2007ur}.

We will start by giving a brief outline of the description of B-type defects in Landau-Ginzburg models by means of matrix factorizations. Then we will review the construction of the respective RG defects from \cite{Brunner:2007ur}. Finally, we will use our construction to realize the IR theories by means of projection defects in the UV theories. In particular, we will show how to realize all the Landau-Ginzburg orbifolds $\cM_d/\mathbb{Z}_d$ in the theory of a free twisted chiral field.

\subsection{B-type defects in Landau-Ginzburg models}
\label{sec:b type defects in LG}

As put forward by Kontsevich, B-type defects in Landau-Ginzburg models can be described in terms of matrix factorizations \cite{Kapustin:2002bi,Brunner:2003dc,Khovanov:2004bc,Brunner:2007qu,orlov2003triangulated}.

A matrix factorization of a polynomial $W\in S=\mathbb{C}[x_1,\ldots,x_n]$ consists of a $\mathbb{Z}_2$-graded free module $D=D_0\oplus D_1$ over the polynomial ring $S$, with an odd endomorphism $d_D:D\rightarrow D$, which squares to $W$ times the identity map, i.e. $d_D^2=W\,\text{id}_D$. One often unfolds matrix factorizations 
into 2-periodic complexes
\uaeq{
	D:D_1\;
		\tikz[baseline=0]{
			\node at (.5,.5) {$\d_{D1}$};
			\node at (.5,-.3) {$\d_{D0}$};
			\draw[arrow position = 1] (0,.2) -- (1,.2);
			\draw[arrow position = 1] (1,0) -- (0,0);
		}\,
	D_0\,,
	\qquad\d_D = \begin{pmatrix}0 & \d_{D1} \\ \d_{D0} & 0\end{pmatrix}.
}
These complexes are twisted by $W$:
$d_{D1}\circ d_{D0}=W\,\text{id}_{D_0}$ and $d_{D0}\circ d_{D1}=W\,\text{id}_{D_1}$.

Now, topological Landau-Ginzburg models are completely specified by their chiral superfields $X_1,\ldots,X_n$ and their superpotential $W(X_1,\ldots,X_n)$. 
B-type defects $D$ between two LG models with superpotential $W(X_1, .., X_n)$ and $V(Z_1, ..., Z_m)$ 
	\uaeq{\tikz{
		\fill[WScolor light] (-4,-1) rectangle (0,1);
		\fill[WScolor] (0,-1) rectangle (4,1);
		\draw[defect, arrow position=.5] (0,-1) -- (0,1);
		\node[DefectColor] at (.2,-.8) {$D$};
		\node at (-2,0) {$V(Z_1, ..., Z_m)$};
		\node at (2,0) {$W(X_1, ..., X_n)$};
	},}
can be described by matrix factorizations of the difference $V-W$ of the respective superpotentials. By abuse of notation we also denote the matrix factorization by $D$ and write $D:W\rightarrow V$.

The space of defect-changing fields $\Hom(D,D')$ between two defects represented by matrix factorizations 
$D,D':W\rightarrow V$ is given by the homology of the induced $\mathbb{Z}_2$-graded complex on the space of homomorphisms $\Hom_S(D,D')$ of the respective $S$-modules\footnote{Note that the $\Hom$-complex is untwisted!}. More precisely, 
\begin{eqnarray*}
&&\Hom(D,D')=H^*_\d(\Hom_S(D,D'))\,,\nonumber\\
&&\quad\text{with differential}\quad
\d \phi = \d_{D'} \circ \phi - (-1)^{\text{deg}} \phi \circ \d_{D}\,,\quad\text{for}\quad
\phi\in\Hom_S(D,D')\,.\nonumber
\end{eqnarray*}
Here $\deg$ denotes the $\mathbb{Z}_2$-degree. The space of defect-changing fields is $\mathbb{Z}_2$-graded with even and odd elements corresponding to bosons and fermions, respectively. The operator product of defect-changing fields is just the composition of homomorphisms.

Defect fusion is described by the tensor product of matrix factorizations \cite{Brunner:2007qu}. Namely, let $U\in\mathbb{C}[X_1,\ldots,X_m]$, $V\in\mathbb{C}[Y_1,\ldots,Y_n]$, $W\in\mathbb{C}[Z_1,\ldots,Z_o]$ and $D:W\rightarrow V$ and $D':V\rightarrow U$ be matrix factorizations of $V-W$ and $U-V$, respectively.
Then the fused defect is given by the tensor product $D'\otimes D$ of matrix factorizations. This is the matrix factorization built on the $\mathbb{Z}_2$-graded $\mathbb{C}[X_1,\ldots,X_m,Z_1,\ldots,Z_o]$-module $D'\otimes_{\mathbb{C}[Y_1,\ldots,Y_n]} D$ with homomorphism
	\uaeq{
		\d_{D' \otimes D} = \d_{D'} \otimes \text{id}_D + \text{id}_{D'} \otimes \d_D\,.
	}
	This differential is to be understood with Koszul signs, meaning that
	\uaeq{
		(\text{id}_{D'}\otimes \d_{D}) (\nu \otimes \omega) = (-1)^{\text{deg}}(\nu) \otimes \d_{D}(\omega).
	}
Since the factorized polynomials add upon taking the tensor product,
this is indeed a matrix factorization of $(U-V)+(V-W) = U-W$, i.e. $D'\otimes D:W\rightarrow U$.\footnote{A priori, tensor product matrix factorizations like this are of infinite rank. It can be shown however, that tensor products of finite-rank matrix factorizations are isomorphic to finite-rank matrix factorizations \cite{Brunner:2007qu}.}

Adjunctions of B-type defects in LG models have been studied in \cite{Carqueville:2010hu,Carqueville:2012st} (see \cite{Carqueville:2013usa} for a nice review). Adjoints are given by
\aeq{\label{eq:eqmfadjoints}
D^\dagger \cong D^\vee[n]\,,\quad
 {}^\dagger D \cong D^\vee[m]\,,
}
where $D^\vee$ is the dual of a matrix factorization $D$, consisting of the dual modules $(D^\vee)_i=(D_i)^\vee$, 
 and the maps
\uaeq{
	\d_{D^\vee} = \begin{pmatrix}0 & \d_{D0}^\vee \\ -\d_{D1}^\vee & 0\end{pmatrix}\,.
}
Moreover, $(\cdot)[m]$ denotes the shift of $\bZ_2$-degree by $m$: 
$(D[m])_i=D_{i+m}$ and $\d_{D[m]i}=(-1)^m\d_{D(i+m)}$.

Indeed, boundary conditions are a special case of defects, namely those with a trivial theory on one side. The trivial LG theory is of course the theory with no chiral fields and zero superpotential. Right (left) B-type boundary conditions of a Landau-Ginzburg theory with superpotential $W$ can therefore be described by matrix factorizations of $W$ ($-W$).

\subsection*{B-type defects in Landau-Ginzburg orbifolds}
\label{sec:b type defects in lg orbifolds}

The description of B-type defects in Landau-Ginzburg models by means of matrix factorizations extends in a straight-forward manner to the context of Landau-Ginzburg orbifolds.  Whenever the polynomial ring $\bC[X_1,\ldots,X_n]$ carries an action of a finite group $G_W$ which leaves a polynomial $W$ invariant, the Landau-Ginzburg model 
defined by $W$ can be orbifolded by $G_W$ leading to a new 2d TQFT which by abuse of notation we denote by $W/G_W$ \cite{Intriligator:1990ua}. 

Now let $V\in\bC[X_1,\ldots,X_n]$ and $W\in\bC[Y_1,\ldots,Y_m]$ be two superpotentials and $G_V$ and $G_W$  orbifold groups. Then B-type defects between the respective LG orbifolds can be described by $G=G_V\times G_W$-equivariant matrix factorizations of $V-W$ \cite{Brunner:2007ur,Ashok:2004zb,Hori:2004ja}.
These are matrix factorizations $D:W\rightarrow V$ as before, which are additionally equipped with a representation $\rho_D$ of $G$. The latter has to be compatible with the module structure on $D$ and has to commute with $\d_D$. Denoting by $\rho$ the representation of $G=G_V\times G_W$ on the combined polynomial ring $S=\bC[X_1,\ldots,X_n,Y_1,\ldots,Y_m]$ this means that for all $g\in G$
\begin{eqnarray*}
&&\rho_D(g)(s\cdot p) = \rho(g)(s) \cdot \rho_D(g)(p), \qquad \forall s\in S, p\in D=D_0\oplus D_1,\\
&&\rho_{D}(g)\circ \d_{D} = \d_{D} \circ \rho_D(g)\,.\nonumber
\end{eqnarray*}
Given two equivariant matrix factorizations $D,D':W\rightarrow V$, the complex $\Hom_S(D,D')$ carries an action of $G=G_V\times G_W$ which commutes with the differential $\d$, inducing a representation on the homology $H_\d^*(\Hom_S(D,D'))$. The space of defect-changing fields in the orbifold theory is then given by the $G$-invariant part $\Hom^G(D,D')=(H_\d^*(\Hom_S(D,D')))^G$. The operator product of defect-changing fields is again just composition of homomorphisms.

Defect fusion carries over from the unorbifolded LG models by taking invariant parts. More precisely, let 
$U\in\mathbb{C}[X_1,\ldots,X_m]$, $V\in\mathbb{C}[Y_1,\ldots,Y_n]$ and $W\in\mathbb{C}[Z_1,\ldots,Z_o]$ be polynomials invariant under actions of groups $G_U,G_V,G_W$ on the respective polynomial rings. And let 
$D:W\rightarrow V$ and $D':V\rightarrow U$ be $G_W\times G_V$-, respectively $G_V\times G_U$-equivariant matrix factorizations. Then the tensor product
$D'\otimes D$ is a $G_U\times G_V\times G_W$-equivariant matrix factorization of $U-W$. Fusion of the defects in the orbifold theory is then given by the $G_V$-invariant part  $D'\otimes_{G_V} D:=\left(D'\otimes D\right)^{G_V}$ of $D'\otimes D$, which is of course $G_U\times G_W$-equivariant. 

Adjunction of defects $D$ in the orbifold theory is given by adjunction \eqref{eq:eqmfadjoints} in the underlying unorbifolded theory, where however the $G$-action on the adjoints is twisted. This can be seen in a systematic way in the generalized orbifold construction \cite{Frohlich:2009gb, Brunner:2013xna, Brunner:2013ota, Carqueville:2012dk}  which offers a completely general framework to describe orbifold theories using defects in the underlying unorbifolded theory. We outline the generalized orbifold procedure in appendix~\ref{sec:generalized orbifold theories}, and in particular spell out the formula for adjoints.

\subsection*{Defects $\cM_{d}/\bZ_d\rightarrow \cM_{d'}/\bZ_{d'}$}
\label{sec:defects in mm orbifolds}

For the case of Landau-Ginzburg orbifolds $\cM_d/\bZ_d$ the discussion simplifies somewhat.
An element $a\in\bZ_d$ of the orbifold group acts on the chiral field $X$ by $X\mapsto e^{\frac{2\pi i a}{d}}X$. 

A defect $D:X^d/\bZ_d\rightarrow Z^{d'}/\bZ_{d'}$ is given by a $G=\bZ_{d'}\times\bZ_d$-equivariant matrix factorization of $Z^{d'}-X^d$. 
Since $G$ is commutative, its representations on $D$ can be specified by $G$-gradings or -charges of the generators 
of the free $S=\bC[Z,X]$-module $D=D_0\oplus D_1$. We will indicate them in square brackets and specify a $G$-equivariant matrix factorizations as
\uaeq{D:S^M\begin{pmatrix}[l_M,r_M]\\ [l_{M+1},r_{M+1}]\\ \vdots\\ [l_{2M-1},r_{2M-1}] \end{pmatrix}\;
		\tikz[baseline=0]{
			\node at (.5,.5) {$\d_{D1}$};
			\node at (.5,-.3) {$\d_{D0}$};
			\draw[arrow position = 1] (0,.2) -- (1,.2);
			\draw[arrow position = 1] (1,0) -- (0,0);
		}\,
	S^M\begin{pmatrix}[l_0,r_0]\\ [l_{1},r_{1}]\\ \vdots\\ [l_{M-1},r_{M-1}]\end{pmatrix}\,,
}
c.f.~\cite{Brunner:2007ur} for more details. 
Adjoints then take the form, c.f. appendix \ref{app:defects and boundaries},
 \aeq{\label{eq:explmfadjoints}\begin{array}{l}
 D^\dagger:S^M\begin{pmatrix}[-r_0+1,-l_0]\\ [-r_{1}+1,-l_{1}]\\ \vdots\\ [-r_{M-1}+1,-l_{M-1}] \end{pmatrix}\;
		\tikz[baseline=0]{
			\node at (.5,.5) {$\d_{D1}^T$};
			\node at (.5,-.3) {$-\d_{D0}^T$};
			\draw[arrow position = 1] (0,.2) -- (1,.2);
			\draw[arrow position = 1] (1,0) -- (0,0);
		}\,
	S^M\begin{pmatrix}[-r_M+1,-l_M]\\ [-r_{M+1}+1,-l_{M+1}]\\ \vdots\\ [-r_{2M-1}+1,-l_{2M-1}]\end{pmatrix}
\\
{}^\dagger D:S^M\begin{pmatrix}[-r_0,-l_0+1]\\ [-r_{1},-l_{1}+1]\\ \vdots\\ [-r_{M-1},-l_{M-1}+1] \end{pmatrix}\;
		\tikz[baseline=0]{
			\node at (.5,.5) {$\d_{D1}^T$};
			\node at (.5,-.3) {$-\d_{D0}^T$};
			\draw[arrow position = 1] (0,.2) -- (1,.2);
			\draw[arrow position = 1] (1,0) -- (0,0);
		}\,
	S^M\begin{pmatrix}[-r_M,-l_M+1]\\ [-r_{M+1},-l_{M+1}+1]\\ \vdots\\ [-r_{2M-1},-l_{2M-1}+1]\end{pmatrix}
\end{array}}
Note that left adjoints differ from right adjoints by a shift in $G$-charges by $[-1,1]$. We write ${}^\dagger D=D^\dagger\{[-1,1]\}$.

An important example is the identity defect $I_d:X^{d}/\bZ_{d}\rightarrow Z^{d}/\bZ_{d}$ which is represented by the following $\bZ_{d}\times\bZ_{d}$-equivariant matrix factorization (c.f. appendix~\ref{sec:app:orbifold identity equivariant})
	\uaeq{
		I_d:S^{d}\begin{pmatrix}[1,0]\\ [2,-1]\\ [3,-2] \\ \vdots \end{pmatrix}
		\tikz[baseline=0]{
			\begin{scope}
				\node at (0,1.8) {$\begin{pmatrix}Z& 0 & ... & 0 & -X \\ -X & Z &&&\\0&-X&Z&&\\\vdots&&\ddots&\ddots&\\0&&&-X&Z\end{pmatrix}$};
				\draw[arrow position = 1] (-3,.2) -- (3,.2);
				\draw[arrow position = 1] (3,0) -- (-3,0);
				\node at (0,-.5) {$\d_{I_d0}$};
			\end{scope}
		}\;S^{d}\begin{pmatrix}[0,0]\\ [1,-1]\\ [2,-2] \\ \vdots \end{pmatrix}.
	}
One easily reads off that this defect is self-adjoint, i.e. $I_d^\dagger\cong I_d\cong {}^\dagger\!I_d$.

\subsection{RG defects in LG orbifolds}
\label{sec:rg defects in LG orbs}

As alluded to above, Landau-Ginzburg orbifolds $\cM_d/\bZ_d$ exhibit relevant perturbations by twisted chiral fields. The corresponding RG flows drive the theory from $\cM_d/\bZ_d$ in the UV  to another orbifold $\cM_{d'}/\bZ_{d'}$ with $d'<d$ in the IR. The associated RG defects have been constructed in \cite{Brunner:2007ur}. They preserve B-type supersymmetry and can therfore be described by $\bZ_{d'}\times\bZ_d$-equivariant matrix factorizations of $Z^{d'}-X^d$. Indeed, due to a singularity in the parameter space, there are different flows from $\cM_d/\bZ_d$ to $\cM_{d'}/\bZ_{d'}$. The corresponding RG defects $R=R(m,n_0,\ldots,n_{d'-1})$ are specified by $m\in\bZ_d$, and integers $n_0,\ldots,n_{d'-1}\geq 1$, such that $n_0+ ... + n_{d'-1}=d$. They 
are represented by matrix factorizations
	\aeq{\label{eq:RG defect}
		R:S^{d'}\begin{psmallmatrix}[1,-m]\\ [2,-m-n_1]\\ [3,-m-n_1-n_2] \\ \vdots \end{psmallmatrix}
		\tikz[baseline=0]{
			\begin{scope}
				\node at (0,1.8) {$\begin{pmatrix}Z& 0 & ... & 0 & -X^{n_0} \\ -X^{n_1} & Z &&&\\0&-X^{n_2}&Z&&\\\vdots&&\ddots&\ddots&\\0&&&-X^{n_{d'-1}}&Z\end{pmatrix}$};
				\draw[arrow position = 1] (-2.8,.2) -- (2.8,.2);
				\draw[arrow position = 1] (2.8,0) -- (-2.8,0);
				\node at (0,-.5) {$\d_{R0}$};
			\end{scope}
		}S^{d'}\begin{psmallmatrix}[0,-m]\\ [1,-m-n_1]\\ [2,-m-n_1-n_2] \\ \vdots \end{psmallmatrix}\,,
	}
where, $S=\bC[X,Z]$. For more details see \cite{Brunner:2007ur}. In the following we will sometimes take the subscripts of the $n_i$ to be elements in $\bZ_{d'}$ by defining $n_{i+z\,d'}=n_i$ for all $z\in\bZ$. 

Using this concrete realization of RG defects, one can now explicitly carry out the construction outlined in section~\ref{sec:rg networks in 2d tqfts} and represent the LG orbifolds $\cM_{d'}/\bZ_{d'}$ in $\cM_{d}/\bZ_{d}$ for any $d'<d$. In order to construct the respective projection defects, we need right and left adjoints of the defects $R$, which can easily be read off from formula~\eqref{eq:explmfadjoints}. They are given by
\uaeq{
		R^\dagger:S^{d'}\left(\begin{smallmatrix}[m+1,0]\\ [m+1+n_1,-1]\\ [m+1+n_1+n_2,-2] \\ \vdots \end{smallmatrix}\right)
		\tikz[baseline=0]{
			\begin{scope}
				\node at (0,1.2) {$\left(\begin{smallmatrix}Z& -X^{n_1} & &  & \\  & Z &-X^{n_2}&&\\&&\ddots&\ddots&\\&&&Z&-X^{n_{d'-1}}\\-X^{n_0}&&&&Z\end{smallmatrix}\right)$};
				\draw[arrow position = 1] (-2.4,.2) -- (2.4,.2);
				\draw[arrow position = 1] (2.4,0) -- (-2.4,0);
				\node at (0,-.5) {$\d_{R^\dagger 0}=\d_{R0}^T$};
			\end{scope}
		}S^{d'}\left(\begin{smallmatrix}[m+1,-1]\\ [m+1+n_1,-2]\\ [m+1+n_1+n_2,-3] \\ \vdots \end{smallmatrix}\right)
	}
	and
	\uaeq{
	{}^\dagger R:S^{d'}\left(\begin{smallmatrix}[m,1]\\ [m+n_1,0]\\ [m+n_1+n_2,-1] \\ \vdots \end{smallmatrix}\right)
		\tikz[baseline=0]{
			\begin{scope}
				\node at (0,1.2) {$\left(\begin{smallmatrix}Z& -X^{n_1} & &  & \\  & Z &-X^{n_2}&&\\&&\ddots&\ddots&\\&&&Z&-X^{n_{d'-1}}\\-X^{n_0}&&&&Z\end{smallmatrix}\right)$};
				\draw[arrow position = 1] (-2.4,.2) -- (2.4,.2);
				\draw[arrow position = 1] (2.4,0) -- (-2.4,0);
				\node at (0,-.5) {$\d_{{}^\dagger\!R 0}=\d_{R0}^T$};
			\end{scope}
		}S^{d'}\left(\begin{smallmatrix}[m,0]\\ [m+n_1,-1]\\ [m+n_1+n_2,-2] \\ \vdots \end{smallmatrix}\right)
	}
A straight-forward calculation presented in appendix~\ref{sec: R otimes R dagger equals A} then shows that indeed
\uaeq{
		R\otimes_{\bZ_d} R^\dagger &\cong I_{d'}\\
		R\otimes_{\bZ_d} {}^\dagger R&\cong I_{d'},
	}
i.e.~the defects $R$ are indeed of RG type. Fusion in the opposite order yields the respective projection defects (see appendix~\ref{app:LG P} for the explicit calculation).
For the unital projection defect  $P=R^\dagger\otimes_{\bZ_{d'}} R$ one obtains
 \uaeq{P:
 S^{d'}\left(\begin{smallmatrix}[m+1,-m] \\ [m+1+n_1,-m- n_1] \\ [m+1+n_1+n_2,-m-n_1-n_2] \\ \vdots \\ [m+1+\sum_{l=1}^{d'-1}n_{l},-m-\sum_{l=1}^{d'-1} n_{l}] \end{smallmatrix}\right)\;		\tikz[baseline=0]{
			\begin{scope}
				\node at (0,0.6) {$\d_{P1}$};
				\draw[arrow position = 1] (-0.6,.2) -- (0.6,.2);
				\draw[arrow position = 1] (0.6,0) -- (-0.6,0);
				\node at (0,-.3) {$\d_{P0}$};
			\end{scope}
		}\;S^{d'}\left(\begin{smallmatrix}[m+1+\sum_{l=1}^{d'-1}n_{l},-m] \\ [m+1,-m- n_1] \\ [m+1+n_1,-m-n_1-n_2] \\ \vdots \\ [m+1+\sum_{l=1}^{d'-2}n_{l},-m-\sum_{l=1}^{d'-1} n_{l}] \end{smallmatrix}\right)\,,
	}
where
\aeq{\label{eqp1}\d_{P1}=\begin{pmatrix}Z^{n_0}& 0 & ... & 0 & -X^{n_0} \\ -X^{n_1} & Z^{n_1} &&&\\0&-X^{n_2}&Z^{n_2}&&\\\vdots&&\ddots&\ddots&\\0&&&-X^{n_{d'-1}}&Z^{n_{d'-1}}\end{pmatrix}\,.}
The counital projection defects $P'={}^\dagger R\otimes_{\bZ_{d'}}  R$ is given by the left adjoint  $P'={}^\dagger P$ of $P$.

The morphism
\uaeq{\phi:=\tikz[anchor=base,baseline=-30]{
		\fill[WScolor light] (-2,-1.3) rectangle (1.5,1.5);
		\fill[WScolor] (-2,-1.3)--(0,-1.3)--(0,.3)..controls(-.4,-1) and (-1,-1) ..(-1,1.5)--(-2,1.5)--cycle;
		\draw[defect, opp arrow position = .5, opp arrow position=.7, opp arrow position=.2] (0,-1.3)--(0,.3)..controls(-.4,-1) and (-1,-1) .. (-1,1.5);
		\node[defect node] at (0,.35) {};
		\draw[densely dashed] (0,.35) -- (-1,1);
		\node[DefectColor] at (.3,-1.1) {${}^\dagger R$};
		\node[DefectColor] at (-1.3,1) {$R^\dagger$};
		\node[DefectColor] at (-.3,0) {$R$};
		\node at (1,1) {IR};
		\node at (-.3,.7) {$I$};
		\node at (-1.5,-1) {UV};
		\begin{scope}[yscale=-1,shift={(0,2)}]
			\fill[WScolor light] (-2,-1.3) rectangle (1.5,1.5);
			\fill[WScolor] (-2,-1.3)--(0,-1.3)--(0,.3)..controls(-.4,-1) and (-1,-1) ..(-1,1.5)--(-2,1.5)--cycle;
			\draw[defect, arrow position = .5, arrow position=.7] (0,-1.3)--(0,.3)..controls(-.4,-1) and (-1,-1) .. (-1,1.5);
			\node[defect node] at (0,.35) {};
			\draw[densely dashed] (0,.35) -- (-1,1);
			\node[DefectColor] at (.3,-.7) {$R^\dagger$};
			\node[DefectColor] at (-1.3,1) {${}^\dagger R$};
			\node[DefectColor] at (-.4,0) {$R$};
		\end{scope}
	}:{}^\dagger R\longrightarrow R^\dagger}
which is used to close right $R$-loops can also be determined explicitly. It is not hard to see that it is given by
\uaeq{
\phi=\left(\begin{array}{cc}\phi_0&0\\0&\phi_1\end{array}\right)\qquad\text{with}\qquad
\phi_0=\phi_1=\left(\begin{array}{cccc}
0&X^{n_1-1}&&\\
&\ddots&\ddots&\\
&&\ddots&\;\;X^{n_{d'}-1}\\
X^{n_0-1}&&&0
\end{array}\right).}

\subsection{\texorpdfstring{Representing $\cM_{d'}/\bZ_{d'}$ in $\cM_{d}/\bZ_{d}$ for $d'<d$}{Representing Md'/Zd' in Md/d' for d' < d}}
\label{sec: LG realization}

The projection defects constructed from RG defects in the previous section can now be used to
represent Landau-Ginzburg orbifolds $\cM_{d'}/\bZ_{d'}$ in orbifolds $\cM_{d}/\bZ_{d}$ for $d'<d$. 

	\subsubsection*{Bulk Hilbert space}
	
	The orbifolds $\cM_{d'}/\bZ_{d'}$ only possess a single bulk chiral field, namely the identity field. Therefore, the bulk Hilbert space in the B-twisted model is trivial, it just contains the vaccuum. One easily checks, that this is also true for $\Hom(P,P)$. Hence, the bulk Hilbert space of $\cM_{d'}/\bZ_{d'}$ agrees with the space of defect fields on the projection defect in $\cM_{d}/\bZ_{d}$.\footnote{Since the bulk Hilbert spaces are trivial, this is not that interesting.
However, there is a way to describe also the twisted chiral fields in the B-twisted LG orbifolds $\cM_{d}/\bZ_{d}$. Namely, being orbifold twist fields, they can be realized as defect changing fields between symmetry defects. This realization then lifts from IR to UV using projection defects,
i.e. one can realize the twisted chiral fields in $\cM_{d'}/\bZ_{d'}$ by defect changing fields in $\cM_{d}/\bZ_{d}$.}

	\subsubsection*{Boundary conditions}

	Next, we demonstrate how to represent the boundary conditions of $\cM_{d'}/\bZ_{d'}$ as $P$-invariant boundary conditions in the models $\cM_{d}/\bZ_{d}$. 

Elementary left boundary conditions in a theory $\cM_{d}/\bZ_d$ are represented by the $\bZ_{d}$-equivariant matrix factorizations
	\uaeq{
		B_{k,N}^d: \bC[X] \begin{pmatrix}[N + k]\end{pmatrix}\,
		\tikz[anchor=base, baseline=0]{
			\draw[->] (0,.2) -- (2,.2);
			\draw[->] (2,0) -- (0,0);
			\node at (1,.3) {$X^k$};
			\node at (1,-.4) {$-X^{d-k}$};
		}\, \bC[X] \begin{pmatrix}[N]\end{pmatrix}
	}
	of $-X^d$, 
	where $k\in\{1,\ldots,d-1\}$ and $N\in\bZ_d$. 
	
	As is shown in appendix \ref{app:B otimes P cong B for LG}, a UV boundary condition $B_\text{UV}=B^{d}_{k,N}$ is invariant under fusion with $P$, i.e. $B_\text{UV}\otimes P \cong B_\text{UV}$ iff
	\uaeq{
		k &= n_{i} + ... + n_{i-l} \\
		\text{and}\quad N &= \left[ -m - \sum_{a=1}^i n_{a} \right] 
	}
	for an $i\in\bZ_{d'}$ and an $l\in\left\{0, ..., d'-2\right\}$. 
	These are of course nothing but the lifts $B_\text{IR}\otimes_{\bZ_{d'}} R$ of IR boundary conditions to the UV. Namely, for $B_\text{IR}=B^{d'}_{l,M}$ one finds \cite{Brunner:2007ur}
	\uaeq{
	B_\text{IR}\otimes_{\bZ_{d'}}R=B^d_{(n_{-M-l+1}+\ldots+n_{-M}),(-m-\sum_{a=1}^{-M}n_a)}\,.
	}

	\subsubsection*{IR symmetries}
\label{sec:uplifting IR symmetries}

The Landau-Ginzburg orbifold model $\cM_{d'}/\bZ_{d'}$ exhibits a $\bZ_{d'}$-symmetry. The action of an element $a\in\bZ_{d'}$ on the theory is described by the symmetry defect ${}_aI_{d'}=I_{d'}\{[a,0]\}\cong I_{d'}\{[0,-a]\}$ obtained by shifting the charges of the identity defect $I_{d'}$ by $[a,0]$ or equivalently by $[0,-a]$. These defects fuse according to the group multiplication in the symmetry group $\bZ_{d'}$:
\uaeq{{}_aI_{d'}\otimes_{\bZ_{d'}}{}_bI_{d'}={}_{a+b}I_{d'}\,,\quad\text{for}\quad a,b\in\bZ_{d'}.}
As any IR defects, they lift into the UV theory $\cM_{d}/\bZ_{d}$ by fusion with RG defects
\uaeq{
{}_aI_{d'}\longmapsto R^\dagger\otimes_{\bZ_{d'}} {}_aI_{d'}\otimes_{\bZ_{d'}} R=:{}_aP.
}
These lifted defects also fuse according to multiplication in the symmetry group, i.e. ${}_aP\otimes_{\bZ_d}{}_bP={}_{a+b}P$, and therefore give a realization of the IR symmetry in the UV. The neutral element of the group however lifts to the defect ${}_0P=P$ and not to the identity defect in the UV. The lifted IR symmetries are therefore not invertible in the full UV theory, and hence are not symmetries of the UV theory.

The explicit form of ${}_aP$ can be easily derived by means of a slight variation of the calculation of $P$ as carried out in appendix \ref{app:LG IR symmetries}. The result is
 \uaeq{{}_aP:
 S^{d'}\left(\begin{smallmatrix}[m+1,-m-\sum_{j=1}^{-a}n_{j}] \\ [m+1+n_1,-m-\sum_{j=1}^{1-a}n_{j}] \\ [m+1+n_1+n_2,-m-\sum_{j=1}^{2-a}n_{j}] \\ \vdots \\ [m+1+\sum_{l=1}^{d'-1}n_{l},-m-\sum_{l=1}^{d'-1} n_{l}] \end{smallmatrix}\right)\;		\tikz[baseline=0]{
			\begin{scope}
				\node at (0,0.6) {$\d_{P1}$};
				\draw[arrow position = 1] (-0.6,.2) -- (0.6,.2);
				\draw[arrow position = 1] (0.6,0) -- (-0.6,0);
				\node at (0,-.3) {$\d_{P0}$};
			\end{scope}
		}\;S^{d'}\left(\begin{smallmatrix}[m+1+\sum_{l=1}^{d'-1}n_{l},-m-\sum_{j=1}^{-a}n_{j}] \\ [m+1,-m-\sum_{j=1}^{1-a}n_{j}] \\ [m+1+n_1,-m-\sum_{j=1}^{2-a}n_{j}] \\ \vdots \\ [m+1+\sum_{l=1}^{d'-2}n_{l},-m-\sum_{l=1}^{d'-1-a} n_{l}] \end{smallmatrix}\right)\,,
	}
where
\uaeq{
	\d_{P1}=
		\begin{pmatrix}
			Z^{n_0}& 0 & ... & 0 & -X^{n_{0-a}} \\
			-X^{n_{1-a}} & Z^{n_1} &&&\\
			0&-X^{n_{2-a}}&Z^{n_2}&&\\
			\vdots&&\ddots&\ddots&\\
			0&&&-X^{n_{d'-1-a}}&Z^{n_{d'-1}}
		\end{pmatrix}\,.
}
For $a=0$, this is the matrix factorization describing $P$. The lifted IR symmetry defects ${}_aP$ are obtained from it by shifting the exponents of $X$ by $a$ steps while keeping left $\bZ_d$-charges fixed and adapting the right ones accordingly.

\subsection{\texorpdfstring{The limit $d\rightarrow \infty$}{The limit d -> infty}}
\label{sec: the limit}

As discussed in the beginning of this section, the RG flows between LG orbifolds $\cM_d/\bZ_d$ are nothing but the mirror versions of flows between LG models $\cM_d$ generated by deformations of the superpotentials $W=X^d$ by lower degree polynomials. Indeed, all the models $\cM_{d'}$ can be obtained as perturbations of the free chiral field theory ($W=0$) by superpotential deformations. Thus, employing our procedure provides a representation of all the models $\cM_{d'}$ inside the theory of a free chiral field, which can be thought of as the limit $\cM_\infty=\lim_{d\to\infty}\cM_d$ of the models $\cM_d$.

In order to make this more explicit we again take the mirror perspective. The representation of the respective RG defects in terms of matrix factorizations then allows us to explicitly realize all the LG orbifolds $\cM_{d'}/\bZ_{d'}$ by means of projection defects in the theory of the free twisted chiral field. The latter can be described as the limit  $\cM_{\infty}/\bZ_{\infty}=\lim_{d\to\infty}\cM_{d}/\bZ_{d}$ and can be thought of as a $U(1)$-equivariant version of the free chiral field.

RG defects between $\cM_{d'}/\bZ_{d'}$ and $\cM_\infty/\bZ_\infty$ can be obtained as limits
of the RG defects (\ref{eq:RG defect}) representing flows $\cM_d/\bZ_d\to\cM_{d'}/\bZ_{d'}$, where one $n_i$ is sent to $\infty$ while the others are kept fixed. Since $d=\sum_in_i$, then also $d\to\infty$. Indeed, we can choose $n_0\to\infty$ and compensate for this choice by allowing a shift of the charges of $R$ by $[k,0]$, $k\in\bZ_{d'}$. In the limit, entries $X^{n_0}$ in the matrix factorization have to be replaced by $0$, and the $\bZ_d$-equivariance turns into a $U(1)$-equivariance. This way, one obtains the $\bZ_{d'}\times U(1)$-equivariant matrix factorizations 
\uaeq{
		R_\infty:S^{d'}\begin{pmatrix}[k+1,-m]\\ [k+2,-m-n_1]\\ [k+3,-m-n_1-n_2] \\ \vdots \end{pmatrix}
		\tikz[baseline=0]{
			\begin{scope}
				\node at (0,.6) {$\d_{R1}$};
				\draw[arrow position = 1] (-1,.2) -- (1,.2);
				\draw[arrow position = 1] (1,0) -- (-1,0);
				\node at (0,-.4) {$\d_{R0}$};
			\end{scope}
		}S^{d'}\begin{pmatrix}[k,-m]\\ [k+1,-m-n_1]\\ [k+2,-m-n_1-n_2] \\ \vdots \end{pmatrix}
	}
of $Z^{d'}$. They are specified by integers $m\in\bZ$,  $n_1, ..., n_{d'-1}\in\mathbb{N}$ and $k\in\bZ_{d'}$. The maps are given by
\uaeq{
	\d_{R1} &= \begin{pmatrix}Z& 0 & ... & 0 & 0 \\ -X^{n_1} & Z &&&\\0&-X^{n_2}&Z&&\\\vdots&&\ddots&\ddots&\\0&&&-X^{n_{d'-1}}&Z\end{pmatrix} \\
	\d_{R0}&=\begin{pmatrix}
		Z^{d'-1}& 0 & 0 & ...  \\
		Z^{d'-2}X^{n_1} & Z^{d'-1} & 0 & ... \\
		Z^{d'-3}X^{n_1+n_2} &Z^{d'-2}X^{n_2}& Z^{d'-1} & \ddots \\
		Z^{d'-4}X^{n_1+n_2+n_3} &Z^{d'-3}X^{n_2+n_3}& Z^{d'-2}X^{n_3} & \ddots \\
		\vdots&\vdots&\vdots&\ddots&
	\end{pmatrix}.
}
These matrix factorizations represent RG defects between $\cM_{d'}/\bZ_{d'}$ and $\cM_\infty/\bZ_\infty$. Indeed $R_\infty\otimes R_\infty^\dagger\cong I_\text{IR}$ as is spelled out in appendix~\ref{app:RG defect fusion infinity}.

In explicit calculations, it is not difficult to see that fusion commutes with the limit $d\to\infty$, at least as long as the theory squeezed between the defects is kept fixed in the limit. In particular, the limit $d\to\infty$ of projection defects is the fusion $P_\infty=R_\infty^\dagger\otimes R_\infty$ of the limit of RG defects. The
 projection defect realizing $\cM_{d'}/\bZ_{d'}$ within the limit theory $\cM_{\infty}/\bZ_{\infty}$  takes the form
 \uaeq{P_\infty:
 S^{d'}\left(\begin{smallmatrix}[m+1,-m] \\ [m+1+n_1,-m- n_1] \\ [m+1+n_1+n_2,-m-n_1-n_2] \\ \vdots \\ [m+1+\sum_{l=1}^{d'-1}n_{l},-m-\sum_{l=1}^{d'-1} n_{l}] \end{smallmatrix}\right)\;		\tikz[baseline=0]{
			\begin{scope}
				\node at (0,0.6) {$\d_{P1}$};
				\draw[arrow position = 1] (-0.6,.2) -- (0.6,.2);
				\draw[arrow position = 1] (0.6,0) -- (-0.6,0);
				\node at (0,-.3) {$\d_{P0}$};
			\end{scope}
		}\;S^{d'}\left(\begin{smallmatrix}[m+1+\sum_{l=1}^{d'-1}n_{l},-m] \\ [m+1,-m- n_1] \\ [m+1+n_1,-m-n_1-n_2] \\ \vdots \\ [m+1+\sum_{l=1}^{d'-2}n_{l},-m-\sum_{l=1}^{d'-1} n_{l}] \end{smallmatrix}\right)\,,
	}
where $S=\bC[Z,X]$ and 
\uaeq{\d_{P1}=\begin{pmatrix}0&  &  &  &  0\\ -X^{n_1} & Z^{n_1} &&&\\&-X^{n_2}&Z^{n_2}&&\\&&\ddots&\ddots&\\&&&-X^{n_{d'-1}}&Z^{n_{d'-1}}\end{pmatrix}\,.}

\section{Conclusion}
\label{sec:discussion}

We conclude with a list of questions for future investigation.
\begin{itemize}
\item 
It would be interesting to apply the construction outlined in this paper to other examples of RG flows and to find representations of more elaborate 2d TQFTs in free theories by means of projectors.\\
The treatment of flows between LG orbifolds discussed in section~\ref{sec:rg networks in lg minimal model orbifolds} easily carries over to flows between orbifolds of free chiral field theories. The latter theories can be obtained from LG orbifolds by setting the superpotentials to zero. The respective RG defects can be described in terms of matrix factorizations and have been worked out in \cite{Becker:2016umn}.  Somewhat more interesting examples are the bulk flows described in terms of gauged linear sigma models in \cite{Vafa:2001ra} and \cite{Moore:2005wp, Moore:2004yt}.
In all these examples, the flows are triggered by twisted chiral perturbations yielding RG defects preserving B-type supersymmetry. It would of course be interesting to apply this method to chiral perturbations as well. Unfortunately much less is known about A-type defects.

\item 
In the example of LG orbifolds $\cM_{d}/\bZ_d$ one could explicitly compare the bulk flows studied in section~\ref{sec:rg defects in LG orbs} with the corresponding flows of the identity defects. 
While the identity defects can be described by means of matrix factorizations, the respective
relevant perturbations are by twisted chiral fields. Such perturbations are not as 
easily treated in the matrix factorization framework as defect perturbations by chiral fields such as the ones discussed in \cite{Brunner:2009zt}.
In the case at hand however, the twisted chiral fields generating the perturbations are orbifold twist fields. As such, they do have a representation in the matrix factorization framework as defect changing fields. This possibly allows for an explicit analysis of the respective defect perturbations in this case. Hence, it might be possible to work out 
in this concrete example how the projection defects $P$ associated to bulk flows arise as perturbations of identity defects.

\item We have argued that bulk RG flows can be interpreted as flows on the identity defect of the UV theory.
Now, bulk flows between $N=2$ SCFTs give rise to $tt^*$ equations \cite{CECOTTI1991359,Cecotti:1992rm}. It would be very interesting to see, whether these equations also have a natural interpretation in terms of the flows on the identity defect.

\item The construction described in this paper heavily relies on topological covariance. So an obvious question is whether it has any bearing on non-topological QFTs (beyond topologically twisted supersymmetric theories). One can of course perturb the identity defect in non-topological QFTs. Also, RG defects exist in more general theories (see e.g.~\cite{Gaiotto:2012np} for an example). However, fusion of non-topological defects is singular in general. Still, at least in some cases it is possible to define a reasonable notion of fusion \cite{Bachas:2007td,Bachas:2012bj}, so that defects $P$ can be constructed from RG defects. The role of these defects is less clear, but it would be very interesting to study them in examples. Perhaps they are related to the line operators appearing in the context of integrable perturbations of conformal field theories \cite{Bazhanov:1994ft,Bazhanov:1996dr,Bazhanov:1998dq,Runkel:2007wd,Runkel:2010ym}.

\item
While the discussion in this paper is restricted  to 2d TQFTs, we expect that bulk perturbations of TQFTs can be described by means of codimension-one projection defects in any dimension.
Indeed, the generalized orbifold construction has been extended from dimension two to higher dimensions in \cite{Carqueville:2017aoe}. 
It would be very interesting to apply the methods described in this paper to higher-dimensional TQFTs.

\end{itemize}

\section*{Acknowledgements}

FK thanks Friedrich-Naumann-Stiftung and the Chinese Scholarship Council for supporting him throughout this project. DR's work is supported by hits. For ease of readability accross the literature, we have borrowed the defect coloring scheme from \cite{Brunner:2013ota}.

\appendix

\section{Properties of projection defects}
\label{app to chapter 3}

		\subsection{IR bulk fields in the UV}
	\label{sec:IR-UV correspondence}
	
	Here, we show that if a projection defect $P$ in the UV factorizes as $P\cong R^\dagger\otimes R$ with $R\otimes R^\dagger\cong I_\text{IR}$, then $P$-bimodule morphisms on $P$ are one-to-one with IR bulk fields ($I_\text{IR}$-bimodule morphisms of $I_\text{IR}$). The latter are mapped into the UV by the homomorphism
	\aeq{\tikz[baseline=25]{
		\fill[WScolor light] (0,0) rectangle (2,2);
		\draw[densely dashed] (1,0) -- (1,2);
		\node[defect node] at (1,1) {};
	}\mapsto\tikz[baseline=25]{
		\fill[WScolor] (0,0) rectangle (2,2);
		\draw[symmetry defect] (1,0) -- (1,2);
		\fill[WScolor light] (1,1) circle (.5);
		\draw[densely dashed] (1,.5) -- (1,1.5);
		\draw[defect] (1,1) circle (.5);
		\node[defect node] at (1,1) {};
	},\label{eq:bulkfieldiso}}
	where the junctions are given by the isomorphisms $P\cong R^\dagger\otimes R$. (To keep the notation light, we refrain from putting arrows on RG defects. We mark UV and IR theory by dark, respectively light background.) By the projection property the right hand side is a $P$-bimodule morphism of $P$.
		We claim that the inverse to the homomorphism (\ref{eq:bulkfieldiso}) is given by
	\uaeq{\tikz[baseline=25]{
		\fill[WScolor] (0,0) rectangle (2,2);
		\draw[symmetry defect] (1,0) -- (1,2);
		\node[symmetry node] at (1,1) {};
	}\mapsto\tikz[baseline=25]{
		\fill[WScolor light] (0,0) rectangle (2,2);
		\draw[densely dashed] (.75,1.5) -- (1,1.75);
		\draw[densely dashed] (1.25,1.5) -- (1,1.75);
		\draw[densely dashed] (1,1.75) -- (1,2);
		\draw[densely dashed] (.75,.5) -- (1,.25);
		\draw[densely dashed] (1.25,.5) -- (1,.25);
		\draw[densely dashed] (1,.25) -- (1,0);
		\fill[WScolor] (1,1.25) arc(0:180:.25) -- (.5,.75) arc(180:360:.25) arc(180:360:.25) -- (1.5,1.25) arc(0:180:.25);
		\draw[symmetry defect] (1,.75) -- (1,1.25);
		\node[symmetry node] at (1,1) {};
		\draw[defect] (1,1.25) arc(0:180:.25) -- (.5,.75) arc(180:360:.25);
		\draw[defect] (1,1.25) arc(180:0:.25) -- (1.5,.75) arc(360:180:.25);
	}.}
	Let us first check that the composition IR $\rightarrow$ UV $\rightarrow$ IR evaluates to the identity on IR bulk fields:\footnote{For readability, $I_\text{IR}$ has been omitted.}
	\uaeq{\tikz[baseline=40]{
		\fill[WScolor light] (0,0) rectangle (2,3);
		\node[defect node] at (1,1.5) {};
	}\mapsto\tikz[baseline=40]{
		\fill[WScolor] (0,0) rectangle (2,3);
		\draw[symmetry defect] (1,0) -- (1,3);
		\fill[WScolor light] (1,1.5) circle (.3);
		\draw[defect] (1,1.5) circle (.3);
		\node[defect node] at (1,1.5) {};
	}\mapsto\tikz[baseline=40]{
		\fill[WScolor light] (0,0) rectangle (2,3);
		\fill[WScolor] (1,.8) arc(360:180:.3) -- (.4,2.2) arc(180:0:.3) -- cycle;
		\fill[WScolor] (1,.8) arc(180:360:.3) -- (1.6,2.2) arc(0:180:.3) -- cycle;
		\draw[symmetry defect] (1,.8) -- (1,1.2);
		\draw[symmetry defect] (1,1.8) -- (1,2.2);
		\draw[defect] (1,.8) arc(360:180:.3) -- (.4,2.2) arc(180:0:.3);
		\draw[defect] (1,.8) arc(180:360:.3) -- (1.6,2.2) arc(0:180:.3);
		\fill[WScolor light] (1,1.5) circle (.3);
		\draw[defect] (1,1.5) circle (.3);
		\node[defect node] at (1,1.5) {};
	}\stackrel{P\cong R^\dagger\otimes R}{=}\tikz[baseline=40]{
		\fill[WScolor light] (0,0) rectangle (2,3);
		\fill[WScolor] (.9,.8) arc(360:180:.25) -- (.4,2.2) arc(180:0:.25) -- (.9,1.75) arc(90:270:.25) -- cycle;
		\draw[defect] (.9,.8) arc(360:180:.25) -- (.4,2.2) arc(180:0:.25) -- (.9,1.75) arc(90:270:.25) -- cycle;
		\fill[WScolor] (1.1,.8) arc(180:360:.25) -- (1.6,2.2) arc(0:180:.25) -- (1.1,1.75) arc(90:-90:.25) -- cycle;
		\draw[defect] (1.1,.8) arc(180:360:.25) -- (1.6,2.2) arc(0:180:.25) -- (1.1,1.75) arc(90:-90:.250000) -- cycle;
		\node[defect node] at (1,1.5) {};
	}\stackrel{R\otimes R^\dagger\cong I_\text{IR}}{=}\tikz[baseline=40]{
		\fill[WScolor light] (0,0) rectangle (2,3);
		\node[defect node] at (1,1.5) {};
	}}
	Similarly, the composition UV $\rightarrow$ IR $\rightarrow$ UV is the identity on $P$-bimodule morphisms of $P$:
	\uaeq{\tikz[baseline=25,scale=0.95]{
		\fill[WScolor] (.5,-.5) rectangle (1.5,2.5);
		\draw[symmetry defect] (1,-.5) -- (1,2.5);
		\node[symmetry node] at (1,1) {};
	}\mapsto\tikz[baseline=25,scale=0.95]{
		\fill[WScolor light] (0,-.5) rectangle (2,2.5);
		\fill[WScolor] (1,1.25) arc(0:180:.25) -- (.5,.75) arc(180:360:.25) arc(180:360:.25) -- (1.5,1.25) arc(0:180:.25);
		\draw[symmetry defect] (1,.75) -- (1,1.25);
		\node[symmetry node] at (1,1) {};
		\draw[defect] (1,1.25) arc(0:180:.25) -- (.5,.75) arc(180:360:.25);
		\draw[defect] (1,1.25) arc(180:0:.25) -- (1.5,.75) arc(360:180:.25);
	}\mapsto\tikz[baseline=25,scale=0.95]{
		\fill[WScolor] (0,-.5) rectangle (2,2.5);
		\draw[symmetry defect] (1,-.5) -- (1,-.2);
		\draw[symmetry defect] (1,2.5) -- (1,2.2);
		\fill[WScolor light] (1,-.2) .. controls (1,.2) and (.3,.2) .. (.3,1) .. controls (.3,1.8) and (1,1.8) .. (1,2.2) .. controls (1,1.8) and (1.7,1.8) .. (1.7,1) .. controls (1.7,.2) and (1,.2) .. (1,-.2) ;
		\draw[defect] (1,-.2) .. controls (1,.2) and (.3,.2) .. (.3,1) .. controls (.3,1.8) and (1,1.8) .. (1,2.2) .. controls (1,1.8) and (1.7,1.8) .. (1.7,1) .. controls (1.7,.2) and (1,.2) .. (1,-.2) ;
		\fill[WScolor] (1,1.25) arc(0:180:.25) -- (.5,.75) arc(180:360:.25) arc(180:360:.25) -- (1.5,1.25) arc(0:180:.25);
		\draw[symmetry defect] (1,.75) -- (1,1.25);
		\node[symmetry node] at (1,1) {};
		\draw[defect] (1,1.25) arc(0:180:.25) -- (.5,.75) arc(180:360:.25);
		\draw[defect] (1,1.25) arc(180:0:.25) -- (1.5,.75) arc(360:180:.25);
	}\stackrel{P\cong R^\dagger\otimes R}{=}\tikz[baseline=25,scale=0.95]{
		\fill[WScolor] (0,-.5) rectangle (2,2.5);
		\draw[symmetry defect] (1,-.5) -- (1,2.5);
		\node[symmetry node] at (1,1) {};
		\draw[symmetry defect] (1,0) .. controls (0,.5) and (0,1.5) .. (1,2);
		\draw[symmetry defect] (1,.5) .. controls (1.5,.75) and (1.5,1.25) .. (1,1.5);
	}\stackrel{\text{bimod. mor.}}{=}\tikz[baseline=25,scale=0.95]{
		\fill[WScolor] (0,-.5) rectangle (2,2.5);
		\draw[symmetry defect] (1,-.5) -- (1,2.5);
		\node[symmetry node] at (1,2.25) {};
		\draw[symmetry defect] (1,0) .. controls (0,.5) and (0,1.5) .. (1,2);
		\draw[symmetry defect] (1,.5) .. controls (1.5,.75) and (1.5,1.25) .. (1,1.5);
	}\stackrel{\text{loop}}{=}\tikz[baseline=25,scale=0.95]{
		\fill[WScolor] (.5,-.5) rectangle (1.5,2.5);
		\draw[symmetry defect] (1,-.5) -- (1,2.5);
		\node[symmetry node] at (1,1) {};
	}}
	Therefore we have established that the map (\ref{eq:bulkfieldiso}) is an isomorphism from the space of IR bulk fields to the space of $P$-bimodule morphisms of $P$.

	\subsection{Bimodule equal bicomodule morphisms}
	\label{app:Algebra equal coalgebra morphisms}

	For any projection defect, bimodule morphisms over itself are automatically cobimodule morphisms and vice versa. In other words, the two types of morphisms are one-to-one:
	\uaeq
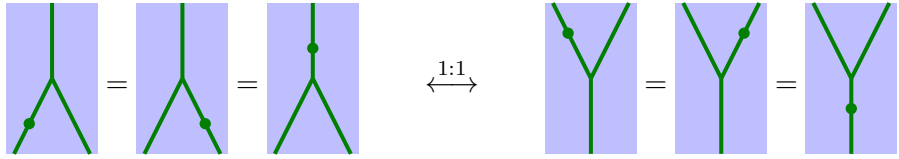
{\tikz[baseline=-5]{
		\fill[WScolor] (-.6,-1) rectangle (.6,1);
		\draw[symmetry defect] (-.5,-1) -- (0,0);
		\draw[symmetry defect] (.5,-1) -- (0,0);
		\draw[symmetry defect] (0,1) -- (0,0);
		\node[symmetry node] at (-.3,-.6) {};
	}=\tikz[baseline=-5]{
		\fill[WScolor] (-.6,-1) rectangle (.6,1);
		\draw[symmetry defect] (-.5,-1) -- (0,0);
		\draw[symmetry defect] (.5,-1) -- (0,0);
		\draw[symmetry defect] (0,1) -- (0,0);
		\node[symmetry node] at (.3,-.6) {};
	}=\tikz[baseline=-5]{
		\fill[WScolor] (-.6,-1) rectangle (.6,1);
		\draw[symmetry defect] (-.5,-1) -- (0,0);
		\draw[symmetry defect] (.5,-1) -- (0,0);
		\draw[symmetry defect] (0,1) -- (0,0);
		\node[symmetry node] at (0,.4) {};
	}\qquad\stackrel{\mathrm{1:1}}{\longleftrightarrow}\qquad\tikz[baseline=-5,yscale=-1]{
		\fill[WScolor] (-.6,-1) rectangle (.6,1);
		\draw[symmetry defect] (-.5,-1) -- (0,0);
		\draw[symmetry defect] (.5,-1) -- (0,0);
		\draw[symmetry defect] (0,1) -- (0,0);
		\node[symmetry node] at (-.3,-.6) {};
	}=\tikz[baseline=-5,yscale=-1]{
		\fill[WScolor] (-.6,-1) rectangle (.6,1);
		\draw[symmetry defect] (-.5,-1) -- (0,0);
		\draw[symmetry defect] (.5,-1) -- (0,0);
		\draw[symmetry defect] (0,1) -- (0,0);
		\node[symmetry node] at (.3,-.6) {};
	}=\tikz[baseline=-5,yscale=-1]{
		\fill[WScolor] (-.6,-1) rectangle (.6,1);
		\draw[symmetry defect] (-.5,-1) -- (0,0);
		\draw[symmetry defect] (.5,-1) -- (0,0);
		\draw[symmetry defect] (0,1) -- (0,0);
		\node[symmetry node] at (0,.4) {};
	}}
	This is easy to see. A bimodule morphism for example (left-hand side above) automatically obeys
	\uaeq
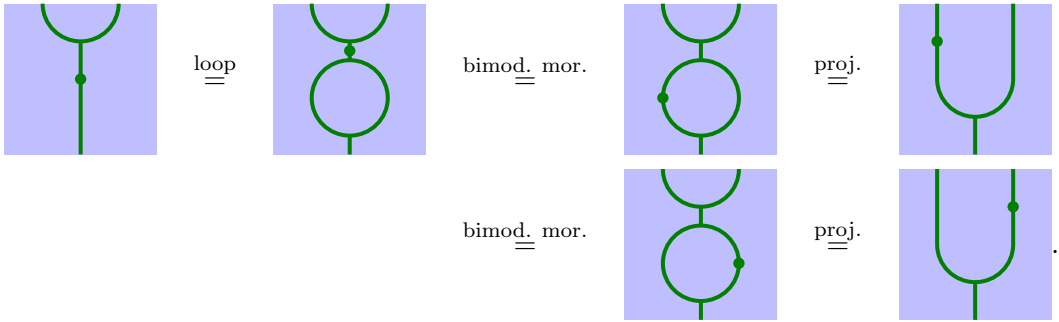
{\tikz[baseline=25]{
		\fill[WScolor] (0,0) rectangle (2,2);
		\draw[symmetry defect] (1,0) -- (1,1.5);
		\draw[symmetry defect] (.5,2) arc(180:360:.5);
		\node[symmetry node] at (1,1) {};
	}\quad\stackrel{\text{loop}}=\quad\tikz[baseline=25]{
		\fill[WScolor] (0,0) rectangle (2,2);
		\draw[symmetry defect] (1,0) -- (1,.25);
		\draw[symmetry defect] (1,1.25) -- (1,1.5);
		\draw[symmetry defect] (.5,2) arc(180:360:.5);
		\draw[symmetry defect] (1,.25) arc(270:270+360:.5);
		\node[symmetry node] at (1,1.375) {};
	}\quad &\stackrel{\text{bimod. mor.}}{=}\quad\tikz[baseline=25]{
		\fill[WScolor] (0,0) rectangle (2,2);
		\draw[symmetry defect] (1,0) -- (1,.25);
		\draw[symmetry defect] (1,1.25) -- (1,1.5);
		\draw[symmetry defect] (.5,2) arc(180:360:.5);
		\draw[symmetry defect] (1,.25) arc(270:270+360:.5);
		\node[symmetry node] at (.5,.75) {};
	}\quad\stackrel{\text{proj.}}=\quad\tikz[baseline=25]{
		\fill[WScolor] (0,0) rectangle (2,2);
		\draw[symmetry defect] (1,0) -- (1,.5);
		\draw[symmetry defect] (.5,2) -- (.5,1) arc(180:360:.5) -- (1.5,2);
		\node[symmetry node] at (.5,1.5) {};
	} \\
	&\stackrel{\text{bimod. mor.}}{=}\quad\tikz[baseline=25]{
		\fill[WScolor] (0,0) rectangle (2,2);
		\draw[symmetry defect] (1,0) -- (1,.25);
		\draw[symmetry defect] (1,1.25) -- (1,1.5);
		\draw[symmetry defect] (.5,2) arc(180:360:.5);
		\draw[symmetry defect] (1,.25) arc(270:270+360:.5);
		\node[symmetry node] at (1.5,.75) {};
	}\quad\stackrel{\text{proj.}}=\quad\tikz[baseline=25]{
		\fill[WScolor] (0,0) rectangle (2,2);
		\draw[symmetry defect] (1,0) -- (1,.5);
		\draw[symmetry defect] (.5,2) -- (.5,1) arc(180:360:.5) -- (1.5,2);
		\node[symmetry node] at (1.5,1.5) {};
	}.}
	The argument that bicomodule morphisms also respect the bimodule structure follows by turning the diagrams above upside down.
	
	Indeed, if the projection defect $P$ comes with a unit then all morphisms of $P$ automatically respect the bicomodule structure, and by the above also the bimodule structure on $P$. (This easily follows from the projection property.) Hence, in this case all morphisms of $P$ are bimodule morphisms and bicomodule morphisms. The same is true if $P$ has a counit.

	\subsection{\texorpdfstring{$P$-modules $B$ and $B\otimes P\cong B$}{P-modules B and B otimes P = B}}
	\label{sec:rephrasing P modules}
	
	In this appendix we show that for a projection defect $P$, any left boundary condition $B$ is a right $P$-module if and only if $B\otimes P\cong B$. This in particular means that left IR boundary conditions can be represented by left boundary conditions $B$ in the UV which are invariant under  fusion with $P$, i.e. $B\otimes P\cong B$. 
This statement extends to right boundaries and defects.
	
	First, a left $P$-module (whose comodule structure is induced by the unit on $P$) obeys $B\otimes P \cong B$:
\uaeq
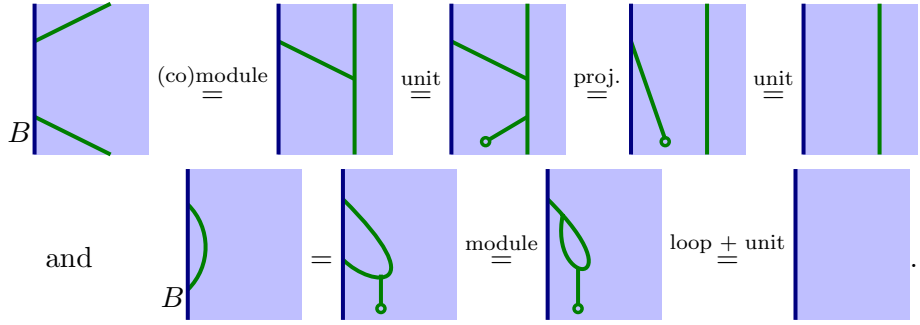
{
	\tikz[baseline=20]{
		\fill[WScolor] (0,0) rectangle (1.5,2);
		\draw[symmetry defect] (0,1.5) -- (1,2);
		\draw[symmetry defect] (0,.5) -- (1,0);
		\draw[defect] (0,0) -- (0,2);
		\node at (-.2,.3) {$B$};
	}\stackrel{\text{(co)module}}=\tikz[baseline=20]{
		\fill[WScolor] (0,0) rectangle (1.5,2);
		\draw[symmetry defect] (1,0) -- (1,2);
		\draw[symmetry defect] (0,1.5) -- (1,1);
		\draw[defect] (0,0) -- (0,2);
	}\stackrel{\text{unit}}=\tikz[baseline=20]{
		\fill[WScolor] (0,0) rectangle (1.5,2);
		\draw[symmetry defect] (1,0) -- (1,2);
		\draw[symmetry defect] (0,1.5) -- (1,1);
		\draw[defect] (0,0) -- (0,2);
		\draw[symmetry defect] (1,.5) -- (.5,.2);
		\node[unit node] at (.45,.17) {};
	}\stackrel{\text{proj.}}=\tikz[baseline=20]{
		\fill[WScolor] (0,0) rectangle (1.5,2);
		\draw[symmetry defect] (1,0) -- (1,2);
		\draw[symmetry defect] (0,1.5) -- (.45,.2);
		\draw[defect] (0,0) -- (0,2);
		\node[unit node] at (.45,.17) {};
	}\stackrel{\text{unit}}=\tikz[baseline=20]{
		\fill[WScolor] (0,0) rectangle (1.5,2);
		\draw[symmetry defect] (1,0) -- (1,2);
		\draw[defect] (0,0) -- (0,2);
	}\\
	\text{and}\qquad
	\tikz[baseline=20]{
		\fill[WScolor] (0,0) rectangle (1.5,2);
		\draw[symmetry defect] (0,.4) arc(-45:45:.8);
		\draw[defect] (0,0) -- (0,2);
		\node at (-.2,.3) {$B$};
	}=\tikz[baseline=20]{
		\fill[WScolor] (0,0) rectangle (1.5,2);
		\draw[symmetry defect] (0,.8) .. controls (.4,.4) and (1.2,.4) .. (0,1.6);
		\draw[symmetry defect] (.5,.6) -- (.5,.2);
		\draw[defect] (0,0) -- (0,2);
		\node[unit node] at (.5,.15) {};
	}\stackrel{\text{module}}=\tikz[baseline=20]{
		\fill[WScolor] (0,0) rectangle (1.5,2);
		\draw[symmetry defect] (.2,1.4) .. controls (0,.4) and (1.2,.4) .. (0,1.6);
		\draw[symmetry defect] (.4,.7) -- (.4,.2);
		\draw[defect] (0,0) -- (0,2);
		\node[unit node] at (.4,.15) {};
	}\stackrel{\text{loop + unit}}=\tikz[baseline=20]{
		\fill[WScolor] (0,0) rectangle (1.5,2);
		\draw[defect] (0,0) -- (0,2);
	}.}
	If on the other hand a left boundary condition $B$ satisfies $B\otimes P\cong B$, 
	$B$ inherits the $P$-module structure of $P$ itself. Namely, there are junctions	
 $B\otimes P\rightarrow B$ and $B\rightarrow B\otimes P$  such that
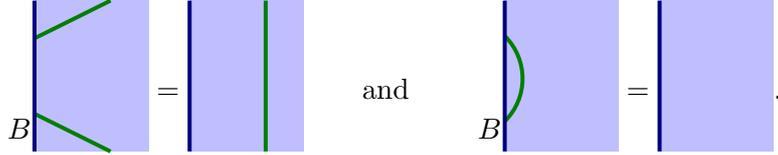
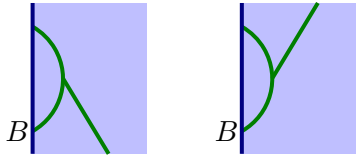
\uaeq{\tikz[baseline=20]{
		\fill[WScolor] (0,0) rectangle (1.5,2);
		\draw[symmetry defect] (0,1.5) -- (1,2);
		\draw[symmetry defect] (0,.5) -- (1,0);
		\draw[defect] (0,0) -- (0,2);
		\node at (-.2,.3) {$B$};
	}=\tikz[baseline=20]{
		\fill[WScolor] (0,0) rectangle (1.5,2);
		\draw[symmetry defect] (1,0) -- (1,2);
		\draw[defect] (0,0) -- (0,2);
	}\qquad\text{and}\qquad
	\tikz[baseline=20]{
		\fill[WScolor] (0,0) rectangle (1.5,2);
		\draw[symmetry defect] (0,.4) arc(-45:45:.8);
		\draw[defect] (0,0) -- (0,2);
		\node at (-.2,.3) {$B$};
	}=\tikz[baseline=20]{
		\fill[WScolor] (0,0) rectangle (1.5,2);
		\draw[defect] (0,0) -- (0,2);
	}.}
This implies that 
	\uaeq{\tikz{
		\fill[WScolor] (0,0) rectangle (1.5,2);
		\draw[symmetry defect] (0,.3) arc(-60:60:.8);
		\draw[defect] (0,0) -- (0,2);
		\node at (-.2,.3) {$B$};
		\draw[symmetry defect] (.4,1) -- (1,0);
	} \qquad\tikz{
		\fill[WScolor] (0,0) rectangle (1.5,2);
		\draw[symmetry defect] (0,.3) arc(-60:60:.8);
		\draw[defect] (0,0) -- (0,2);
		\node at (-.2,.3) {$B$};
		\draw[symmetry defect] (.4,1) -- (1,2);
	}}
	define $P$-module, respectively $P$-comodule structures on $B$, which  are also inverse to each other and hence provide isomorphisms
	$B\otimes P\cong P$.

	\subsection{\texorpdfstring{$P$-adjunction}{P-adjunction}}
	\label{app:P-adjoints}
	
	In this appendix we show that the adjunction of IR defects is lifted to the UV by the following formulas
		\uaeq{
		D^{\dagger_P} &= P \otimes D^\dagger\otimes P \\
		{}^{\dagger_P} D &= P\otimes {}^\dagger D \otimes P.
	}
	where $P$ is the corresponding projection defect, and $D$ is a defect in the UV theory representing an IR defect. We will only consider the first equation and will furthermore restrict to the case that $P$ is unital. The arguments for the second equation and the counital case are similar. 
	The IR right adjoints have to satisfy the following Zorro move identities	
	\uaeq
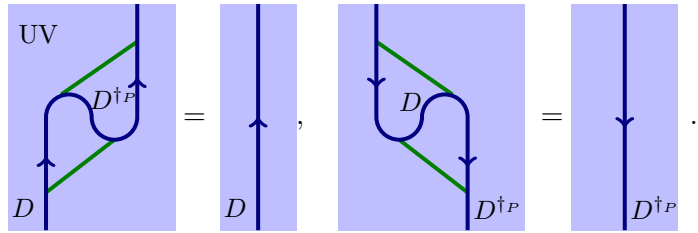
{\tikz[baseline=40]{
		\fill[WScolor] (0,0) rectangle (2.2,3);
		\draw[symmetry defect] (.7,1.8) -- (1.7,2.5);
		\draw[symmetry defect] (.5,.5) -- (1.4,1.2);
		\draw[defect, arrow position=.2, arrow position = .8] (.5,0) -- (.5,1.5) arc(180:0:.3) arc(180:360:.3) -- (1.7,3);
		\node at (.2,.3) {\small $D$};
		\node at (.4,2.6) {\small UV};
		\node at (1.4,1.75) {\small $D^{\dagger_P}$};
	}=\tikz[baseline=40]{
		\fill[WScolor] (0,0) rectangle (1,3);
		\draw[defect, arrow position=.5] (.5,0) -- (.5,3);
		\node at (.2,.3) {\small $D$};
	},\quad\tikz[baseline=40,xscale=-1]{
		\fill[WScolor] (-0.25,0) rectangle (2.2,3);
		\draw[symmetry defect] (.7,1.8) -- (1.7,2.5);
		\draw[symmetry defect] (.5,.5) -- (1.4,1.2);
		\draw[defect, opp arrow position=.2, opp arrow position = .8] (.5,0) -- (.5,1.5) arc(180:0:.3) arc(180:360:.3) -- (1.7,3);
		\node at (0.1,.3) {\small $D^{\dagger_P}$};
		\node at (1.25,1.7) {\small $D$};
	}=\tikz[baseline=40]{
		\fill[WScolor] (-0.2,0) rectangle (1.2,3);
		\draw[defect, opp arrow position = .5] (.5,0) -- (.5,3);
		\node at (.9,.3) {\small $D^{\dagger_P}$};
	}.}
	These are satisfied for $D^{\dagger_P}$ above when we choose the following natural (co-)evaluation maps
		\uaeq
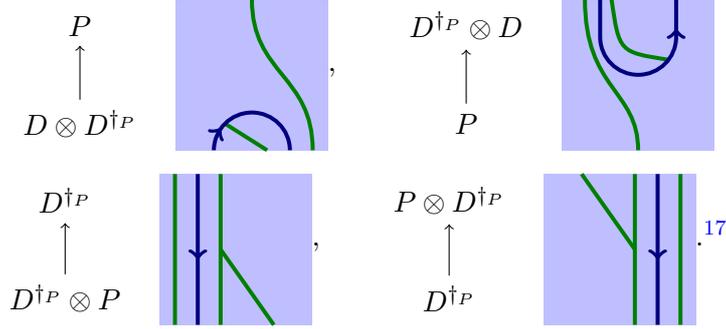
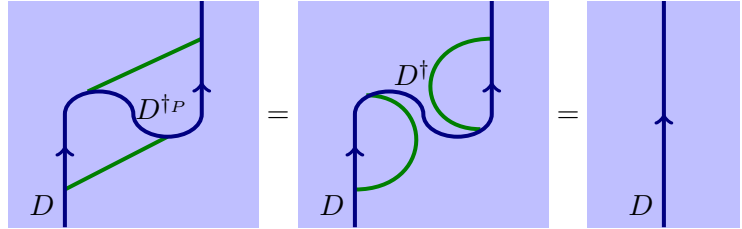
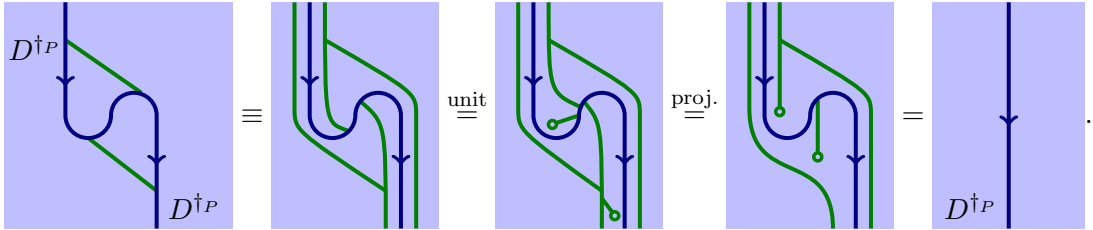
{\tikz[baseline=20]{
			\node (start) at (0,0) {$D\otimes D^{\dagger_P}$};
			\node (end) at (0,1.3) {$P$};
			\draw[->] (start) -- (end);
		}\quad\tikz[baseline=30]{
		\fill[WScolor] (0,0) rectangle (2,2);
		\draw[symmetry defect] (1.8,0) .. controls (1.8,1) and (1,1) .. (1,2); 
		\draw[symmetry defect] (1.2,0) -- (.65,.35); 
		\draw[defect, arrow position = .2] (.5,0) arc (180:0:.5);
	},\qquad\tikz[baseline=20]{
			\node (start) at (0,1.3) {$D^{\dagger_P}\otimes D$};
			\node (end) at (0,0) {$P$};
			\draw[<-] (start) -- (end);
		}\quad\tikz[baseline=30]{
		\fill[WScolor] (0,0) rectangle (2,2);
		\draw[symmetry defect] (1.4,1.2) .. controls (.75,1.3) .. (.65,2); 
		\draw[symmetry defect] (.3,2) .. controls (.3,.8) and (1,.8) .. (1,0);
		\draw[defect, arrow position = .85] (.5,2) -- (.5,1.5) arc (180:360:.5) -- (1.5,2);
	}}
	\uaeq{\tikz[baseline=20]{
			\node (start) at (0,0) {$D^{\dagger_P}\otimes P$};
			\node (end) at (0,1.3) {$D^{\dagger_P}$};
			\draw[->] (start) -- (end);
		}\quad\tikz[baseline=30]{
		\fill[WScolor] (0,0) rectangle (2,2);
		\draw[defect, opp arrow position = .5] (.5,0) -- (.5,2); 
		\draw[symmetry defect] (.8,0) -- (.8,2);
		\draw[symmetry defect] (.8,1) -- (1.5,0);
		\draw[symmetry defect] (.2,0) -- (.2,2);
	},\qquad\tikz[baseline=20]{
			\node (start) at (0,0) {$D^{\dagger_P}$};
			\node (end) at (0,1.3) {$P\otimes D^{\dagger_P}$};
			\draw[->] (start) -- (end);
		}\quad\tikz[baseline=30,xscale=-1]{
		\fill[WScolor] (0,0) rectangle (2,2);
		\draw[defect, opp arrow position = .5] (.5,0) -- (.5,2); 
		\draw[symmetry defect] (.8,0) -- (.8,2);
		\draw[symmetry defect] (.8,1) -- (1.5,2);
		\draw[symmetry defect] (.2,0) -- (.2,2);
	}.\footnotemark}\footnotetext{The lower maps follow as natural generalizations from the generalized orbifold procedure \cite[Prop. 4.7]{Carqueville:2012dk}. Namely,
	\uaeq{\tikz[baseline=30]{
		\fill[WScolor] (0,0) rectangle (2.5,2);
		\draw[symmetry defect] (.3,2) .. controls (0,0) and (1.7,.2) .. (1.7,0);
		\draw[symmetry defect] (.7,2) .. controls (.7,1.5) and (2.5,2) .. (2.1,0);
		\draw[symmetry defect] (.9,.95) -- (.9,.85);
		\node[unit node] at (.9,.8) {};
		\draw[symmetry defect] (.9,1.8) .. controls (.6,.8) and (1,.8) .. (1.2,1.1);
		\draw[defect, arrow position = .57] (.5,2) -- (.5,1) arc(180:360:.35) arc(180:0:.35) -- (1.9,0);
		\draw[symmetry defect] (2.1,1) -- (2.5,.5);
	}=\tikz[baseline=30]{
		\fill[WScolor] (0,0) rectangle (2,2);
		\draw[defect, arrow position = .5] (1,2) -- (1,0);
		\draw[symmetry defect] (.5,2) -- (.5,0);
		\draw[symmetry defect] (1.5,2) -- (1.5,0);
		\draw[symmetry defect] (1.5,1) -- (2,.5);
	}\qquad\tikz[baseline=30]{
			\fill[WScolor] (0,0) rectangle (2,2);
			\draw[symmetry defect] (1.7,0) -- (1.7,1.5) .. controls (1.7,2) and (.7,1.8) .. (.7,2);
			\draw[symmetry defect] (1.3,0) .. controls (1.3,.4) and (.3,.4) .. (.3,2);
			\draw[symmetry defect] (.85,1.05) -- (.85,.8);
			\node[unit node] at (.85,.77) {};
			\draw[symmetry defect] (1,1.4) .. controls (1,.9) and (.7,.9) .. (.35,1.4);
			\draw[symmetry defect] (.6,.8) .. controls (.6,.5) and (0,.5) .. (0,1.5);
			\draw[symmetry defect] (.4,.6) -- (.4,.4);
			\node[unit node] at (.4,.35) {};
			\draw[defect, arrow position = .37] (.5,2) -- (.5,1.5) arc(180:360:.25) arc(180:0:.25) -- (1.5,0);
		}=\tikz[baseline=30,xscale=-1]{
			\fill[WScolor] (0,0) rectangle (2,2);
			\draw[defect, opp arrow position = .5] (.5,0) -- (.5,2); 
			\draw[symmetry defect] (.8,0) -- (.8,2);
			\draw[symmetry defect] (.8,1) -- (1.5,2);
			\draw[symmetry defect] (.2,0) -- (.2,2);
		}}}
	Namely,
	\uaeq{\tikz[baseline=40,xscale=1.5]{
		\fill[WScolor] (0,0) rectangle (2.2,3);
		\draw[symmetry defect] (.7,1.8) -- (1.7,2.5);
		\draw[symmetry defect] (.5,.5) -- (1.4,1.2);
		\draw[defect, arrow position=.2, arrow position = .8] (.5,0) -- (.5,1.5) arc(180:0:.3) arc(180:360:.3) -- (1.7,3);
		\node at (1.35,1.6) {$D^{\dagger_P}$};
		\node at (.3,.3) {$D$};
	}=\tikz[baseline=40,xscale=1.5]{
		\fill[WScolor] (0,0) rectangle (2.2,3);
		\draw[symmetry defect] (1.6,1.3) .. controls (1,1.3) and (1,2.5) .. (1.7,2.5);
		\draw[symmetry defect] (.5,.5) .. controls (1.2,.5) and (1.2,1.75) .. (.6,1.75);
		\draw[defect, arrow position=.2, arrow position = .8] (.5,0) -- (.5,1.5) arc(180:0:.3) arc(180:360:.3) -- (1.7,3);
		\node at (1,2.05) {$D^{\dagger}$};
		\node at (.3,.3) {$D$};
	}=\tikz[baseline=40]{
		\fill[WScolor] (-.5,0) rectangle (1.5,3);
		\draw[defect, arrow position=.5] (.5,0) -- (.5,3);
		\node at (.2,.3) {$D$};
	}}
	and
	\uaeq{\tikz[baseline=40,xscale=-1]{
		\fill[WScolor] (-.5,0) rectangle (2.5,3);
		\draw[symmetry defect] (.7,1.8) -- (1.7,2.5);
		\draw[symmetry defect] (.5,.5) -- (1.4,1.2);
		\draw[defect, opp arrow position=.2, opp arrow position = .8] (.5,0) -- (.5,1.5) arc(180:0:.3) arc(180:360:.3) -- (1.7,3);
		\node at (2.1,2.4) {$D^{\dagger_P}$};
		\node at (0,.3) {$D^{\dagger_P}$};
	}\equiv\tikz[baseline=40,xscale=-1]{
		\fill[WScolor] (0,0) rectangle (2.2,3);
		\draw[symmetry defect] (1.5,3) .. controls (1.5,1.4) .. (1.2,1.3);
		\draw[symmetry defect] (.3,0) -- (.3,1.5) .. controls (.3,1.8) .. (1.5,2.5);
		\draw[symmetry defect] (.7,0) .. controls (.7,1.4) .. (1.05,1.7);
		\draw[symmetry defect] (1.9,3) -- (1.9,1.6) .. controls (1.9,1.3) .. (.7,.5);
		\draw[defect, opp arrow position=.2, opp arrow position = .8] (.5,0) -- (.5,1.5) arc(180:0:.3) arc(180:360:.3) -- (1.7,3);
	}\stackrel{\text{unit}}=\tikz[baseline=40,xscale=-1]{
		\fill[WScolor] (0,0) rectangle (2.2,3);
		\draw[symmetry defect] (1.5,3) .. controls (1.5,1.8) .. (1.05,1.6);
		\draw[symmetry defect] (1.1,1.5) -- (1.4,1.4);
		\node[unit node] at (1.45,1.38) {};
		\draw[symmetry defect] (.3,0) -- (.3,1.5) .. controls (.3,1.8) .. (1.5,2.5);
		\draw[symmetry defect] (.8,0) .. controls (.8,1.4) .. (1.05,1.7);
		\draw[symmetry defect] (1.9,3) -- (1.9,1.6) .. controls (1.9,1.3) .. (.8,.5);
		\draw[symmetry defect] (.8,.4) -- (.65,.2);
		\node[unit node] at (.63,.17) {};
		\draw[defect, opp arrow position=.2, opp arrow position = .8] (.5,0) -- (.5,1.5) arc(180:0:.3) arc(180:360:.3) -- (1.7,3);
	}\stackrel{\text{proj.}}=\tikz[baseline=40,xscale=-1]{
		\fill[WScolor] (0,0) rectangle (2.2,3);
		\draw[symmetry defect] (1.5,3) -- (1.5,1.6);
		\node[unit node] at (1.5,1.55) {};
		\draw[symmetry defect] (.3,0) -- (.3,1.5) .. controls (.3,1.8) .. (1.5,2.5);
		\draw[symmetry defect] (1,1.7) -- (1,1);
		\node[unit node] at (1,.95) {};
		\draw[symmetry defect] (1.9,3) -- (1.9,1.6) .. controls (1.9,.5) and (.8,1.1) .. (.8,0);
		\draw[defect, opp arrow position=.2, opp arrow position = .8] (.5,0) -- (.5,1.5) arc(180:0:.3) arc(180:360:.3) -- (1.7,3);
	}=\tikz[baseline=40]{
		\fill[WScolor] (-.5,0) rectangle (1.5,3);
		\draw[defect, opp arrow position = .5] (.5,0) -- (.5,3);
		\node at (0,.3) {$D^{\dagger_P}$};
	}.}
	using unit condition, Zorro-move and the definition of $D^{\dagger_P}$ in the last step. For counital $P$ and left-adjoints the above diagrams have to be flipped appropriately.
	
	The defect $P$ is a $P$-module and a $P$-comodule, so it can be regarded as an IR defect. As such, it should be selfadjoint, and, using the above notion of IR adjunction one finds that this is indeed the case:  $P^{\dagger_P}\cong P \cong {}^{\dagger_P}P$. If for instance $P$ is unital, the two maps 
		\uaeq
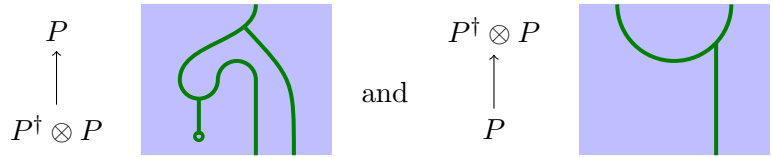
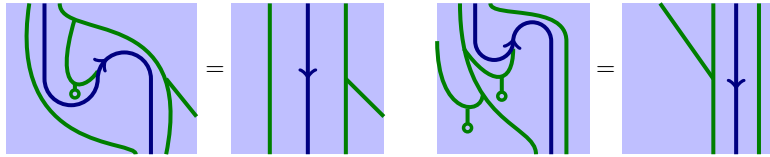
{\tikz[baseline=10]{
			\node (start) at (0,0) {$P^\dagger\otimes P$};
			\node (end) at (0,1.3) {$P$};
			\draw[->] (start) -- (end);
		}\quad\tikz[baseline=20]{
		\fill[WScolor] (0,0) rectangle (2.5,2);
		\draw[symmetry defect] (1.5,0) -- (1.5,1) arc(0:180:.25) arc(360:180:.25) .. controls (.5,1.5) and (1.5,1.5) .. (1.5,2);
		\draw[symmetry defect] (2,0) .. controls (2,1) .. (1.35,1.7);
		\draw[symmetry defect] (.75,.27) -- (.75,.75);
		\node[unit node] at (.75,.25) {};
	}\quad\text{and}\quad\tikz[baseline=10]{
			\node (start) at (0,1.3) {$P^\dagger\otimes P$};
			\node (end) at (0,0) {$P$};
			\draw[<-] (start) -- (end);
		}\quad\tikz[baseline=20]{
			\fill[WScolor] (0,0) rectangle (2.5,2);
			\draw[symmetry defect] (.5,2) arc(180:360:.75);
			\draw[symmetry defect] (1.8,0) -- (1.8,1.5);
	}}
	are inverse to each other and hence provide isomorphisms $P\cong P^\dagger\otimes P$. 
	The projection property of $P$, $P\otimes P\cong P$ implies that $P^{\dagger_P}=P\otimes P^\dagger\otimes P\cong P$.
	The argument for counital $P$ is analogous.

	\subsection{Projections with unit and counit}
	\label{app:Projections with unit and counit}
	
	If a projection defect has a unit as well as a counit, it is automatically self-adjoint as there are natural (co)evaluation maps:
	\uaeq
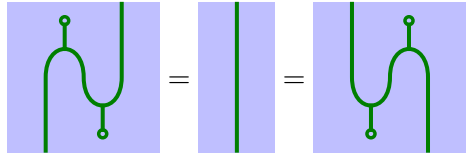
{\tikz[baseline=25]{
		\fill[WScolor] (0,0) rectangle (2,2);
		\draw[symmetry defect] (.5,0) -- (.5,1) .. controls (.5,1.5) and (1,1.5) .. (1,1) .. controls (1,.5) and (1.5,.5) .. (1.5,1) -- (1.5,2);
		\draw[symmetry defect] (.75,1.35) -- (.75,1.725);
		\node[unit node] at (.75,1.75) {};
		\draw[symmetry defect] (1.25,.65) -- (1.25,2-1.725);
		\node[unit node] at (1.25,.25) {};
	}=\tikz[baseline=25]{
		\fill[WScolor] (0,0) rectangle (1,2);
		\draw[symmetry defect] (.5,0) -- (.5,2);
	}=\tikz[baseline=25,xscale=-1]{
		\fill[WScolor] (0,0) rectangle (2,2);
		\draw[symmetry defect] (.5,0) -- (.5,1) .. controls (.5,1.5) and (1,1.5) .. (1,1) .. controls (1,.5) and (1.5,.5) .. (1.5,1) -- (1.5,2);
		\draw[symmetry defect] (.75,1.35) -- (.75,1.725);
		\node[unit node] at (.75,1.75) {};
		\draw[symmetry defect] (1.25,.65) -- (1.25,2-1.725);
		\node[unit node] at (1.25,.25) {};
	}}
	The equalities follow from the two Frobenius and (co)unit properties. If $P$ has a unit, any $P$-module automatically carries a $P$-comodule structure. Vice versa, any $P$-comodule is automatically a $P$-module in case $P$ has a counit. If $P$ has both, a unit and a counit, these two constructions are compatible:
Starting with a $P$-module, a $P$-comodule structure is induced which in turn induces a $P$-module structure. This $P$-module structure is identical to the original one:
	\uaeq
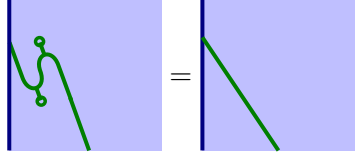
{\tikz[baseline=25]{
		\fill[WScolor] (0,0) rectangle (2,2);
		\begin{scope}[scale=.5,xscale=-1,shift={(-2.1,1.5)},rotate around={-20:(0,0)}]
			\draw[symmetry defect] (.5,0) -- (.5,1) .. controls (.5,1.5) and (1,1.5) .. (1,1) .. controls (1,.5) and (1.5,.5) .. (1.5,1) -- (1.5,2);
			\draw[symmetry defect] (.75,1.35) -- (.75,1.725);
			\node[unit node] at (.75,1.75) {};
			\draw[symmetry defect] (1.25,.65) -- (1.25,2-1.725);
			\node[unit node] at (1.25,.25) {};
		\end{scope}
		\draw[defect] (0,0) -- (0,2);
		\draw[symmetry defect] (.8,.7) -- (1.05,0);
	}=\tikz[baseline=25]{
		\fill[WScolor] (0,0) rectangle (2,2);
		\draw[defect] (0,0) -- (0,2);
		\draw[symmetry defect] (0,1.5) -- (1,0);
	}}
	
	As discussed in section~\ref{sec:rg defects from projections}, all projection defects $P$ factorize into RG type defects $P\cong R^\dagger\otimes R$.  For example, unitary $P$ factorize as
	\uaeq{
		\left(P_{\text{IR}|\text{UV}}\right)^{\dagger} &= P_{\text{UV}|\text{IR}} \\
		{}^{\dagger}\!\left(P_{\text{IR}|\text{UV}}\right) &= \left({}^\dagger P\right)_{\text{UV}|\text{IR}}\,.
	}
	For selfadjoint projection defects $P$, the respective RG type defects $R$ then satisfy ${}^\dagger R \cong R^\dagger$.

\subsection{Equivalence of (co)multiplication and Frobenius properties for projections}
\label{app:Equivalence of (co)multiplication and Frobenius properties for projections}

Here, we show the equivalence of associativity, coassociativity and the two Frobenius properties for projection defects and how they follow from the existence of a (co)unit. The identities in question are:
	\uaeq
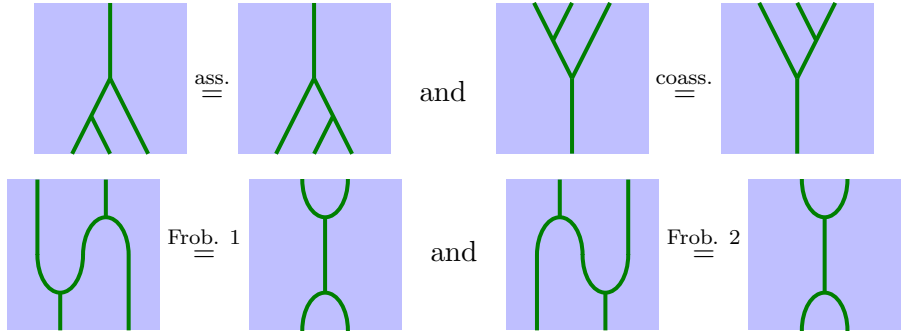
{\tikz[baseline=20]{
		\fill[WScolor] (0,0) rectangle (2,2);
		\draw[symmetry defect] (.5,0) -- (1,1);
		\draw[symmetry defect] (1,1) -- (1,2);
		\draw[symmetry defect] (1,0) -- (.75,.5);
		\draw[symmetry defect] (1.5,0) -- (1,1);
	}\stackrel{\text{ass.}}=\tikz[baseline=20,xscale=-1]{
		\fill[WScolor] (0,0) rectangle (2,2);
		\draw[symmetry defect] (.5,0) -- (1,1);
		\draw[symmetry defect] (1,1) -- (1,2);
		\draw[symmetry defect] (1,0) -- (.75,.5);
		\draw[symmetry defect] (1.5,0) -- (1,1);
	}\quad\text{and}\quad\tikz[baseline=-37,yscale=-1]{
		\fill[WScolor] (0,0) rectangle (2,2);
		\draw[symmetry defect] (.5,0) -- (1,1);
		\draw[symmetry defect] (1,1) -- (1,2);
		\draw[symmetry defect] (1,0) -- (.75,.5);
		\draw[symmetry defect] (1.5,0) -- (1,1);
	}\stackrel{\text{coass.}}=\tikz[baseline=-37,xscale=-1,yscale=-1]{
		\fill[WScolor] (0,0) rectangle (2,2);
		\draw[symmetry defect] (.5,0) -- (1,1);
		\draw[symmetry defect] (1,1) -- (1,2);
		\draw[symmetry defect] (1,0) -- (.75,.5);
		\draw[symmetry defect] (1.5,0) -- (1,1);
	}}
	\uaeq{\tikz[anchor=base, baseline=12]{
			\fill[WScolor] (-1,-.5) rectangle (1,1.5);
			\begin{scope}[yscale=-1,xscale=1, shift={(-.3,-.5)}]
				\clip (-.4,0) rectangle (.4,.6);
				\draw[symmetry defect] (0,0) ellipse (.3 and .5);
			\end{scope}
			\begin{scope}[shift={(.3,.5)}]
				\clip (-.4,0) rectangle (.4,.6);
				\draw[symmetry defect] (0,0) ellipse (.3 and .5);
			\end{scope}
			\draw[symmetry defect] (.6,-.5) -- (.6,.5);
			\draw[symmetry defect] (.3,1) -- (.3,1.5);
			\draw[symmetry defect] (-.6,.5) --(-.6, 1.5);
			\draw[symmetry defect] (-.3,0) --(-.3, -.5);
		}
		\stackrel{\text{Frob. 1}}=
		\tikz[anchor=base, baseline=12]{
			\fill[WScolor] (-1,-.5) rectangle (1,1.5);
			\begin{scope}[yscale=-1,xscale=1, shift={(0,-1.5)}]
				\clip (-.4,0) rectangle (.4,.6);
				\draw[symmetry defect] (0,0) ellipse (.3 and .5);
			\end{scope}
			\draw[symmetry defect] (0,0) -- (0,1);
			\begin{scope}[shift={(0,-.5)}]
				\clip (-.4,0) rectangle (.4,.6);
				\draw[symmetry defect] (0,0) ellipse (.3 and .5);
			\end{scope}
		}
		\quad\text{and}\quad
		\tikz[anchor=base, baseline=12]{\begin{scope}[yscale=1,xscale=-1]
			\fill[WScolor] (-1,-.5) rectangle (1,1.5);
			\begin{scope}[yscale=-1,xscale=1, shift={(-.3,-.5)}]
				\clip (-.4,0) rectangle (.4,.6);
				\draw[symmetry defect] (0,0) ellipse (.3 and .5);
			\end{scope}
			\begin{scope}[shift={(.3,.5)}]
				\clip (-.4,0) rectangle (.4,.6);
				\draw[symmetry defect] (0,0) ellipse (.3 and .5);
			\end{scope}
			\draw[symmetry defect] (.6,-.5) -- (.6,.5);
			\draw[symmetry defect] (.3,1) -- (.3,1.5);
			\draw[symmetry defect] (-.6,.5) --(-.6, 1.5);
			\draw[symmetry defect] (-.3,0) --(-.3, -.5);
		\end{scope}
		}\stackrel{\text{Frob. 2}}=
		\tikz[anchor=base, baseline=12]{
			\fill[WScolor] (-1,-.5) rectangle (1,1.5);
			\begin{scope}[yscale=-1,xscale=1, shift={(0,-1.5)}]
				\clip (-.4,0) rectangle (.4,.6);
				\draw[symmetry defect] (0,0) ellipse (.3 and .5);
			\end{scope}
			\draw[symmetry defect] (0,0) -- (0,1);
			\begin{scope}[shift={(0,-.5)}]
				\clip (-.4,0) rectangle (.4,.6);
				\draw[symmetry defect] (0,0) ellipse (.3 and .5);
			\end{scope}
		}
	}
Equivalence is shown by the following chain of implications
	\uaeq{
		\text{associativity}\Rightarrow \text{Frob. 2}\Rightarrow \text{coassociativity}\Rightarrow\text{Frob. 1} \Rightarrow\text{associativity}.
	}
	\subsubsection*{associativity $\Rightarrow$ Frob. 2:}
	\uaeq
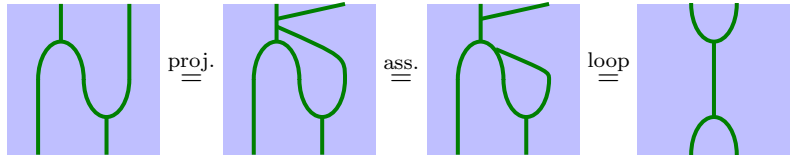
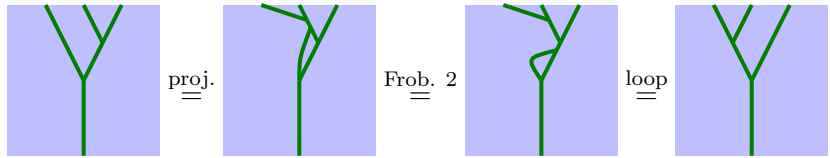
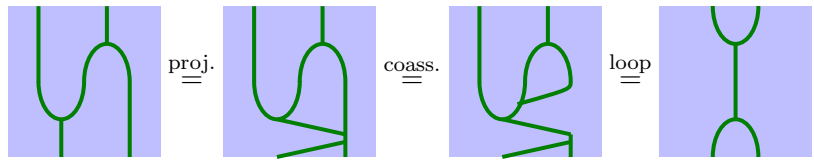
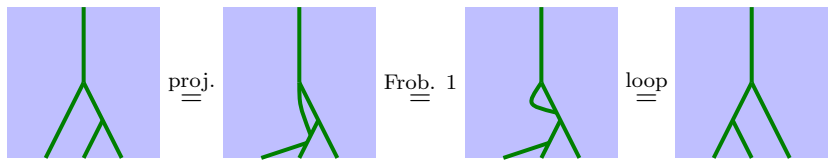
{
		\tikz[anchor=base, baseline=12]{\begin{scope}[yscale=1,xscale=-1]
			\fill[WScolor] (-1,-.5) rectangle (1,1.5);
			\begin{scope}[yscale=-1,xscale=1, shift={(-.3,-.5)}]
				\clip (-.4,0) rectangle (.4,.6);
				\draw[symmetry defect] (0,0) ellipse (.3 and .5);
			\end{scope}
			\begin{scope}[shift={(.3,.5)}]
				\clip (-.4,0) rectangle (.4,.6);
				\draw[symmetry defect] (0,0) ellipse (.3 and .5);
			\end{scope}
			\draw[symmetry defect] (.6,-.5) -- (.6,.5);
			\draw[symmetry defect] (.3,1) -- (.3,1.5);
			\draw[symmetry defect] (-.6,.5) --(-.6, 1.5);
			\draw[symmetry defect] (-.3,0) --(-.3, -.5);
		\end{scope}
		}\stackrel{\text{proj.}}=\tikz[anchor=base, baseline=12]{\begin{scope}[,xscale=-1]
			\fill[WScolor] (-1,-.5) rectangle (1,1.5);
			\begin{scope}[yscale=-1,xscale=1, shift={(-.3,-.5)}]
				\clip (-.4,0) rectangle (.4,.6);
				\draw[symmetry defect] (0,0) ellipse (.3 and .5);
			\end{scope}
			\begin{scope}[shift={(.3,.5)}]
				\clip (-.4,0) rectangle (.4,.6);
				\draw[symmetry defect] (0,0) ellipse (.3 and .5);
			\end{scope}
			\draw[symmetry defect] (.6,-.5) -- (.6,.5);
			\draw[symmetry defect] (.3,1) -- (.3,1.5);
			\draw[symmetry defect] (-.6,.5) .. controls (-.6,.8) .. (.3, 1.2);
			\draw[symmetry defect] (.3,1.3) -- (-.6, 1.5);
			\draw[symmetry defect] (-.3,0) --(-.3, -.5);
		\end{scope}
		}\stackrel{\text{ass.}}=\tikz[anchor=base, baseline=12]{\begin{scope}[,xscale=-1]
			\fill[WScolor] (-1,-.5) rectangle (1,1.5);
			\begin{scope}[yscale=-1,xscale=1, shift={(-.3,-.5)}]
				\clip (-.4,0) rectangle (.4,.6);
				\draw[symmetry defect] (0,0) ellipse (.3 and .5);
			\end{scope}
			\begin{scope}[shift={(.3,.5)}]
				\clip (-.4,0) rectangle (.4,.6);
				\draw[symmetry defect] (0,0) ellipse (.3 and .5);
			\end{scope}
			\draw[symmetry defect] (.6,-.5) -- (.6,.5);
			\draw[symmetry defect] (.3,1) -- (.3,1.5);
			\draw[symmetry defect] (-.6,.5) .. controls (-.6,.6) .. (.1, .9);
			\draw[symmetry defect] (.3,1.3) -- (-.6, 1.5);
			\draw[symmetry defect] (-.3,0) --(-.3, -.5);
		\end{scope}
		}\stackrel{\text{loop}}=
		\tikz[anchor=base, baseline=12]{
			\fill[WScolor] (-1,-.5) rectangle (1,1.5);
			\begin{scope}[yscale=-1,xscale=1, shift={(0,-1.5)}]
				\clip (-.4,0) rectangle (.4,.6);
				\draw[symmetry defect] (0,0) ellipse (.3 and .5);
			\end{scope}
			\draw[symmetry defect] (0,0) -- (0,1);
			\begin{scope}[shift={(0,-.5)}]
				\clip (-.4,0) rectangle (.4,.6);
				\draw[symmetry defect] (0,0) ellipse (.3 and .5);
			\end{scope}
		}
	}
	\subsubsection*{Frob. 2 $\Rightarrow$ coassociativity:}
	
	\uaeq{\tikz[baseline=-37,xscale=-1,yscale=-1]{
		\fill[WScolor] (0,0) rectangle (2,2);
		\draw[symmetry defect] (.5,0) -- (1,1);
		\draw[symmetry defect] (1,1) -- (1,2);
		\draw[symmetry defect] (1,0) -- (.75,.5);
		\draw[symmetry defect] (1.5,0) -- (1,1);
	}\stackrel{\text{proj.}}=\tikz[baseline=-37,xscale=-1,yscale=-1]{
		\fill[WScolor] (0,0) rectangle (2,2);
		\draw[symmetry defect] (.5,0) -- (1,1);
		\draw[symmetry defect] (1,1) -- (1,2);
		\draw[symmetry defect] (1,0) -- (.75,.5);
		\draw[symmetry defect] (.85,.3) .. controls (1,.7) .. (1,1);
		\draw[symmetry defect] (1.5,0) -- (.9,.2);
	}\stackrel{\text{Frob. 2}}=\tikz[baseline=-37,xscale=-1,yscale=-1]{
		\fill[WScolor] (0,0) rectangle (2,2);
		\draw[symmetry defect] (.5,0) -- (1,1);
		\draw[symmetry defect] (1,1) -- (1,2);
		\draw[symmetry defect] (1,0) -- (.75,.5);
		\draw[symmetry defect] (.8,.6) .. controls (1.2,.7) .. (1,1);
		\draw[symmetry defect] (1.5,0) -- (.9,.2);
	}\stackrel{\text{loop}}=\tikz[baseline=-37,yscale=-1]{
		\fill[WScolor] (0,0) rectangle (2,2);
		\draw[symmetry defect] (.5,0) -- (1,1);
		\draw[symmetry defect] (1,1) -- (1,2);
		\draw[symmetry defect] (1,0) -- (.75,.5);
		\draw[symmetry defect] (1.5,0) -- (1,1);
	}}
	
	\subsubsection*{coassociativity $\Rightarrow$ Frob. 1:}
	
	\uaeq{\tikz[anchor=base, baseline=12]{
			\fill[WScolor] (-1,-.5) rectangle (1,1.5);
			\begin{scope}[yscale=-1,xscale=1, shift={(-.3,-.5)}]
				\clip (-.4,0) rectangle (.4,.6);
				\draw[symmetry defect] (0,0) ellipse (.3 and .5);
			\end{scope}
			\begin{scope}[shift={(.3,.5)}]
				\clip (-.4,0) rectangle (.4,.6);
				\draw[symmetry defect] (0,0) ellipse (.3 and .5);
			\end{scope}
			\draw[symmetry defect] (.6,-.5) -- (.6,.5);
			\draw[symmetry defect] (.3,1) -- (.3,1.5);
			\draw[symmetry defect] (-.6,.5) --(-.6, 1.5);
			\draw[symmetry defect] (-.3,0) --(-.3, -.5);
		}\stackrel{\text{proj.}}=\tikz[anchor=base, baseline=12]{
			\fill[WScolor] (-1,-.5) rectangle (1,1.5);
			\begin{scope}[yscale=-1,xscale=1, shift={(-.3,-.5)}]
				\clip (-.4,0) rectangle (.4,.6);
				\draw[symmetry defect] (0,0) ellipse (.3 and .5);
			\end{scope}
			\begin{scope}[shift={(.3,.5)}]
				\clip (-.4,0) rectangle (.4,.6);
				\draw[symmetry defect] (0,0) ellipse (.3 and .5);
			\end{scope}
			\draw[symmetry defect] (.6,-.5) -- (.6,.5);
			\draw[symmetry defect] (.3,1) -- (.3,1.5);
			\draw[symmetry defect] (-.6,.5) --(-.6, 1.5);
			\draw[symmetry defect] (-.3,0) -- (.6, -.2);
			\draw[symmetry defect] (.6,-.3) -- (-.3, -.5);
		}\stackrel{\text{coass.}}=\tikz[anchor=base, baseline=12]{
			\fill[WScolor] (-1,-.5) rectangle (1,1.5);
			\begin{scope}[yscale=-1,xscale=1, shift={(-.3,-.5)}]
				\clip (-.4,0) rectangle (.4,.6);
				\draw[symmetry defect] (0,0) ellipse (.3 and .5);
			\end{scope}
			\begin{scope}[shift={(.3,.5)}]
				\clip (-.4,0) rectangle (.4,.6);
				\draw[symmetry defect] (0,0) ellipse (.3 and .5);
			\end{scope}
			\draw[symmetry defect] (.6,.5) .. controls (.6,.4) .. (-.1,.2);
			\draw[symmetry defect] (.6,-.5) -- (.6,-.2);
			\draw[symmetry defect] (.3,1) -- (.3,1.5);
			\draw[symmetry defect] (-.6,.5) --(-.6, 1.5);
			\draw[symmetry defect] (-.3,0) -- (.6, -.2);
			\draw[symmetry defect] (.6,-.3) -- (-.3, -.5);
		}\stackrel{\text{loop}}=\tikz[anchor=base, baseline=12]{
			\fill[WScolor] (-1,-.5) rectangle (1,1.5);
			\begin{scope}[yscale=-1,xscale=1, shift={(0,-1.5)}]
				\clip (-.4,0) rectangle (.4,.6);
				\draw[symmetry defect] (0,0) ellipse (.3 and .5);
			\end{scope}
			\draw[symmetry defect] (0,0) -- (0,1);
			\begin{scope}[shift={(0,-.5)}]
				\clip (-.4,0) rectangle (.4,.6);
				\draw[symmetry defect] (0,0) ellipse (.3 and .5);
			\end{scope}
		}}
	
	\subsubsection*{Frob. 1 $\Rightarrow$ associativity:}
	
	\uaeq{\tikz[baseline=20,xscale=-1,yscale=1]{
		\fill[WScolor] (0,0) rectangle (2,2);
		\draw[symmetry defect] (.5,0) -- (1,1);
		\draw[symmetry defect] (1,1) -- (1,2);
		\draw[symmetry defect] (1,0) -- (.75,.5);
		\draw[symmetry defect] (1.5,0) -- (1,1);
	}\stackrel{\text{proj.}}=\tikz[baseline=20,xscale=-1,yscale=1]{
		\fill[WScolor] (0,0) rectangle (2,2);
		\draw[symmetry defect] (.5,0) -- (1,1);
		\draw[symmetry defect] (1,1) -- (1,2);
		\draw[symmetry defect] (1,0) -- (.75,.5);
		\draw[symmetry defect] (.85,.3) .. controls (1,.7) .. (1,1);
		\draw[symmetry defect] (1.5,0) -- (.9,.2);
	}\stackrel{\text{Frob. 1}}=\tikz[baseline=20,xscale=-1,yscale=1]{
		\fill[WScolor] (0,0) rectangle (2,2);
		\draw[symmetry defect] (.5,0) -- (1,1);
		\draw[symmetry defect] (1,1) -- (1,2);
		\draw[symmetry defect] (1,0) -- (.75,.5);
		\draw[symmetry defect] (.8,.6) .. controls (1.2,.7) .. (1,1);
		\draw[symmetry defect] (1.5,0) -- (.9,.2);
	}\stackrel{\text{loop}}=\tikz[baseline=20,yscale=1]{
		\fill[WScolor] (0,0) rectangle (2,2);
		\draw[symmetry defect] (.5,0) -- (1,1);
		\draw[symmetry defect] (1,1) -- (1,2);
		\draw[symmetry defect] (1,0) -- (.75,.5);
		\draw[symmetry defect] (1.5,0) -- (1,1);
	}}
	Next, we show how the existence of a unit for a projection defect implies coassociativity:
	\subsubsection*{unit $\Rightarrow$ coassociativity:}
	
	\uaeq
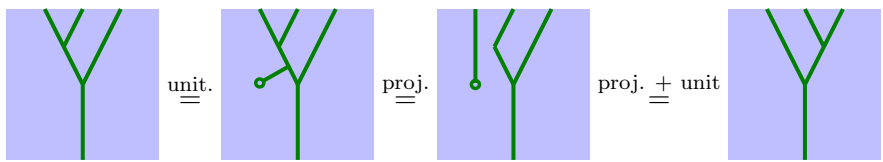
{\tikz[baseline=-37,yscale=-1]{
		\fill[WScolor] (0,0) rectangle (2,2);
		\draw[symmetry defect] (.5,0) -- (1,1);
		\draw[symmetry defect] (1,1) -- (1,2);
		\draw[symmetry defect] (1,0) -- (.75,.5);
		\draw[symmetry defect] (1.5,0) -- (1,1);
	}\stackrel{\text{unit.}}=\tikz[baseline=-37,yscale=-1]{
		\fill[WScolor] (0,0) rectangle (2,2);
		\draw[symmetry defect] (.5,0) -- (1,1);
		\draw[symmetry defect] (1,1) -- (1,2);
		\draw[symmetry defect] (1,0) -- (.75,.5);
		\draw[symmetry defect] (1.5,0) -- (1,1);
		\draw[symmetry defect] (.9,.75) -- (.55,.95);
		\node[unit node] at (.5,.98) {};
	}\stackrel{\text{proj.}}=\tikz[baseline=-37,yscale=-1]{
		\fill[WScolor] (0,0) rectangle (2,2);
		\draw[symmetry defect] (.5,0) -- (.5,.98);
		\node[unit node] at (.5,1) {};
		\draw[symmetry defect] (.75,.5) -- (1,1);
		\draw[symmetry defect] (1,1) -- (1,2);
		\draw[symmetry defect] (1,0) -- (.75,.5);
		\draw[symmetry defect] (1.5,0) -- (1,1);
	}\stackrel{\text{proj. + unit}}=\tikz[baseline=-37,xscale=-1,yscale=-1]{
		\fill[WScolor] (0,0) rectangle (2,2);
		\draw[symmetry defect] (.5,0) -- (1,1);
		\draw[symmetry defect] (1,1) -- (1,2);
		\draw[symmetry defect] (1,0) -- (.75,.5);
		\draw[symmetry defect] (1.5,0) -- (1,1);
	}}
	In the last step we applied the projection property to the left and the lower defect.
	Turning these diagrams upside down shows how associativity follows from the existence of a counit.

	\subsection{Adjoints of induced RG defects}
	\label{app:Adjoints of induced RG defects}
	
	Right and left adjoints of the induced RG defect $P_{\text{IR}|\text{UV}}$ must satisfy the Zorro move identities
	\uaeq{\tikz[baseline=40]{
		\fill[WScolor] (0,0) rectangle (2.2,3);
		\draw[symmetry defect] (.7,1.8) -- (1.7,2.5);
		\draw[densely dashed] (.5,.5) -- (1.4,1.2);
		\draw[defect, arrow position=.2, arrow position = .8] (.5,0) -- (.5,1.5) arc(180:0:.3) arc(180:360:.3) -- (1.7,3);
		\node at (.5,-.25) {$P_{\text{IR}|\text{UV}}$};
	}=\tikz[baseline=40]{
		\fill[WScolor] (0,0) rectangle (1,3);
		\draw[defect, arrow position=.5] (.5,0) -- (.5,3);
		\node at (.5,-.25) {$P_{\text{IR}|\text{UV}}$};
	}\quad\text{and}\quad\tikz[baseline=40,xscale=-1]{
		\fill[WScolor] (0,0) rectangle (2.2,3);
		\draw[symmetry defect] (.7,1.8) -- (1.7,2.5);
		\draw[densely dashed] (.5,.5) -- (1.4,1.2);
		\draw[defect, opp arrow position=.2, opp arrow position = .8] (.5,0) -- (.5,1.5) arc(180:0:.3) arc(180:360:.3) -- (1.7,3);
		\node at (.5,-.25) {$\left(P_{\text{IR}|\text{UV}}\right)^\dagger$};
	}=\tikz[baseline=40,xscale=-1]{
		\fill[WScolor] (0,0) rectangle (1,3);
		\draw[defect, opp arrow position=.5] (.5,0) -- (.5,3);
		\node at (.5,-.25) {$\left(P_{\text{IR}|\text{UV}}\right)^\dagger$};
	},}
	\uaeq{\tikz[baseline=40,xscale=-1]{
		\fill[WScolor] (0,0) rectangle (2.2,3);
		\draw[densely dashed] (.7,1.8) -- (1.7,2.5);
		\draw[symmetry defect] (.5,.5) -- (1.4,1.2);
		\draw[defect, arrow position=.2, arrow position = .8] (.5,0) -- (.5,1.5) arc(180:0:.3) arc(180:360:.3) -- (1.7,3);
		\node at (.5,-.25) {$P_{\text{IR}|\text{UV}}$};
	}=\tikz[baseline=40,xscale=-1]{
		\fill[WScolor] (0,0) rectangle (1.2,3);
		\draw[defect, arrow position=.5] (.5,0) -- (.5,3);
		\node at (.5,-.25) {$P_{\text{IR}|\text{UV}}$};
	}\quad\text{and}\quad\tikz[baseline=40]{
		\fill[WScolor] (0,0) rectangle (2.2,3);
		\draw[densely dashed] (.7,1.8) -- (1.7,2.5);
		\draw[symmetry defect] (.5,.5) -- (1.4,1.2);
		\draw[defect, opp arrow position=.2, opp arrow position = .8] (.5,0) -- (.5,1.5) arc(180:0:.3) arc(180:360:.3) -- (1.7,3);
		\node at (.5,-.25) {${}^\dagger\left(P_{\text{IR}|\text{UV}}\right)$};
	}=\tikz[baseline=40]{
		\fill[WScolor] (-.5,0) rectangle (1.5,3);
		\draw[defect, opp arrow position=.5] (.5,0) -- (.5,3);
		\node at (.5,-.25) {${}^\dagger\left(P_{\text{IR}|\text{UV}}\right)$};
	}.}
	We will discuss the case that $P$ has a unit. The counital case can be treated analogously. Indeed, it is easy to see that
	\uaeq{
		\left(P_{\text{IR}|\text{UV}}\right)^\dagger &= P_{\text{UV}|\text{IR}},
	}
	i.e. $P$ regarded as defect from IR to UV is the right adjoint of $P$ regarded as a defect from UV to IR.
The evaluation map is just given by the algebra  $P\otimes P\rightarrow P$
and the coevaluation map is induced by the unit $I_\text{UV}\rightarrow P\rightarrow P\otimes P$. (The Zorro identities immediately follow from associativity and the unit condition.) It is slightly more involved to see that the left adjoint is given by
	\uaeq{
		{}^\dagger\left(P_{\text{IR}|\text{UV}}\right) &= \left({}^\dagger P\right)_{\text{UV}|\text{IR}},
	}
	the left adjoint of  $P$ regarded as a defect from the IR to the UV theory. Evaluation and coevaluation are given by the maps
	\uaeq{\tikz[baseline=35]{
			\fill[WScolor] (0,0) rectangle (2,2.6);
			\draw[symmetry defect] (.5,0) -- (.5,2.6);
			\draw[symmetry defect] (.5,1) -- (1.5,0);
			\node at (.25,.2) {${}^\dagger P$};
			\node at (.25,2.4) {${}^\dagger P$};
			\node at (1.2,.55) {$P$};
	}\quad:=\quad\tikz[baseline=35]{
			\fill[WScolor] (0,0) rectangle (2.5,2.6);
			\draw[symmetry defect] (.8,0) -- (.8,1) arc(180:0:.35) arc(180:360:.35) -- (2.2,2.6);
			\draw[symmetry defect] (.4,0) -- (.4,1.325) arc(180:0:.375);
			\node at (.25,1.7) {${}^\dagger P$};
			\node at (1.9,2.4) {${}^\dagger P$};
			\node at (1.1,.3) {$P$};
		}\quad\text{and}\quad\tikz[baseline=35]{
	\fill[WScolor] (0,0) rectangle (2,2.6);
	\draw[symmetry defect] (.5,0) -- (.5,2.6);
	\draw[symmetry defect] (.5,1) -- (1.5,2.6);
	\node at (.3,.2) {$P$};
	\node at (.3,2.4) {$P$};
	\node at (1.2,1.6) {${}^\dagger P$};
	}\quad :=\quad\tikz[baseline=-20, scale=1.3,yscale=-1,xscale=-1]{\begin{scope}[xscale=-1]
			\fill[WScolor] (-1,-.5) rectangle (1,1.5);
			\begin{scope}[yscale=-1,xscale=1, shift={(-.3,-.5)}]
				\clip (-.4,0) rectangle (.4,.6);
				\draw[symmetry defect] (0,0) ellipse (.3 and .5);
			\end{scope}
			\begin{scope}[shift={(.3,.5)}]
				\clip (-.4,0) rectangle (.4,.6);
				\draw[symmetry defect] (0,0) ellipse (.3 and .5);
			\end{scope}
			\draw[symmetry defect] (.6,-.5) -- (.6,.5);
			\draw[symmetry defect] (-.6,.5) --(-.6, 1.5);
			\draw[symmetry defect] (-.3,0) --(-.3, -.5);
		\end{scope}
			\node at (.4,1.2) {$P$};
			\node at (-.8,-.3) {${}^\dagger P$};
			\node at (.5,-.3) {$P$};
			\node at (-.2,.3) {$P$};
		},}
which define the right $P$-module structure of ${}^\dagger P$ and the right ${}^\dagger P$-comodule structure of $P$, respectively. The first Zorro identity then follows from the UV Zorro move and loop omission, while the second one additionally requires associativity and the unit property.

\section{Generalized orbifold theories}
\label{sec:generalized orbifold theories}
	
	The generalized orbifold procedure \cite{Frohlich:2009gb, Brunner:2013xna, Brunner:2013ota, Carqueville:2012dk} is a method to produce a new theory out of a given 2d TQFT $T$ by inserting networks of an endo-defect 
	$A:T\rightarrow T$ into its correlation functions. These modified correlation functions are well-defined if the defect $A$ satisfies the following special properties. It has to come with 
(co)multiplication and (co)unit fields
	\uaeq{
		\tikz[anchor=base, baseline=12]{
			\begin{scope}
				\clip (-.4,0) rectangle (.4,.6);
				\draw[symmetry defect] (0,0) ellipse (.3 and .5);
			\end{scope}
			\node[symmetry node] at (0, .5) {};
			\draw[symmetry defect] (0,.5) -- (0,1);
		}\;
		A\otimes A\rightarrow A,\qquad
		\tikz[anchor=base, baseline=12]{\begin{scope}[yscale=-1,xscale=1,shift={(0,-1)}]
			\begin{scope}
				\clip (-.4,0) rectangle (.4,.6);
				\draw[symmetry defect] (0,0) ellipse (.3 and .5);
			\end{scope}
			\node[symmetry node] at (0, .5) {};
			\draw[symmetry defect] (0,.5) -- (0,1);
		\end{scope}}\;
		A\rightarrow A\otimes A,\qquad
		\tikz[anchor=base, baseline=18]{
			\node[unit node] at (0, 1.05) {};
			\draw[symmetry defect] (0,.5) -- (0,1);
		}\;
		A\rightarrow I,\qquad
		\tikz[anchor=base, baseline=18]{
			\node[unit node] at (0, .45) {};
			\draw[symmetry defect] (0,.5) -- (0,1);
		}\;
		I\rightarrow A
	}
	which make $A$ into a separable Frobenius algebra, i.e. it obeys the (co)associativity and (co)unit conditions
	\uaeq{
		\tikz[anchor=base, baseline=18]{
			\draw[symmetry defect] (.6,0) -- (.6,.5);
			\begin{scope}
				\clip (-.4,0) rectangle (.4,.6);
				\draw[symmetry defect] (0,0) ellipse (.3 and .5);
			\end{scope}
			\node[symmetry node] at (0, .5) {};
			\begin{scope}[shift={(.3,.5)}]
				\clip (-.4,0) rectangle (.4,.6);
				\draw[symmetry defect] (0,0) ellipse (.3 and .5);
			\end{scope}
			\node[symmetry node] at (.3, 1) {};
			\draw[symmetry defect] (.3,1) -- (.3,1.5);
		}
		=
		\tikz[anchor=base, baseline=18]{\begin{scope}[yscale=1,xscale=-1]
			\draw[symmetry defect] (.6,0) -- (.6,.5);
			\begin{scope}
				\clip (-.4,0) rectangle (.4,.6);
				\draw[symmetry defect] (0,0) ellipse (.3 and .5);
			\end{scope}
			\node[symmetry node] at (0, .5) {};
			\begin{scope}[shift={(.3,.5)}]
				\clip (-.4,0) rectangle (.4,.6);
				\draw[symmetry defect] (0,0) ellipse (.3 and .5);
			\end{scope}
			\node[symmetry node] at (.3, 1) {};
			\draw[symmetry defect] (.3,1) -- (.3,1.5);
		\end{scope}}
		,\quad
		\tikz[anchor=base, baseline=18]{
			\draw[symmetry defect] (0,0) -- (0,1.5);
			\node[symmetry node] at (0,1.2) {};
			\draw[symmetry defect] (-.4,.4) .. controls (-.4,.7) .. (0,1.2);
			\node[unit node] at (-.4,.35) {};
		}
		=
		\tikz[anchor=base, baseline=18]{
			\draw[symmetry defect] (0,0) -- (0,1.5);
		}
		=
		\tikz[anchor=base, baseline=18]{\begin{scope}[yscale=1,xscale=-1]
			\draw[symmetry defect] (0,0) -- (0,1.5);
			\node[symmetry node] at (0,1.2) {};
			\draw[symmetry defect] (-.4,.4) .. controls (-.4,.7) .. (0,1.2);
			\node[unit node] at (-.4,.35) {};
		\end{scope}}
		,\quad
		\tikz[anchor=base, baseline=18]{\begin{scope}[yscale=-1,xscale=1, shift={(0,-1.5)}]
			\draw[symmetry defect] (.6,0) -- (.6,.5);
			\begin{scope}
				\clip (-.4,0) rectangle (.4,.6);
				\draw[symmetry defect] (0,0) ellipse (.3 and .5);
			\end{scope}
			\node[symmetry node] at (0, .5) {};
			\begin{scope}[shift={(.3,.5)}]
				\clip (-.4,0) rectangle (.4,.6);
				\draw[symmetry defect] (0,0) ellipse (.3 and .5);
			\end{scope}
			\node[symmetry node] at (.3, 1) {};
			\draw[symmetry defect] (.3,1) -- (.3,1.5);
		\end{scope}}
		=
		\tikz[anchor=base, baseline=18]{\begin{scope}[yscale=-1,xscale=-1, shift={(0,-1.5)}]
			\draw[symmetry defect] (.6,0) -- (.6,.5);
			\begin{scope}
				\clip (-.4,0) rectangle (.4,.6);
				\draw[symmetry defect] (0,0) ellipse (.3 and .5);
			\end{scope}
			\node[symmetry node] at (0, .5) {};
			\begin{scope}[shift={(.3,.5)}]
				\clip (-.4,0) rectangle (.4,.6);
				\draw[symmetry defect] (0,0) ellipse (.3 and .5);
			\end{scope}
			\node[symmetry node] at (.3, 1) {};
			\draw[symmetry defect] (.3,1) -- (.3,1.5);
		\end{scope}}
		,\quad
		\tikz[anchor=base, baseline=18]{\begin{scope}[yscale=-1,xscale=1, shift={(0,-1.5)}]
			\draw[symmetry defect] (0,0) -- (0,1.5);
			\node[symmetry node] at (0,1.2) {};
			\draw[symmetry defect] (-.4,.4) .. controls (-.4,.7) .. (0,1.2);
			\node[unit node] at (-.4,.35) {};
		\end{scope}}
		=
		\tikz[anchor=base, baseline=18]{
			\draw[symmetry defect] (0,0) -- (0,1.5);
		}
		=
		\tikz[anchor=base, baseline=18]{\begin{scope}[yscale=-1,xscale=-1, shift={(0,-1.5)}]
			\draw[symmetry defect] (0,0) -- (0,1.5);
			\node[symmetry node] at (0,1.2) {};
			\draw[symmetry defect] (-.4,.4) .. controls (-.4,.7) .. (0,1.2);
			\node[unit node] at (-.4,.35) {};
		\end{scope}}
	}
	as well as the Frobenius and loop-omission properties:
	\uaeq{
		\tikz[anchor=base, baseline=12]{
			\begin{scope}[yscale=-1,xscale=1, shift={(-.3,-.5)}]
				\clip (-.4,0) rectangle (.4,.6);
				\draw[symmetry defect] (0,0) ellipse (.3 and .5);
			\end{scope}
			\node[symmetry node] at (-.3,0) {};
			\begin{scope}[shift={(.3,.5)}]
				\clip (-.4,0) rectangle (.4,.6);
				\draw[symmetry defect] (0,0) ellipse (.3 and .5);
			\end{scope}
			\node[symmetry node] at (.3, 1) {};
			\draw[symmetry defect] (.6,-.5) -- (.6,.5);
			\draw[symmetry defect] (.3,1) -- (.3,1.5);
			\draw[symmetry defect] (-.6,.5) --(-.6, 1.5);
			\draw[symmetry defect] (-.3,0) --(-.3, -.5);
		}
		=
		\tikz[anchor=base, baseline=12]{
			\begin{scope}[yscale=-1,xscale=1, shift={(0,-1.5)}]
				\clip (-.4,0) rectangle (.4,.6);
				\draw[symmetry defect] (0,0) ellipse (.3 and .5);
			\end{scope}
			\node[symmetry node] at (0,1) {};
			\draw[symmetry defect] (0,0) -- (0,1);
			\node[symmetry node] at (0, 0) {};
			\begin{scope}[shift={(0,-.5)}]
				\clip (-.4,0) rectangle (.4,.6);
				\draw[symmetry defect] (0,0) ellipse (.3 and .5);
			\end{scope}
		}
		=
		\tikz[anchor=base, baseline=12]{\begin{scope}[yscale=1,xscale=-1]
			\begin{scope}[yscale=-1,xscale=1, shift={(-.3,-.5)}]
				\clip (-.4,0) rectangle (.4,.6);
				\draw[symmetry defect] (0,0) ellipse (.3 and .5);
			\end{scope}
			\node[symmetry node] at (-.3,0) {};
			\begin{scope}[shift={(.3,.5)}]
				\clip (-.4,0) rectangle (.4,.6);
				\draw[symmetry defect] (0,0) ellipse (.3 and .5);
			\end{scope}
			\node[symmetry node] at (.3, 1) {};
			\draw[symmetry defect] (.6,-.5) -- (.6,.5);
			\draw[symmetry defect] (.3,1) -- (.3,1.5);
			\draw[symmetry defect] (-.6,.5) --(-.6, 1.5);
			\draw[symmetry defect] (-.3,0) --(-.3, -.5);
		\end{scope}}
		,\quad
		\tikz[anchor=base, baseline=12]{
			\draw[symmetry defect] (0,.5) ellipse (.3 and .5);
			\node[symmetry node] at (0,1) {};
			\draw[symmetry defect] (0,1) -- (0,1.5);
			\draw[symmetry defect] (0,-.5) -- (0,0);
			\node[symmetry node] at (0, 0) {};
		}
		=
		\tikz[anchor=base, baseline=12]{
			\draw[symmetry defect] (0,-.5) -- (0,1.5);
		}.
	}
	The respective orbifold theory is denoted by $(T,A)$. An obvious example for a defect satsifying the above conditions is the identity defect $A=I$ in any TQFT. Orbifolding by $I$ of course just gives back the original theory, $(T,I)\cong T$.
	In the following we will briefly outline how objects in the orbifold $(T,A)$ are defined in terms of objects in $A$.
	
	For any two TQFTs $T$ and $T'$ with defects $A$ and $A'$ as above an $A$-$A'$-bimodule $D$ is a defect $D:T'\rightarrow T$ with junctions $A\otimes D \rightarrow D$, $D\otimes A' \rightarrow D$ such that
	\uaeq{
		\tikz[anchor=base, baseline=18]{
			\draw[defect] (0,0) -- (0,1.5);
			\node[symmetry node] at (0,1.2) {};
			\node[symmetry node] at (0,.8) {};
			\draw[symmetry defect] (-.8,0) .. controls (-.8,.4) .. (0,1.2);
			\draw[symmetry defect] (-.4,0) .. controls (-.4,.3) .. (0,.8);
		}
		=
		\tikz[anchor=base, baseline=18]{
			\draw[defect] (0,0) -- (0,1.5);
			\node[symmetry node] at (0,1.2) {};
			\node[symmetry node] at (-.5,.5) {};
			\draw[symmetry defect] (-.5,.5) .. controls (-.4,.8) .. (0,1.2);
			\begin{scope}[shift={(-.5,0)}]
				\clip (-.4,0) rectangle (.4,.6);
				\draw[symmetry defect] (0,0) ellipse (.3 and .5);
			\end{scope}
		},
		\qquad
		\tikz[anchor=base, baseline=18]{
			\draw[defect] (0,0) -- (0,1.5);
			\node[symmetry node] at (0,1.2) {};
			\draw[symmetry defect] (-.4,.4) .. controls (-.4,.7) .. (0,1.2);
			\node[unit node] at (-.4,.35) {};
		}
		=
		\tikz[anchor=base, baseline=18]{
			\draw[defect] (0,0) -- (0,1.5);
		}
		=
		\tikz[anchor=base, baseline=18]{\begin{scope}[yscale=1,xscale=-1]
			\draw[defect] (0,0) -- (0,1.5);
			\node[symmetry node] at (0,1.2) {};
			\draw[symmetry defect] (-.4,.4) .. controls (-.4,.7) .. (0,1.2);
			\node[unit node] at (-.4,.35) {};
		\end{scope}}
		,\qquad
		\tikz[anchor=base, baseline=18]{\begin{scope}[yscale=1,xscale=-1]
			\draw[defect] (0,0) -- (0,1.5);
			\node[symmetry node] at (0,1.2) {};
			
			\node[symmetry node] at (0,.8) {};
			\draw[symmetry defect] (-.8,0) .. controls (-.8,.4) .. (0,1.2);
			\draw[symmetry defect] (-.4,0) .. controls (-.4,.3) .. (0,.8);
		\end{scope}}
		=
		\tikz[anchor=base, baseline=18]{\begin{scope}[yscale=1,xscale=-1]
			\draw[defect] (0,0) -- (0,1.5);
			\node[symmetry node] at (0,1.2) {};
			\node[symmetry node] at (-.5,.5) {};
			\draw[symmetry defect] (-.5,.5) .. controls (-.4,.8) .. (0,1.2);
			\begin{scope}[shift={(-.5,0)}]
				\clip (-.4,0) rectangle (.4,.6);
				\draw[symmetry defect] (0,0) ellipse (.3 and .5);
			\end{scope}
		\end{scope}}.
	}
	For two such bimodules $D$ and $\tilde D$, $\Hom_{A, A'}(D, \tilde D)$ denotes the space of all defect changing fields $D\rightarrow \tilde D$ commuting with the bimodule structure, i.e.
	\uaeq{
		\tikz[anchor=base, baseline=18]{
			\draw[defect] (0,0) -- (0,1.5);
			\node[symmetry node] at (0,1.2) {};
			\draw[symmetry defect] (-.8,0) .. controls (-.8,.4) .. (0,1.2);
			\node[defect node] at (0,1) {};
		}
		=
		\tikz[anchor=base, baseline=18]{
			\draw[defect] (0,0) -- (0,1.5);
			\node[symmetry node] at (0,.8) {};
			\node[defect node] at (0,1) {};
			\draw[symmetry defect] (-.4,0) .. controls (-.4,.3) .. (0,.8);
		}
		,\qquad
		\tikz[anchor=base, baseline=18]{\begin{scope}[,xscale=-1]
			\draw[defect] (0,0) -- (0,1.5);
			\node[symmetry node] at (0,1.2) {};
			\node[defect node] at (0,1) {};
			\draw[symmetry defect] (-.8,0) .. controls (-.8,.4) .. (0,1.2);
		\end{scope}}
		=
		\tikz[anchor=base, baseline=18]{\begin{scope}[xscale=-1]
			\draw[defect] (0,0) -- (0,1.5);
			\node[symmetry node] at (0,.8) {};
			\node[defect node] at (0,1) {};
			\draw[symmetry defect] (-.4,0) .. controls (-.4,.3) .. (0,.8);
		\end{scope}}.
	}
	Via the unit, such modules are automatically also comodules, c.f.~\cite[eqn. (4.1)]{Carqueville:2012dk}:
	\uaeq{
		\tikz[anchor=base, baseline=-25]{\begin{scope}[yscale=-1]
			\draw[defect] (0,0) -- (0,1.5);
			\node[symmetry node] at (0,.8) {};
			\draw[symmetry defect] (-.4,0) .. controls (-.4,.3) .. (0,.8);
		\end{scope}} :=
		\tikz[anchor=base, baseline=18]{
			\draw[defect] (0,0) -- (0,1.5);
			\node[symmetry node] at (0,1.3) {};
			\node[symmetry node] at (-.5,.5) {};
			\draw[symmetry defect] (-.2,1) .. controls (-.2,1.1) .. (0,1.3);
			\begin{scope}[yscale=-1,shift={(-.5,-1)}]
				\clip (-.4,0) rectangle (.4,.6);
				\draw[symmetry defect] (0,0) ellipse (.3 and .5);
			\end{scope}
			\draw[symmetry defect] (-.5,.2) -- (-.5,.5);
			\node[unit node] at (-.5,.15) {};
			\draw[symmetry defect] (-.8,1) -- (-.8,1.5);
		},\qquad
		\tikz[anchor=base, baseline=-25]{\begin{scope}[yscale=-1,xscale=-1]
			\draw[defect] (0,0) -- (0,1.5);
			\node[symmetry node] at (0,.8) {};
			\draw[symmetry defect] (-.4,0) .. controls (-.4,.3) .. (0,.8);
		\end{scope}}:=
		\tikz[anchor=base, baseline=18]{\begin{scope}[xscale=-1]
			\draw[defect] (0,0) -- (0,1.5);
			\node[symmetry node] at (0,1.3) {};
			\node[symmetry node] at (-.5,.5) {};
			\draw[symmetry defect] (-.2,1) .. controls (-.2,1.1) .. (0,1.3);
			\begin{scope}[yscale=-1,shift={(-.5,-1)}]
				\clip (-.4,0) rectangle (.4,.6);
				\draw[symmetry defect] (0,0) ellipse (.3 and .5);
			\end{scope}
			\draw[symmetry defect] (-.5,.2) -- (-.5,.5);
			\node[unit node] at (-.5,.15) {};
			\draw[symmetry defect] (-.8,1) -- (-.8,1.5);
		\end{scope}}.
	}
	With these notations at hand, one can now represent objects of the generalized orbifold theory (T, A) in terms of objects of $T$ as follows:
		\begin{enumerate}[label=\roman*)]
		\item Its invisible defect is $A$.
		\item Its bulk Hilbert space is $\text{Hom}_{A,A}(A,A)$, the space of $A-A$-bimodule endomorphisms of $A$.
		\item Boundary conditions $B$ of $(T,A)$ are those boundary conditions $B$ of $T$ carrying an appropriate $A$-module structure.
		\item The space of boundary condition changing fields between boundary conditions $B$ and $\tilde B$ is given by $\Hom_A(B,\tilde B)$, the space of $A$-module morphisms from $B$ to $\tilde B$.
\item Defects $D$ from $(T', A')$ to $(T, A)$ are $A$-$A'$-bimodules.
\item The space of defect changing fields from defects $D$ to $\tilde D$ is given by $\Hom_{A, A'}(D, \tilde D)$, the space of $A-A'$-bimodule morphisms from $D$ to $\tilde D$.
\item The fusion product $D \otimes_A \tilde D$ in the orbifold theory $(T,A)$  of two defects $D$ and $\tilde D$ is given by the image of the fusion $D \otimes \tilde D$ in the unorbifolded theory $T$ under
			\uaeq
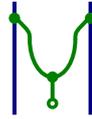
{
				\tikz[anchor=base, baseline=18]{
					\draw[defect] (0,0) -- (0,1.5);
					\node[symmetry node] at (0,1.28) {};
					\draw[symmetry defect] (-.2,1) .. controls (-.2,1.1) .. (0,1.3);
					\begin{scope}[yscale=-1,shift={(-.5,-1)}]
						\clip (-.4,0) rectangle (.4,.6);
						\draw[symmetry defect] (0,0) ellipse (.3 and .5);
					\end{scope}
					\node[symmetry node] at (-.5,.5) {};
					\draw[symmetry defect] (-.5,.2) -- (-.5,.5);
					\node[unit node] at (-.5,.15) {};
					\draw[symmetry defect] (-.8,1) .. controls (-.8,1.1) .. (-1,1.3);
					\draw[defect] (-1,0) -- (-1,1.5);
					\node[symmetry node] at (-1,1.28) {};
				}
			}
		\item  \label{item:induced charges} The adjoints of defects $D$ in the orbifold theory are defined in the following way in terms of the adjoints in the unorbifolded theory. The $A$-actions on any defect $D$ in $(T,A)$ induce actions on its non-orbifold adjoints ${}^\dagger D$ and $D^\dagger$:
			\uaeq
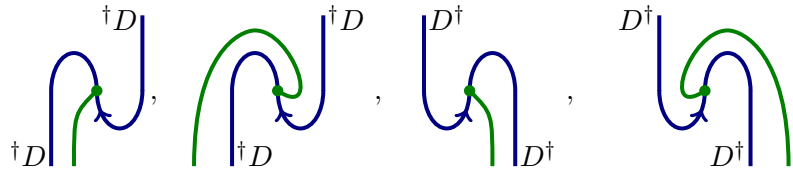
{
				\tikz[anchor=base, baseline=-3]{
					\draw[defect] (-.3,-1) -- (-.3,0);
					\begin{scope}
						\clip (-.4,0) rectangle (.4,.6);
						\draw[defect] (0,0) ellipse (.3 and .5);
					\end{scope}
					\begin{scope}[yscale=-1, shift={(.6,0)}]
						\clip (-.4,0) rectangle (.4,.6);
						\draw[defect, arrow position=.4] (0,0) ellipse (.3 and .5);
					\end{scope}
					\draw[defect] (.9,0) -- (.9,1);
					\node at (-.6,-1) {${}^\dagger D$};
					\node at (.6,.8) {${}^\dagger D$};
					\draw[symmetry defect] (0,-1) .. controls (0,-.3) .. (.3,0);
					\node[symmetry node] at (.3,0) {};
			}
			,\quad
			\tikz[anchor=base, baseline=-3]{
				\draw[defect] (-.3,-1) -- (-.3,0);
				\begin{scope}
					\clip (-.4,0) rectangle (.4,.6);
					\draw[defect] (0,0) ellipse (.3 and .5);
				\end{scope}
				\begin{scope}[yscale=-1, shift={(.6,0)}]
					\clip (-.4,0) rectangle (.4,.6);
					\draw[defect, arrow position=.4] (0,0) ellipse (.3 and .5);
				\end{scope}
				\draw[defect] (.9,0) -- (.9,1);
				\node at (0,-1) {${}^\dagger D$};
				\node at (1.2,.8) {${}^\dagger D$};
				\draw[symmetry defect] (-.8,-1) .. controls (-.8,.5) and (-.2,.8) .. (0,.8) .. controls (.5,.8) and (.9, -.4) .. (.3,0);
				\node[symmetry node] at (.3,0) {};
			}
			,\quad
				\tikz[anchor=base, baseline=-3]{\begin{scope}[xscale=-1]
					\draw[defect] (-.3,-1) -- (-.3,0);
					\begin{scope}
						\clip (-.4,0) rectangle (.4,.6);
						\draw[defect] (0,0) ellipse (.3 and .5);
					\end{scope}
					\begin{scope}[yscale=-1, shift={(.6,0)}]
						\clip (-.4,0) rectangle (.4,.6);
						\draw[defect, arrow position=.4] (0,0) ellipse (.3 and .5);
					\end{scope}
					\draw[defect] (.9,0) -- (.9,1);
				\node at (-.6,-1) {$D^\dagger$};
				\node at (.6,.8) {$D^\dagger$};
				\draw[symmetry defect] (0,-1) .. controls (0,-.3) .. (.3,0);
				\node[symmetry node] at (.3,0) {};
			\end{scope}}
			,\quad
			\tikz[anchor=base, baseline=-3]{\begin{scope}[xscale=-1]
				\draw[defect] (-.3,-1) -- (-.3,0);
				\begin{scope}
					\clip (-.4,0) rectangle (.4,.6);
					\draw[defect] (0,0) ellipse (.3 and .5);
				\end{scope}
				\begin{scope}[yscale=-1, shift={(.6,0)}]
					\clip (-.4,0) rectangle (.4,.6);
					\draw[defect, arrow position=.4] (0,0) ellipse (.3 and .5);
				\end{scope}
				\draw[defect] (.9,0) -- (.9,1);
				\node at (0,-1) {$D^\dagger$};
				\node at (1.2,.8) {$D^\dagger$};
				\draw[symmetry defect] (-.8,-1) .. controls (-.8,.5) and (-.2,.8) .. (0,.8) .. controls (.5,.8) and (.9, -.4) .. (.3,0);
				\node[symmetry node] at (.3,0) {};
			\end{scope}}
			}
			The action of an algebra $A$ on any module can be twisted by an algebra			automorphism $\alpha:A\rightarrow A$. So for any defect $D$ in the orbifold theory, one can define twisted defects $_{\alpha}(D)$ and $(D)_\alpha$ by twisting the left, respectively right $A$-action:
			\uaeq
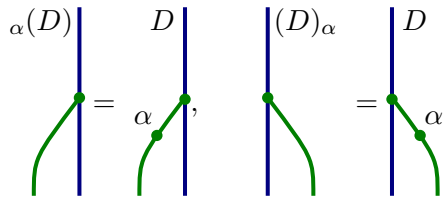
{
				\tikz[anchor=base, baseline=18]{
					\draw[defect] (0,-.5) -- (0,2);
					\node[symmetry node] at (0,.8) {};
					\draw[symmetry defect] (-.6,-.5) .. controls (-.6,0) .. (0,.8);
					\node at (-.5,1.7) {${}_{\alpha}(D)$};
				}
				=
				\tikz[anchor=base, baseline=18]{
					\draw[defect] (0,-.5) -- (0,2);
					\node[symmetry node] at (0,.78) {};
					\draw[symmetry defect] (-.6,-.5) .. controls (-.6,0) .. (0,.8);
					\node[symmetry node] at (-.37,.3) {};
					\node at (-.55,.45) {$\alpha$};
					\node at (-.3,1.7) {$D$};
				}
				,\qquad
				\tikz[anchor=base, baseline=18]{\begin{scope}[xscale=-1]
					\draw[defect] (0,-.5) -- (0,2);
					\node[symmetry node] at (0,.8) {};
					\draw[symmetry defect] (-.6,-.5) .. controls (-.6,0) .. (0,.8);
					\node at (-.5,1.7) {$(D)_{\alpha}$};
				\end{scope}}
				=
				\tikz[anchor=base, baseline=18]{\begin{scope}[xscale=-1]
					\draw[defect] (0,-.5) -- (0,2);
					\node[symmetry node] at (0,.78) {};
					\draw[symmetry defect] (-.6,-.5) .. controls (-.6,0) .. (0,.8);
					\node[symmetry node] at (-.37,.3) {};
					\node at (-.55,.45) {$\alpha$};
					\node at (-.3,1.7) {$D$};
				\end{scope}}
			}
			Left and right adjoints in the orbifold theory can be obtained by twisting 
			the respective adjoints in the unorbifolded theory by the 
						Nakayama automorphism 
			\uaeq{\gamma_A =
				
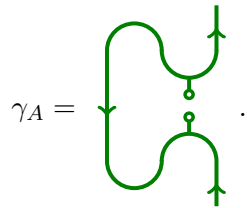
\tikz[anchor=base, baseline=.6cm, scale=.6]{
				\draw[symmetry defect, arrow position=.5] (1.8,-1) -- (1.8,0);
					\begin{scope}
						\clip (-1.1,-1.2) rectangle (1.1,0);
						\draw[symmetry defect] (0,0) circle (.6);
					\end{scope}
					\begin{scope}[shift={(1.2,0)}]
						\clip (-1.1,0) rectangle (1.1,1.2);
						\draw[symmetry defect] (0,0) circle (.6);
					\end{scope}
					\draw[symmetry defect] (1.2,.6) -- (1.2,.9);
					\node[unit node] at (1.2,.97) {};
					\draw[symmetry defect, opp arrow position=.5] (-.6,0) -- (-.6,2.44);
					\node[unit node] at (1.2,1.47) {};
					\draw[symmetry defect] (1.2,1.54) -- (1.2,1.84);
					\begin{scope}[shift={(1.2,2.44)}]
						\clip (-1.1,-1.2) rectangle (1.1,0);
						\draw[symmetry defect] (0,0) circle (.6);
					\end{scope}
					\begin{scope}[shift={(0,2.44)}]
						\clip (-1.1,0) rectangle (1.1,1.2);
						\draw[symmetry defect] (0,0) circle (.6);
					\end{scope}
					\draw[symmetry defect, arrow position=.5] (1.8,2.44) -- (1.8,3.44);
				}.}
				More precisely, the left and right adjoints in the orbifold theory\footnote{In this appendix we denote adjunction in the orbifold by `${}^*$' to distinguish it from the adjunction `${}^\dagger$' in the unorbidolded theory.} are given by \cite[Prop. 4.7]{Carqueville:2012dk}
				\aeq{\label{eqn:orbifold adjoints}
					{}^\ast D = {}_{\gamma_A^{-1}}\left( {}^\dagger D \right), \qquad
					D^\ast = \left( D^\dagger \right)_{\gamma_{A'}}.
				}
			For ${}^\ast D$, the (co)evaluation maps are given by
			\uaeq{
				\ev_D = \;\tikz[baseline=10]{
					\begin{scope}
						\clip (-.7,0) rectangle (.7,1.1);
						\draw[defect, arrow position = .25] (0,0) ellipse (.6 and 1);
					\end{scope}
					\draw[symmetry defect] (.3,.85) .. controls (1,.7) .. (1,.4);
					\node[symmetry node] at (.3,.85) {};
					\begin{scope}[yscale=-1,shift={(1.3,-.4)}]
						\clip (-.4,0) rectangle (.4,.6);
						\draw[symmetry defect] (0,0) ellipse (.3 and .5);
					\end{scope}
					\draw[symmetry defect] (1.6,.4) -- (1.6,1);
					\draw[symmetry defect] (1.3,-.1) -- (1.3,-.3);
					\node[unit node] at (1.3,-.33) {};
					\node at (-.7,-.2) {${}^\ast D$};
					\node at (.6,-.2) {$D$};
				}
				\circ \xi, \qquad \coev_D = \vartheta\circ\;\tikz[baseline=-12]{
					\begin{scope}[yscale=-1]
						\clip (-.7,0) rectangle (.7,1.1);
						\draw[defect] (0,0) ellipse (.6 and 1);
					\end{scope}
					\draw[symmetry defect] (-.54,-.44) .. controls (-1,-.7) .. (-1,-1);
					\node[symmetry node] at (-.54,-.44) {};
					\node at (-.9,0) {$D$};
					\node at (.9,0) {${}^\ast D$};
				}
			}
			with the inclusion and projections maps $\xi: {}^\ast D \otimes_A D \rightarrow {}^\ast D \otimes D$ and $\vartheta: D \otimes {}^\ast D \rightarrow D \otimes_A {}^\ast D$. There are similar formulas for the (co)evaluation maps for  $D^\ast$.
	\end{enumerate}

\section{Orbifold minimal models as generalized orbifolds}
\label{sec:orbifold minimal models as generalized orbifolds}

Here, we construct the Landau-Ginzburg orbifolds $\cM_{d} / \bZ_d$ as generalized orbifolds following \cite[Chapter 7]{Carqueville:2012dk}. To distinguish objects 
in the orbifold from objects in the unorbifolded theory, we adopt the following notation, which is different from the one used in the main text: the identity defect in $\cM_d$ is denoted by $I$, whereas the identity defect in the orbifold theory is represented by the orbifold defect $A$. Also, adjunction in the orbifold is denoted by `${}^*$' to distinguish it from adjunction `${}^\dagger$' in the unorbifolded theory. This notation is only used in this appendix. In section~\ref{sec:rg networks in lg minimal model orbifolds} of the main text, we do not explicitly refer to the orbifold construction and therefore do not need this distinction. There, $I$ denotes the identity defect and `${}^\dagger$' the adjunction in the orbifold theory $\cM_{d} / \bZ_d$.

	\subsection{Orbifold identity defect}
	\label{sec:app:orbifold identity}
	
	The models $\cM_{d} / \bZ_d$ are standard orbifolds. In this case the defect $A$ is given by the direct sum of the defects implementing the respective actions of all the symmetries in the orbifold group: $A = \oplus_{g\in\bZ_d} \left({}_g I\right)$. The symmetry defects ${}_g I$ can be represented by the rank-one matrix factorizations ($\eta = e^{2 \pi i / d}$)
	\aeq{\label{eq:twisted identity defect}
		{}_g I: \bC[Z,X] e_{d+[g]}\,
		\tikz[anchor=base, baseline=0]{
			\draw[->] (0,.2) -- (2,.2);
			\draw[->] (2,0) -- (0,0);
			\node at (1,.4) {$\eta^g Z - X$};
			\node at (1,-.5) {$\frac{Z^d - {X}^d}{\eta^g Z - X}$};
		}\, \bC[Z,X] e_{[g]}.
	}
	where $[g]$ denotes the representative of $g\in\mathbb{Z}_d$ in $\{0,\ldots,d-1\}$, and the $e_a$, $a\in\{0,\ldots, 2d-1\}$ are the generators of the respective rank-one free modules $\left({}_gI\right)_{0,1}$. ${}_gI$ is the right twist of the identity defect $I$ in $\cM_d$ by $g\in\bZ_d$.
	
	Since $A$ is the direct sum of the ${}_gI$, the modules $A_0$ and $A_1$ are rank-$d$ free modules generated by $e_0,\ldots, e_{d-1}$ and $e_d,\ldots,e_{2d-1}$, respectively.
In the basis $(e_a)$, the differential of the matrix factorization $A$ takes the form
	\uaeq{
		\left( \d_A\right)_{ab} = \delta_{a, b-d} (\eta^a Z - X) + \delta_{a-d,b} \sum_{l=1}^d \eta^{-l\cdot a} Z^{d-l} {X}^{l-1}
	}
	for $a, b = 0, ..., 2d-1$.
	
	The following maps give $A$ a separable Frobenius structure \cite[Prop. 7.1]{Carqueville:2012dk}:
	\begin{enumerate}
		\item The unit $I={}_0I\hookrightarrow A$ is given by the obvious inclusion while the counit is the projection multiplied by d
		\uaeq{A \twoheadrightarrow I\,,\qquad e_i\mapsto d\cdot\,\left\{\begin{array}{cc}e_i\,,&i\in\{0,d\}\\0\,,&\text{otherwise}\end{array}\right.}
		\item Multiplication and comultiplication 
			\uaeq{\tikz[baseline=5]{
				\begin{scope}
					\clip (-.8,0) rectangle (.8,.8);
					\draw[symmetry defect] (0,0) circle (.7);
				\end{scope}
				\draw[symmetry defect] (0,.7) -- (0,1.2);
				\node at (0,.25) {${Y}^d$};
				\node at (-.8,.9) {${Z}^d$};
				\node at (1,.9) {${X}^d$};
				\node[symmetry node] at (0,.7) {};
			},\qquad\tikz[baseline=-30]{\begin{scope}[yscale=-1]
				\begin{scope}
					\clip (-.8,0) rectangle (.8,.8);
					\draw[symmetry defect] (0,0) circle (.7);
				\end{scope}
				\draw[symmetry defect] (0,.7) -- (0,1.2);
				\node at (0,.25) {$Y^d$};
				\node at (-.8,.9) {$Z^d$};
				\node at (1,.9) {$X^d$};
				\node[symmetry node] at (0,.7) {};
			\end{scope}}}
			are given by			
			\uaeq{
				A&\otimes A &&\rightarrow A \\
				e_{[g]} &\otimes Y^q e_{[h]} &&\mapsto \left(\eta^g Z\right)^q e_{[g+h]} \\
				e_{[g]}&\otimes Y^q e_{d+[h]} &&\mapsto \left(\eta^g Z\right)^q e_{d+[g+h]}\\
				e_{[g]+d}&\otimes Y^q e_{[h]}&&\mapsto 0\\
				e_{[g]+d}&\otimes Y^q e_{[h]+d}&&\mapsto 0
			}
		 and 
			\aeq{\label{eqn:explicit comultiplication}
				\Delta: A&\rightarrow A\otimes A \\
				e_{[g]} &\mapsto \frac 1 d \sum_{h\in\bZ_d} \left[ e_{[g-h]}\otimes e_{[h]} + e_{d+[g-h]}\otimes \left\{ \partial^{Z,Y} \frac{Z^d-X^d}{\eta^h Z - X} \right\}^{Z\rightarrow \eta^{g-h} Z} e_{d+[h]} \right] \\
				e_{d+[g]} &\mapsto \frac 1 d \sum_{h\in\bZ_d} \left[ e_{[g-h]}\otimes e_{d+[h]} + \eta^h e_{d+[g-h]} \otimes e_{[h]} \right],
			}
where $g,h\in\bZ_d$ and $q\in\bN$.
Moreover, $\partial^{Z,Y} Z^i=\frac{Z^i-Y^i}{Z-Y}$ and $\{\ldots\}^{Z\rightarrow \eta^{g-h} Z}$
means that  all instances of $Z$ within the brakets have to be replaced by $\eta^{g-h} Z$ after performing all operations. These formulas can be obtained from the natural junctions of symmetry defects ${}_h I$ with the identity defect $I={}_0I$. The calculation for the comultiplication is sketched in appendix~\ref{sec:comultiplication} below.
	\end{enumerate}
	In appendix \ref{sec:app:orbifold identity equivariant} below, we will reexpress the orbifold identity defect $A$ using an equivariant basis.
	
	\subsection{Nakayama automorphism}
	
	The Nakayama automorphism (c.f. appendix~\ref{sec:generalized orbifold theories}) takes the form \cite[Example 3.1]{Brunner:2013ota}
	\uaeq{
		\gamma_A= \sum_{g\in\bZ_d} \det(g)\cdot 1_{{}_gI}
	}
	where $\det(g)$ denotes the matrix representing the action of $g$ on the chiral fields of the model to the right of $A$. In our case, $g$ acts on $X$ by multiplication with $\eta^g$ and hence $\gamma_A$ reduces to
	\uaeq{
		\gamma_A : A&\rightarrow A \\
		e_a &\mapsto \eta^a e_a.
	}
	A Frobenius algebra $B$ is symmetric iff $\gamma_B=\id_B$ \cite{Carqueville:2012dk,FS}. Since $\gamma_A\neq\id_A$, $A$ is not symmetric. Therefore, left and right adjoints of defects in the orbifold theory (see equation~(\ref{eqn:orbifold adjoints})) generally differ
	\uaeq{
		\left( D^\dagger \right)_{\gamma_{A'}} = D^\ast \not\cong {}^\ast D={}_{\gamma_A^{-1}}\left( {}^\dagger D \right).
	}
					
	This means that we do not have a general prescription of how to close defect loops in $\cM_d/\bZ_d$. However, loops of RG defects can be closed with an explicit natural morphism.

	\subsection{Bulk space}
	
	The $(c,c)$-bulk space of the orbifold $\cM_{d}/\bZ_d$ contains only the identity $\text{id}:A\rightarrow A$. However, in the unorbifolded theory the defect $A$ carries additional fields -- one for each $g\neq 0$:
	\uaeq{
		\psi_g: A &\rightarrow A \\
		e_{[h]} &\mapsto\frac{Z^d - {X}^d}{(\eta^i Z-X)(\eta^{h+g} Z - X)} e_{d+[g+h]}\\
		e_{d+[h]}&\mapsto e_{[g+h]}
	}
	These correspond to the twist fields in the orbifold theory.
	
	\subsection{Defects and their adjoints}
	\label{app:defects and boundaries}
	
	Consider a rank-$M$ $\bZ_{d'}\times\bZ_d$-equivariant matrix factorization $D$ of $Z^{d'}-X^d$ with equivariant generators $f_0,\ldots,f_{M-1}$ of $D_0$ and $f_M,\ldots,f_{2M-1}$. Let 
$[l_k, r_k]$ be the $\bZ_{d'}\times\bZ_d$-charges of $f_k$.
 As discussed in the second part of \ref{sec:calculations}, these charges determine the $A$-action on $D$. Denoting the chiral fields as in 
	\uaeq{\tikz[baseline=20]{
		\draw[symmetry defect] (-1.5,0) -- (0,1.5);
		\draw[defect] (0,0) -- (0,2);
		\node at (-1.5,1) {$Z$};
		\node at (-.5,.5) {$Y$};
		\node at (.5,1) {$X$};
	}\qquad\text{and}\qquad\tikz[baseline=20]{
		\draw[symmetry defect] (1.5,0) -- (0,1.5);
		\draw[defect] (0,0) -- (0,2);
		\node at (1.5,1) {$X$};
		\node at (.5,.5) {$Y$};
		\node at (-.5,1) {$Z$};
	},}
one obtains
	\uaeq{
		A \otimes D &\rightarrow D \\
		e_a \otimes Y^p f_k &\mapsto \delta_{|e_a|, 0} \cdot (\epsilon^a Z)^p \cdot \epsilon^{a\cdot l_k} \cdot  f_k.
	}
	and
	\uaeq{
		 D \otimes A &\rightarrow D \\
		f_k \otimes Y^p e_a &\mapsto \delta_{|e_a|, 0} \cdot f_k \cdot \eta^{- a\cdot r_k} (\eta^{-a} X)^p
	}
	where $\epsilon=e^{\frac{2\pi i}{d'}}$ and $\eta=e^{\frac{2\pi i}{d}}$ are elementary $d'$th, respectively $d$th roots of unity.
	
	In appendix~\ref{sec:app:orbifold identity equivariant} below, we will define equivariant generators for $A$ itself, and will reexpress the $A$-action on $D$ in terms of these generators.
		
	The adjoints in the orbifold theory are given by ${}^\ast D = {}_{\gamma_A^{-1}}\left( {}^\dagger D \right)$ and $D^\ast = \left( D^\dagger \right)_{\gamma_{A'}}$, see equation~(\ref{eqn:orbifold adjoints}).
	Here $D^\dagger \cong D^\vee[1]\cong{}^\dagger D$ denotes adjunction in the unorbifolded models, c.f. section~\ref{sec:b type defects in LG}. 
	An explicit calculation carried out in the last part of appendix~\ref{sec:calculations} 
	determines the induced $A$-action on $D^\dagger$, c.f. equation~\eqref{eq:induced charge on non-orbifold adjoint}. From this, one can read off the $\bZ_d\times\bZ_{d'}$ charges of the equivariant generators $f^\dagger_k$ and ${}^\dagger f_k$ of $D^\dagger$ and ${}^\dagger D$ to be
	\uaeq{
		[-r_{k+M} + 1, -l_{k+M} + 1].
	}
	Here, we have extended the range of indices of the charges $r$ and $l$  to $\bZ$ by identification modulo $2M$, i.e.
	$r_{i+2Mz}=r_i$ and $l_{i+2Mz}=l_i$ for $i\in\{0,\ldots, 2M-1\}$ and $z\in\bZ$.
	
	Twisting by the Nakayama automorphism one then obtains the charges of the generators $f_k^\ast$ and ${}^\ast f_k$ of the matrix factorizations describing the orbifold adjoints $D^\ast$ and ${}^\ast D$. They are given by 
	\uaeq{
		&[-r_{k+M}, -l_{k+M} + 1] \\
		\text{and}\qquad &[-r_{k+M} + 1, -l_{k+M}],
	}
	respectively. By construction, ${}^\ast D$ and $D^\ast$ obey the Zorro moves whose building blocks are provided in \ref{sec:explicit (co)evaluation maps}.
	
	\subsection{Left boundary conditions and their adjoints}
	
	As special case of defects, a left boundary condition  in $\cM_d/\bZ_d$ is a $\bZ_d$-equivariant matrix factorization $B$ of $-X^d$. Using the same notation as in appendix~\ref{app:defects and boundaries}, we denote 
the generators of the modules $B_0$ and $B_1$ by $f_k$ and their $\bZ_d$-charges by $[r_k]$ which as in the general case determine the $A$-action on $B$. The induced charges on the right and left adjoint generators $f^\dagger_k$ of $B^\dagger$ and ${}^\dagger f_k$ of ${}^\dagger B$ are $[- r_{k+M}+1]$ and $[-r_k+1]$, respectively. Using $B^\ast = B^\dagger$ and ${}^\ast B = {}_{\gamma^{-1}}({}^\dagger B)$, the charges of the adjoint generators $f^\ast_k$ and ${}^\ast f_k$ become
	\uaeq{
		f^\ast_k&: [- r_{k+M}+1] \\
		{}^\ast f_k&: [-r_k]
	}
	The explicit expressions of the relevant (co-)evaluation maps for defects as well as boundaries are given in appendix \ref{sec:explicit (co)evaluation maps}. 
	
	\subsection{Equivariant generators of the orbifold identity defect}
	\label{sec:app:orbifold identity equivariant}
	
	One can define equivariant generators of the orbifold identity matrix factorization $A$ (c.f. section~\ref{sec:app:orbifold identity}) by
	\uaeq{e_b' = \frac 1 d \sum_c \delta_{|e_b|, |e_c|} \eta^{-(b + |e_b|) c} e_c,}
	where the original generators $e_c$ are expressed in terms of the equivariant ones as
	 \uaeq{
		e_c=\sum_b \delta_{|e_b|, |e_c|} \eta^{c(b + |e_c|)} e_b'
	}
	In this basis, the matrix factorization $A$ takes the equivariant form 
	\uaeq{A:\bC[Z,X]\begin{pmatrix}[1,0]\\ [2,-1]\\\vdots \end{pmatrix}
		\tikz{
			\begin{scope}[shift={(0,15)}]
				\node at (0,2) {$\begin{pmatrix}Z& 0 & ... & 0 & -X \\ -X & \ddots &&&\\0&\vdots&\ddots&&\\\vdots&&\vdots&\vdots&\\0&&&-X&Z\end{pmatrix}$};
				\draw[arrow position = 1] (-3,.2) -- (3,.2);
				\draw[arrow position = 1] (3,0) -- (-3,0);
			\end{scope}
		}
		\;\bC[Z,X]\begin{pmatrix}[0,0]\\ [1,-1]\\ \vdots \end{pmatrix}.
	}
(This is the form used in \cite{Brunner:2007ur}.) The $A$-action on equivariant matrix factorizations determined in appendix~\ref{sec:calculations} and used in appendix~\ref{app:defects and boundaries} simplifies in this basis.

Consider a   $\bZ_d\times\bZ_d$-equivariant matrix factorization $D$ of $Z^d-X^d$. 
Let $f_0,\ldots,f_{M-1}$ and $f_M,\ldots,f_{2M-1}$ be $\bZ_d\times\bZ_d$-equivariant generators of 
$D_0$ and $D_1$, respectively. Denote the  $\bZ_d\times \bZ_d$-charges of $f_k$ by $[l_k,r_k]$.
In terms of the equivariant generators $e_i'$ of $A$, the $A$-action 
	\aeq{\label{img:choice of variables}\tikz[baseline=20]{
		\draw[symmetry defect] (-1.5,0) -- (0,1.5);
		\draw[defect] (0,0) -- (0,2);
		\node at (-1.5,1) {$Z$};
		\node at (-.5,.5) {$Y$};
		\node at (.5,1) {$X$};
	}\qquad\text{and}\qquad\tikz[baseline=20]{
		\draw[symmetry defect] (1.5,0) -- (0,1.5);
		\draw[defect] (0,0) -- (0,2);
		\node at (1.5,1) {$X$};
		\node at (.5,.5) {$Y$};
		\node at (-.5,1) {$Z$};
	}}
becomes
	\uaeq{
		A \otimes D &\rightarrow D, &\quad e_a' \otimes Y^p f_k &\mapsto \delta_{a, [p+l_k]}\, Z^p\, f_k \\
		 D \otimes A &\rightarrow D, &\quad f_k \otimes Y^p e_a' &\mapsto \delta_{a,[-r_k-p]} \, f_k \, {X}^p.}
		
\subsection{Important calculations}
\label{sec:calculations}

In this appendix we sketch some calculations used in the main text and the previous sections of this appendix.

	\subsubsection*{Comultiplication of identity defect $A$}
	\label{sec:comultiplication}
	
	Following \cite{Carqueville:2012dk,Carqueville:2012st}, 
	we define $\lambda^{-1}_{{}_hI}: {}_h I \rightarrow I \otimes {}_h I$ to be the natural junction of the identity defect with the symmetry defect ${}_hI$. It is given by 
	\uaeq{
		e_{[h]} &\mapsto 1\otimes e_{[h]} + \theta\otimes \left\{ \partial^{Z,Y} \frac{Z^d-X^d}{\eta^h Z-X} \right\} e_{d+[h]} \\
		e_{d+[h]} &\mapsto 1\otimes e_{d+[h]} + \eta^h \theta \otimes e_{[h]}.
	}
	Here $X$ and $Z$ are the chiral fields of the models to the right, respectively left of the defects, and $Y$ is the chiral field of the model sandwiched between the defects $I$ and ${}_hI$.
	Twisting by $g$ from the left (i.e. fusion by ${}_gI$ from the left)
one obtains junction fields 	
 $\Delta_{g,h} := {}_g\left(\lambda^{-1}_{{}_hI}\right): {}_{g+h}I\rightarrow {}_gI\otimes{}_hI$:
	\uaeq{
		e_{[g+h]} &\mapsto e_{[g]}\otimes e_{[h]} + e_{d+[g]}\otimes \left\{ \partial^{Z,Y} \frac{Z^d-X^d}{\eta^h Z-X} \right\}^{Z\rightarrow \eta^g Z} e_{d+[h]} \\
		e_{d+[g+h]} &\mapsto e_{[g]}\otimes e_{d+[h]} + \eta^h e_{d+[g]} \otimes e_{[h]}.
	}
	(Here, the notation $\{\ldots\}^{Z\rightarrow \eta^g Z}$ means that that all instances of $Z$ in the brackets have to be replaced by $\eta^g Z$ after performing all calculations.)
	
	Summing up all the $\Delta_{g,h}$ yields the comultiplication 
		\uaeq{
		\Delta = \frac 1 d \sum_{g,h} \Delta_{g,h}:A&\rightarrow A\otimes A \\
		e_{[g]} &\mapsto \frac 1 d \sum_{h\in\bZ_d} \Delta_{g-h,h}(e_{[g]}) \\
		e_{d+[g]} &\mapsto \frac 1 d \sum_{h\in\bZ_d} \Delta_{g-h,h}(e_{d+[g]})
	}
	of the identity defect $A$ in the orbifold. It is spelled out completely in 
equation~\eqref{eqn:explicit comultiplication}.
	
	\subsubsection*{$A$ actions on equivariant defect}
	
	According to \cite[Section 7.1]{Carqueville:2012dk}, the data of a $G\times H$-equivariant defect is encoded in its left and right fusion with the symmetry defects $A_G$ and $A_H$. Namely, it is described by a matrix factorization together with isomorphisms
	\begin{itemize}
		\item $\varphi_g:{}_gD\rightarrow D$ such that $\varphi_e = \id_D$ and $\varphi_{g_1} \circ {}_{g_1}(\varphi_{g_2}) = \varphi_{g_1+g_2}$ and
		\item $\phi_h:D_h\rightarrow D$ such that $\phi_e = \id_D$ and $\phi_{h_1} \circ (\phi_{h_2})_{h_1} = \phi_{h_1+h_2}$.
	\end{itemize}
	Here, one can think of ${}_gD$ as the matrix factorization where all variables $Z_i$ to the left of $D$ have been replaced by $g(Z_i)$, see for example ${}_gI$ in \eqref{eq:twisted identity defect}. Also, for some morphism $\alpha:D\rightarrow D'$ of matrix factorizations, ${}_g(\alpha)_h:{}_gD_h\rightarrow{}_g{D'}_h$ is the same morphism considered as a morphism between the respective twisted matrix factorizations. However, special attention has to be paid to morphisms including an identification of variables. For example, the left and right $I$-actions $\lambda_D:I\otimes D\rightarrow D$ and $\rho_D:D\otimes I\rightarrow D$ identify the middle variable with the one on the left or right, respectively. The identification of variables in the twisted versions ${}_g(\lambda_D)$ and $(\rho_D)_{-h}$ must respect the twist.
	
	Following the proof of Thm. 7.2 in \cite{Carqueville:2012dk}, the above data determine the left $A_G$-action on $D$:
	\uaeq{
		\sum_{g\in G} \left( A_G \otimes D \twoheadrightarrow {}_g I\otimes D\xrightarrow{{}_g(\lambda_D)} {}_g D\xrightarrow{\varphi_g} D \right).
	}
	The right action includes the canonical isomorphism ${}_hI\rightarrow I_{-h}$ which we will comment on later:
	\aeq{\label{eq:left A action with canonical iso}
		\sum_{h\in H} \left( D \otimes A_H \twoheadrightarrow D\otimes {}_h I \rightarrow D\otimes I_{-h} \xrightarrow{(\rho_D)_{-h}} D_{-h}\xrightarrow{\phi_{-h}} D \right).
	}
	
	Turning to our example, set $G=\bZ_{d'}$ and $H=\bZ_d$ and consider a $G\times H$-equivariant defect $D$, i.e. a $\bZ_{d'}\times\bZ_d$-equivariant matrix factorization $D$ of $Z^{d'}-X^d$. Let $f_0,\ldots,f_{M-1}$ and $f_M,\ldots,f_{2M-1}$ be equivariant generators of $D_0$, respectively $D_1$. Denote the $\bZ_{d'}\times\bZ_d$-charges of $e_k$ by 
	\uaeq{
		[l_k, r_k].
	}
	In other words the action of $(g,h)\in \bZ_{d'}\times\bZ_d$ is given by
	\aeq{\label{eq:expected action}
		Z^p f_k X^q \;\mapsto\;(\epsilon^g Z)^p \cdot \epsilon^{g\cdot l_k} f_k \eta^{h\cdot r_k} \cdot (\eta^h X)^q
	}
	where $\epsilon = e^{\frac{2\pi i}{d'}}, \eta = e^{\frac{2\pi i}{d}}$.
	
	We now reformulate this group action in terms of left and right $A$-actions. It is not hard to see that in our case the above isomorphisms are given by\footnote{$\varphi_g\circ {}_g(\varphi_h) = \varphi_{g+h}$ is trivial and $\d_D\circ\varphi_g = \varphi_g \circ {}_g(\d_D)$ amounts to $\d_D$ being a degree zero map, i.e. $e_k$ and $\d_D(e_k)$ carrying the same $\bZ_{d'}\times\bZ_d$ charges.}
	\uaeq{
		\varphi_g: {}_gD &\rightarrow D\qquad & \phi_h: D_h &\rightarrow D \\
		f_k &\rightarrow \epsilon^{g\cdot l_k} f_k\qquad & f_k &\rightarrow  f_k \eta^{h\cdot r_k}.
	}
	The explicit form of the left $A$-action on $D$ then turns out to be
	\uaeq{
		A \otimes D &\rightarrow D \\
		e_a \otimes Y^p f_k &\mapsto \delta_{|e_a|, 0} \cdot (\epsilon^a Z)^p \cdot \epsilon^{a\cdot l_k} \cdot  f_k
	}
	for the same choice of variables as in \eqref{img:choice of variables} ($|\cdot |$ denotes the  $\bZ_2$-charge). This coincides with the expected action \eqref{eq:expected action}. The right $A$-action on the other hand takes the form
	\uaeq{
		 D \otimes A &\rightarrow D \\
		f_k \otimes Y^p e_a &\mapsto \delta_{|e_a|, 0} \cdot f_k \cdot \eta^{- a\cdot r_k} (\eta^{-a} X)^p
	}
	where we emphasize the crucial minus sign for the right charges which differs from the expected \eqref{eq:expected action}. It originates from the fact that the symmetry defect $A$ was defined as the direct sum of the \emph{left} twisted identity morphisms which requires us to include the canonical isomorphism
	\uaeq{
		{}_hI &\rightarrow I_{-h} \\
		e_i &\mapsto \eta^{h \cdot |e_i|} e_{i}, \qquad i = 0,1
	}
	in the construction \eqref{eq:left A action with canonical iso}.

	\subsubsection*{Left $A$-action on right adjoint}
	\label{sec:left A action on right adjoint}
	
	As explained in appendix~\ref{sec:generalized orbifold theories} item \ref{item:induced charges}, adjoints of defects in the orbifold theory are induced by their non-orbifold counterpart. Here, given an equivarant matrix factorization $D$ of $Z^{d^\prime}-X^d$, we explicitely calculate the induced left $A$-action (i.e. the left charges, see previous calculation) on the non-orbifold adjoint $D^\dagger$. This leads to the charges of the right adjoint in the orbifold theory as $D^\ast \cong (D^\dagger)_\gamma$.

Let $f_0,\ldots, f_{M-1}$ be equivariant generators of $D_0$ and $f_M,\ldots,f_{2M-1}$ equivariant generators of $D_1$. We denote the $\bZ_{d'}\times\bZ_d$-charges of $f_k$  by $[l_k,r_k]$. Then, $D^\dagger_{0}$ and $D^\dagger_1$ are generated by $f_0^\dagger,\ldots,f_{M-1}^\dagger$ and $f^\dagger_M,\ldots,f_{2M-1}^\dagger$ respectively, where $f_i^\dagger = f^\vee_{i+M}$ for $i<M$ and $f_i^\dagger = f^\vee_{i-M}$ for $i\geq M$. (`${}^\vee$' denotes the dual.)

	From the explicit expressions of the (co-)evaluation maps \cite{Carqueville:2012st} we obtain
	\aeq{\label{eq:left A action formula}
		A \otimes D^\dagger &\rightarrow D^\dagger \\
		e_a \otimes f_i^\dagger &\mapsto - \delta_{|e_a|, 0} \delta_{|f_i^\dagger|,0} \sum_{k=0}^{M-1} \text{ Res} \left[ \frac{ \left(\substack{[\partial_X \d_0]^{X\rightarrow X''} \cdot {\eta}^{-a}_{r, 0} \cdot \partial^{X, \eta^{-a} X''} \d_1 + \\
				 	+ \eta^a [\partial^{X, X''} \d_0]^{X\rightarrow \eta^a X} \cdot [\partial_X \d_1]^{X\rightarrow X''} \cdot {\eta}^{-a}_{r, 1} }
			\right)_{i, k}^{Z\rightarrow Z'} \d X'' }{d \cdot {X''}^{d-1}}\right] f_k^\dagger \\
		&\quad- \delta_{|e_a|, 0} \delta_{|f_i^\dagger|,1} \sum_{k=M}^{2M-1} \text{ Res} \left[ \frac{ \left(\substack{[\partial_X \d_1 ]^{X\rightarrow X''} \cdot {\eta}^{-a}_{r, 1} \cdot \partial^{X, \eta^{-a} X''} \d_0 +\\+ \eta^a [\partial^{X, X''} \d_1]^{X\rightarrow \eta^a X} \cdot [\partial_X \d_0]^{X\rightarrow X''} \cdot {\eta}^{-a}_{r, 0}}\right)_{i-M, k-M}^{Z\rightarrow Z'} \d X'' }{d \cdot {X''}^{d-1}}\right] f_k^\dagger
	}
	for the following choice of variables
	\uaeq
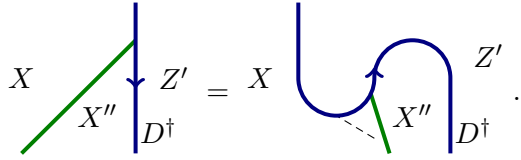
{\tikz[baseline=20]{
		\draw[symmetry defect] (-1.5,0) -- (0,1.5);
		\draw[defect, opp arrow position =.5] (0,0) -- (0,2);
		\node at (-1.5,1) {$X$};
		\node at (-.5,.5) {$X''$};
		\node at (.5,1) {$Z'$};
		\node at (.3,.3) {$D^\dagger$};
	}
	=\tikz[baseline=20]{
		\draw[densely dashed] (-1,.2) -- (-1.5,.5);
		\draw[symmetry defect] (-.8,0) -- (-1.05,.8);
		\draw[defect, opp arrow position =.5] (0,0) -- (0,1) arc(0:180:.5) arc(0:-180:.5) -- (-2,2);
		\node at (-2.5,1) {$X$};
		\node at (-.5,.5) {$X''$};
		\node at (.5,1.3) {$Z'$};
		\node at (.3,.3) {$D^\dagger$};
	}.
	}
	Here $\eta=e^{2\pi i/d}$, and  $\eta_{r,0}$ and $\eta_{r,1}$ are the diagonal matrices 
	\uaeq{&\eta_{r,0}=\text{diag}(\eta^{r_0}, ..., \eta^{r_{M-1}})\\ 
	&\eta_{r,1}=\text{diag}(\eta^{ r_M}, ..., \eta^{r_{2M-1}}).}
	Moreover, $\partial^{X,X''}$ is the divided difference operator which is defined as
	$\partial^{X,X''}g(X,\ldots)=\frac{g(X,\ldots)-g(X'',\ldots)}{X-X''}$ on any polynomial $g$, and the residue $\Res{\frac{g\cdot \d X''}{{X''}^{d-1}}}$ picks out the prefactor of ${X''}^{d-2}$ in the polynomial $g\in\bC[Z',X,X'']$.

	We now simplify expression~\eqref{eq:left A action formula} by calculating the ${X''}^{d-2}$-term in the numerator. We first derive a few identities which follow from the very definition of a graded matrix factorization. 
	
	From the basic property of matrix factorizations $\d_0 \d_1 = (Z^{d'} - X^d) \mathbb 1$ one can deduce 
	\uaeq{
		\partial^{X,X''}(-d\cdot X^{d-1}) &= \left[\partial_X\d_0\right\vert^{X\rightarrow X''}\cdot\partial^{X,X''}\d_1 + \partial^{X,X''}\d_0 \cdot \left[\partial_X\d_1\right\vert^{X\rightarrow X''}  \\
		&\quad+\left\{\left(\partial^{X,X''}\partial_X\d_0\right)\cdot \d_1 + \d_0\cdot\left(\partial^{X,X''}\partial_X\d_1\right)\right\}.
	}
	Now, we will simplify the derivation by assuming that the matrices $d_0$ and $d_1$ do not contain terms $X^n$ for $n\geq d$ . This is certainly true for all the matrix factorizations relevant in this paper, namely the ones associated to RG and projection defects, boundary conditions etc. Under this assumption, the curly bracket part of the last equation does not contain a term $\sim {X''}^{d-2}$, and hence
	\aeq{\label{eq:A action 1}
		\left\{ [\partial_X \d_0]^{X\rightarrow X''} \cdot \partial^{X, X''} \d_1 + \partial^{X, X''}  \d_0 \cdot [\partial_X \d_1|^{X\rightarrow X''} \right\}_{ik} = -d {X''}^{d-2} \delta_{ik} + ...
	}
	where $...$ contains only powers $(X'')^n$ with $n<d-2$. In order to make contact with equation~\eqref{eq:left A action formula} we replace $\partial^{X, X''} \d_0$ in \eqref{eq:A action 1} by $\partial^{X, X''} \d_0\vert^{X\rightarrow \eta^a X}$ which does not alter the leading $X''$-term:
	\aeq{\label{eq:A action 2}
		\left\{ [\partial_X \d_0]^{X\rightarrow X''} \cdot \partial^{X, X''} \d_1 + \partial^{X, X''} \d_0\vert^{X\rightarrow \eta^a X} \cdot [\partial_X \d_1|^{X\rightarrow X''} \right\}_{ik} = -d {X''}^{d-2} \delta_{ik} + ...
	}
	Also, since $\d_1$ is of grade zero
	\uaeq{
		\eta^{-a}_{r,0} \cdot [\d_1\vert^{X\rightarrow \eta^{-a} X}=\d_1 \cdot \eta^{-a}_{r,1}
	}
	which together with the definition of the divided difference operator yields
	\aeq{\label{eq:grading permutation}
		\left.{\eta}^{-a}_{r, 0} \cdot \partial^{X, \eta^{-a} X''} \d_1 \right\vert_{\substack{\text{leading}\\ X''\text{-term}}} = \left. \partial^{X, X''} \d_1\right\vert_{\substack{\text{leading}\\ X''\text{-term}}} \cdot \eta^a \cdot{\eta}^{-a}_{r, 1} .
	}
	Here, $\left.\partial^{X,\eta^{-a}X''}\d_1\right\vert_{\substack{\text{leading}\\ X''\text{-term}}}$ is the matrix $\d_1$ with all entries $X^{p+1}$ replaced by $(\eta^{-a}\cdot {X''})^p$ and similarly for $\left. \partial^{X, X''} \d_1\right\vert_{\substack{\text{leading}\\ X''\text{-term}}}$. Finally, we evaluate the first summand of \eqref{eq:left A action formula}:
	\uaeq{
		&\quad- \delta_{|e_a|, 0} \delta_{|f_i^\dagger|,0} \sum_{k=0}^{M-1} \text{ Res} \left[ \frac{ \left(\substack{[\partial_X \d_0]^{X\rightarrow X''} \cdot {\eta}^{-a}_{r, 0} \cdot \partial^{X, \eta^{-a} X''} \d_1 + \\
				 	+ \eta^a [\partial^{X, X''} \d_0]^{X\rightarrow \eta^a X} \cdot [\partial_X \d_1]^{X\rightarrow X''} \cdot {\eta}^{-a}_{r, 1} }
			\right)_{i, k} \d X}{d \cdot {X''}^{d-1}}\right]^{Z\rightarrow Z'} f_k^\dagger
	}\uaeq{
		&\stackrel{\eqref{eq:grading permutation}}= - \delta_{|e_a|, 0} \delta_{|f_i^\dagger|,0} \sum_{k=0}^{M-1} \text{ Res} \left[ \frac{ \left(\substack{[\partial_X \d_0]^{X\rightarrow X''} \cdot \partial^{X, X''} \d_1 \cdot\eta^a\cdot {\eta}^{-a}_{r, 1} +  ...\\
				 	+ \eta^a [\partial^{X, X''} \d_0]^{X\rightarrow \eta^a X} \cdot [\partial_X \d_1]^{X\rightarrow X''} \cdot {\eta}^{-a}_{r, 1} }
			\right)_{i, k} \d X }{d \cdot {X''}^{d-1}}\right]^{Z\rightarrow Z'} f_k^\dagger
	}\uaeq{
		&\;\;= - \delta_{|e_a|, 0} \delta_{|f_i^\dagger|,0} \sum_{k=0}^{M-1} \text{ Res} \left[ \frac{ \left(\substack{[\partial_X \d_0]^{X\rightarrow X''} \cdot \partial^{X, X''} \d_1 + \\
				 	+ [\partial^{X, X''} \d_0]^{X\rightarrow \eta^a X} \cdot [\partial_X \d_1]^{X\rightarrow X''} }
			\right)_{i, k} \cdot \eta^{-a(r_{k+M}-1)} \d X }{d \cdot {X''}^{d-1}}\right]^{Z\rightarrow Z'} f_k^\dagger \\
		&\stackrel{\eqref{eq:A action 2}}= - \delta_{|e_a|, 0} \delta_{|f_i^\dagger|,0} \sum_{k=0}^{M-1} \text{ Res} \left[ \frac{ (-d {X''}^{d-2} \delta_{ik} + ...)\d X }{d \cdot {X''}^{d-1}}\right]^{Z\rightarrow Z'} \cdot \eta^{-a(r_{k+M}-1)} f_k^\dagger \\
		&\;\;= \delta_{|e_a|, 0} \delta_{|f_i^\dagger|,0} \eta^{a(-r_{i+M}+1)} f_i^\dagger.
	}
	Here, with 
	`$...$' we indicate that we omitted terms which do not contribute to the residue.
	
		The second summand in equation~\eqref{eq:left A action formula} can be determined in a similar way and also takes a similar form. We find that the left charges of $D^\dagger$ are the negative right charges of $D$  shifted by $+1$:
	\aeq{\label{eq:induced charge on non-orbifold adjoint}
		A \otimes D^\dagger &\rightarrow D^\dagger \\
		e_a \otimes f_i^\dagger &\mapsto \delta_{|e_a|,0}\eta^{a ( - r_{i+M} + 1)} f_i^\dagger.
	}
	
	\subsection{(Co)evaluation maps}
	\label{sec:explicit (co)evaluation maps}
	Finally, we provide the explicit (co)evaluation maps used in calculations in the main text. They follow from the generalized orbifold construction, (c.f.~appendix~\ref{sec:generalized orbifold theories}) and the expressions of \cite{Carqueville:2012st}. Throughout, $[...]$ denotes the representative in $\{0,\ldots, d-1\}$ modulo d. Furthermore, $\partial^{Z,X}Z^i=\frac{Z^i-X^i}{Z-X}$, $\sigma=\begin{pmatrix}\mathbb{1}&0\\0&-\mathbb{1} \end{pmatrix}$, $\eta=e^{\frac{2\pi i}{d}}$, $\epsilon=e^{\frac{2\pi i}{d'}}$, $\eta_r = \text{diag}(\eta^{r_0}, \eta^{r_1}, \eta^{r_2}, ...)$, $\epsilon_l = \text{diag}(\epsilon^{l_0}, \epsilon^{l_1}, \epsilon^{l_2}, ...)$ and $\epsilon_{1+l} = \text{diag}(\epsilon^{1+l_0}, \epsilon^{1+l_1}, \epsilon^{1+l_2}, ...)$.
	
	\subsubsection*{Orbifold evaluation map (left)}
	
	\uaeq{
		\ev_D = \;
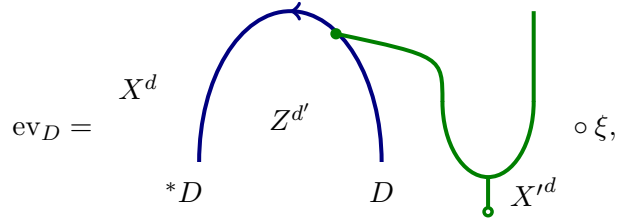
\tikz[baseline=10, scale=2]{
			\begin{scope}
				\clip (-.7,0) rectangle (.7,1.1);
				\draw[defect, arrow position = .25] (0,0) ellipse (.6 and 1);
			\end{scope}
			\draw[symmetry defect] (.3,.85) .. controls (1,.7) .. (1,.4);
			\node[symmetry node] at (.3,.85) {};
			\begin{scope}[yscale=-1,shift={(1.3,-.4)}]
				\clip (-.4,0) rectangle (.4,.6);
				\draw[symmetry defect] (0,0) ellipse (.3 and .5);
			\end{scope}
			\draw[symmetry defect] (1.6,.4) -- (1.6,1);
			\draw[symmetry defect] (1.3,-.1) -- (1.3,-.3);
			\node[unit node] at (1.3,-.33) {};
			\node at (-.7,-.2) {${}^\ast D$};
			\node at (.6,-.2) {$D$};
			\node at (0,.3) {$Z^{d'}$};
			\node at (-1,.5) {${X}^d$};
			\node at (1.6,-.2) {${X'}^{d}$};
		}
		\circ \xi,
	}
	where $\xi:{}^\ast D\otimes_
	A D\rightarrow {}^\ast D\otimes D$ is the inclusion.
	\uaeq{
		ev_D: {}^\ast D &\otimes_A D \rightarrow A \\
		{}^\ast f_k &\otimes Z^n f_i \mapsto \frac 1{d} \sum_{h\in\bZ_d} \Res{ \frac{ Z^n \left( \sigma\cdot\partial_Z\d_D \cdot\eta^h_r\right)_{(k+M),i} }{d' \cdot Z^{d'-1}}} e_{[h]} \\
		&+ \frac 1{d} \sum_j \sum_{h\in\bZ_d} \Res{ \frac{ Z^n \left( \sigma\cdot\partial_Z\d_D \cdot \eta^h_r \cdot \left[\partial^{X,X'}\d_D\right]^{X\rightarrow \eta^h X} \cdot\sigma \right)_{(k+M),i} }{d' \cdot Z^{d'-1}}} e_{d+[h]}
	}
	
	\subsubsection*{Orbifold evaluation map (right)}
	
	\uaeq{
		\tev_D = \;\tikz[baseline=10, scale=2]{\begin{scope}[xscale=-1]
			\begin{scope}
				\clip (-.7,0) rectangle (.7,1.1);
				\draw[defect, arrow position = .25] (0,0) ellipse (.6 and 1);
			\end{scope}
			\draw[symmetry defect] (.3,.85) .. controls (1,.7) .. (1,.4);
			\node[symmetry node] at (.3,.85) {};
			\begin{scope}[yscale=-1,shift={(1.3,-.4)}]
				\clip (-.4,0) rectangle (.4,.6);
				\draw[symmetry defect] (0,0) ellipse (.3 and .5);
			\end{scope}
			\draw[symmetry defect] (1.6,.4) -- (1.6,1);
			\draw[symmetry defect] (1.3,-.1) -- (1.3,-.3);
			\node[unit node] at (1.3,-.33) {};
			\node at (-.7,-.2) {$D^\ast$};
			\node at (.6,-.2) {$D$};
			\node at (0,.3) {$X^{d'}$};
			\node at (-1,.5) {${Z'}^d$};
			\node at (1.6,-.2) {${Z}^{d}$};
		\end{scope}}
		\circ \xi
	}
	where $\xi:D\otimes_A D^\ast \rightarrow D\otimes D^\ast$ is the inclusion.
	\uaeq{
		\tev_D: D \otimes_A D^\ast &\rightarrow A \\
		f_i \otimes X^n f_k^\ast &\mapsto - \frac 1{d'} \sum_{h\in\bZ_{d'}} e_{[-h]} \Res{ \frac{ \left(\partial_X\d_D \cdot \epsilon^h_l \right)_{(k+M),i}^{Z\rightarrow Z'} X^n }{d\cdot X^{d-1}}} \\
		&- \frac 1{d'} \sum_{h\in\bZ_{d'}} e_{d'+[-h]} \Res{ \frac{ \left(\partial_X\d_D^{Z\rightarrow Z'} \cdot \epsilon^h_{1+l} \cdot \partial^{Z,\epsilon^h Z'} \d_D \right)_{(k+M),i} X^n }{d\cdot X^{d-1}}}
	}
	
	\subsubsection*{Orbifold coevaluation map (left)}
	
	\uaeq{
		\coev_D = \vartheta\circ \;\tikz[baseline=-30, scale=2]{
			\begin{scope}[yscale=-1, xscale=-1]
				\clip (-.7,0) rectangle (.7,1.1);
				\draw[defect, arrow position = .25] (0,0) ellipse (.6 and 1);
			\end{scope}
			\draw[symmetry defect] (-.52,-.5) .. controls (-.9,-.6) .. (-1,-1);
			\node[symmetry node] at (-.52,-.5) {};
			\node at (-.7,0) {$D$};
			\node at (.7,0) {${}^\ast D$};
			\node at (0,-.5) {$X^{d}$};
			\node at (-1,-.4) {${Z}^{d'}$};
			\node at (1,-.5) {${Z'}^{d'}$};
		}
	}
	where $\vartheta:D\otimes {}^\ast D \rightarrow D\otimes_A {}^\ast D$ is the projection.
	\uaeq{
		\coev_D: A &\rightarrow D\otimes_A {}^\ast D \\
		e_a &\mapsto \delta_{|e_a|,0} \sum_{ij} (-1)^{|e_j|} \left(\left\{ \partial^{Z,Z'} \d_D \right\}^{Z\mapsto \epsilon^a Z}\right)_{ij} \epsilon^{a l_i} f_i\otimes f^\ast_{(j+M)} \\
		&\quad+ \delta_{|e_a|,1} \sum_i (-1)^{|e_i|} \epsilon^{a l_i} f_i \otimes f_{(i+M)}^\ast
	}
	
	\subsubsection*{Orbifold coevaluation map (right)}
	
	\uaeq{
		\tcoev_D = \vartheta\circ \;\tikz[baseline=-30, scale=2]{
			\begin{scope}[yscale=-1, xscale=1]
				\clip (-.7,0) rectangle (.7,1.1);
				\draw[defect, arrow position = .25] (0,0) ellipse (.6 and 1);
			\end{scope}
			\draw[symmetry defect] (.52,-.5) .. controls (.9,-.6) .. (1,-1);
			\node[symmetry node] at (.52,-.5) {};
			\node at (-.7,0) {$D^\ast$};
			\node at (.7,0) {$D$};
			\node at (0,-.5) {$Z^{d}$};
			\node at (-.8,-.7) {${X}^{d'}$};
			\node at (1,-.3) {${X'}^{d'}$};
		}
	}
	where $\vartheta:D\otimes D^\ast \rightarrow D\otimes_A D^\ast$ is the projection. 
	\uaeq{
		\tcoev_D: A &\rightarrow D^\ast \otimes D  \\
		e_a &\mapsto \delta_{|e_a|,0}\sum_{ij} \left( \partial^{X,\eta^{-a}X'}\d_D \right)_{ji} f_{(i+M)}^\ast \otimes f_j \eta^{-a r_j} \\
		&\quad+ \delta_{|e_a|,1 } \eta^a \sum_{i} (-1)^{|e_i|} f_{(i+M)}^\ast \otimes f_i \eta^{-a r_i}
	}

	\subsubsection*{Orbifold evaluation map (right) for boundaries}
	
	\uaeq{
		\tev_B = \;\tikz[baseline=10, scale=2]{\begin{scope}[xscale=-1]
			\begin{scope}
				\clip (-.7,0) rectangle (.7,1.1);
				\draw[defect, arrow position = .2] (0,0) ellipse (.6 and 1);
			\end{scope}
			\node at (-.7,-.2) {$B^\ast$};
			\node at (.6,-.2) {$B$};
			\node at (0,.3) {$X^d$};
		\end{scope}}
	}
	\uaeq{\tev_B: 
		B \otimes_A B^\ast &\rightarrow \bC \\
		f_i \otimes X^p f^\ast_k &\mapsto - \Res{\frac{ X^p \left( \partial_X\d_B \right)_{(k+M),i} \d X} {d \cdot X^{d-1}}} 
	}
	
	\subsubsection*{Orbifold coevaluation map (left) for boundaries}
	
	\uaeq{
		\coev_B = \;\tikz[baseline=-30, scale=2]{
			\begin{scope}[yscale=-1, xscale=-1]
				\clip (-.7,0) rectangle (.7,1.1);
				\draw[defect, opp arrow position = .2] (0,0) ellipse (.6 and 1);
			\end{scope}
			\node at (-.7,0) {$B$};
			\node at (.7,0) {${}^\ast B$};
			\node at (0,-.5) {$X^{d}$};
		}
	}
	\uaeq{\coev_B:
		\bC&\rightarrow B \otimes_A {}^\ast B \\
		1 &\mapsto \sum_i f_i \otimes {}^\ast f_i
	}

	\section{Explicit calculations for RG defects in LG orbifolds}
	
	In this appendix we explicitly check that the RG defects $R$ between LG orbifolds presented in section~\ref{sec:rg networks in lg minimal model orbifolds} satisfy the RG property that $R\otimes R^\dagger\cong I$ (appendix~\ref{sec: R otimes R dagger equals A}) and determine the corresponding projection defects $P = R^\dagger \otimes R$ (appendix~\ref{app:LG P}). We show how IR boundary conditions and symmetries are realized in the UV (appendices \ref{app:B otimes P cong B for LG} and \ref{app:LG IR symmetries}) and we perform the calculation $R_\infty\otimes R_\infty^\dagger\cong I_\text{IR}$ (appendix \ref{app:RG defect fusion infinity}). For the purpose of this appendix we again adopt the generalized orbifold notation of appendix~\ref{sec:orbifold minimal models as generalized orbifolds}.
	
	\subsection{\texorpdfstring{$R \otimes_A R^\ast \cong A$}{R otimes R dagger = A}}
	\label{sec: R otimes R dagger equals A}
	
	Here, we show that $R \otimes_A R^\ast \cong A$. (In this appendix we adopt the following notation from appendix~\ref{sec:orbifold minimal models as generalized orbifolds}: $\otimes_A$ denotes fusion in the generalized orbifold theory defined by $A$, while $\otimes$ denotes the fusion in the unorbifolded theory. Moreover, ${}^\ast$ denotes adjunction in the orbifold theory, while ${}^\dagger$ refers to adjunction in the unorbifolded theory.)  Fusion of B-type defects has been discussed in \cite{Brunner:2007qu}, for the orbifold version see \cite{Brunner:2007ur}.
	
	As explained in those papers, matrix factorizations of $W$ over a polynomal ring $R$ are related to finitely generated modules over $\hat R := R/(W)$ as free resolutions of such modules always turn two-periodic after finitely many steps \cite{Eisenbud}. The two-periodic part then gives a matrix factorization of $W$.
	
	In order to calculate $R \otimes_A R^\ast$, we fix the coordinates on all three parts of the worldsheet to be $Z$, $X$ and $Y$:
	\uaeq
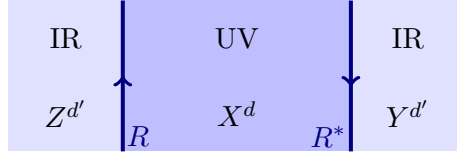
{
	\tikz[baseline=\offset]{
		\fill[WScolor light] (-3,-1) rectangle (-1.5,1);
		\fill[WScolor] (-1.5,-1) rectangle (1.5,1);
		\fill[WScolor light] (1.5,-1) rectangle (3,1);
		\draw[defect, arrow position=.5] (-1.5,-1) -- (-1.5,1);
		\draw[defect, opp arrow position=.5] (1.5,-1) -- (1.5,1);
		\node[DefectColor] at (-1.3,-.8) {$R$};
		\node[DefectColor] at (1.2,-.8) {$R^\ast$};
		\node at (-2.25,.5) {IR};
		\node at (0,.5) {UV};
		\node at (2.25,.5) {IR};
		\node at (-2.25,-.5) {$Z^{d'}$};
		\node at (0,-.5) {$X^d$};
		\node at (2.25,-.5) {$Y^{d'}$};
	}}
	The matrix factorization describing $R$ is given by
	\uaeq{
		R:R_1
		\tikz[baseline=0]{
			\begin{scope}
				\node at (0,2) {$\d_{R1}=\begin{pmatrix}Z& 0 & ... & 0 & -X^{n_0} \\ -X^{n_1} & Z &&&\\0&-X^{n_2}&Z&&\\\vdots&&\ddots&\ddots&\\0&&&-X^{n_{d'-1}}&Z\end{pmatrix}$};
				\draw[arrow position = 1] (-3.4,.2) -- (3.4,.2);
				\draw[arrow position = 1] (3.4,0) -- (-3.4,0);
				\node at (0,-.5) {$\d_{R0}$};
			\end{scope}
		}R_0,
	}
	see section~\ref{sec:rg defects in LG orbs}. The generators $f_{[i]}, i\in\bZ_{d'},$ of $R_0$ carry $\bZ_{d'}\times\bZ_d$-charges $[i,-m-\sum_{l=1}^i n_{l}]$ while the generators $e_{d'+[i]}$ of $R_1$ have charges $[i+1,-m-\sum_{l=1}^i n_{l}]$.
	
	According to section~\ref{sec:defects in mm orbifolds}, the right adjoint $R^\ast$ is given by the matrix factorization
	\uaeq{
		R^\ast:R^\ast_1
		\tikz[baseline=0]{
			\begin{scope}
				\node at (0,2) {$\d_{R^\ast 1}=\begin{pmatrix}Y& -X^{n_1} & 0 & ... & 0 \\ 0 & Y &-X^{n_2}&&\\\vdots&&Y&\ddots&\\0&&&\ddots&-X^{n_{d'-1}}\\-X^{n_0}&&&&Y\end{pmatrix}$};
				\draw[arrow position = 1] (-3.4,.2) -- (3.4,.2);
				\draw[arrow position = 1] (3.4,0) -- (-3.4,0);
				\node at (0,-.5) {$\d_{R^\ast 0}$};
			\end{scope}
		}R^\ast_0,
	}
	The generators $f_{[k]}^\ast$, $k\in\bZ_{d'}$, of $R^\ast_0$ carry $\bZ_d\times\bZ_{d'}$-charges
$[+m+\sum_{l=1}^k n_{l}+1,-k-1]$, 
and the generators $f_{d'+[k]}^\ast$ of $R^\ast_1$ carry charges $[+m+\sum_{l=1}^k n_{l}+1,-k-1]$. 
	
	Following the tensor product formula of section \ref{sec:b type defects in LG}, the matrix factorization describing $R \otimes_A R^\ast$ is the one associated to the $\bZ_d$-invariant part of the $\bC[Z,Y]/(Z^{d^\prime}-Y^{d^\prime})\bC[Z,Y]$-module
	\uaeq{M:=
		\text{coker}\begin{pmatrix} \id_{R0}\otimes \d_{R^\ast 1} & \d_{R1}\otimes\id_{R^\ast 0} \\
		\d_{R0}\otimes\id_{R^\ast 1} & -\id_{R1} \otimes \d_{R^\ast 0}\end{pmatrix}.
	}
	The two-periodic resolution of $M$ is isomorphic to the two-periodic part of the resolution of
	\uaeq{
		 M':=\text{coker}\left( \d_{R1}\otimes\id_{R^\ast0}, \id_{R0}\otimes\d_{R^\ast1} \right).
	}
The module $M'$ is generated by $f_{[i],[k]}^l := f_{[i]}\otimes X^l f_{[k]}^\ast$. They satisfy the relations
	\uaeq{
		Z f_{[i]} = X^{n_{i+1}} f_{[i+1]} \quad\text{and}\quad Y f^\ast_{[k]} = X^{n_{[k]}} f^\ast_{[k-1]},
	}
	which allow to reduce the generators to 
 $f_{[i],[k]}^l$ with $0\leq l<\text{min}(n_{i},n_{k+1})$. These carry $\bZ_{d'}\times\bZ_d\times\bZ_{d'}$-charges
	\uaeq{
		[i,-m-\sum_{j=1}^i n_{j} + l +m+\sum_{j=1}^k n_{j}+1,-k-1].
	}
	The $\bZ_d$-invariant part $(M')^{\bZ_d}$ is generated by the $\bZ_d$-invariant generators of $M'$, which are given by $\hat{f}_{[i]}:=f_{[i],[i-1]}^{n_i-1}$. They carry $\bZ_{d'}\times\bZ_{d'}$-charges $[i,-i]$ and satisfy the relations
	\uaeq{
		Z \hat{f}_{[i]} = Y \hat{f}_{[i+1]}.
	}
	Hence, $(M')^{\bZ_d}$ is isomorphic to the module $\text{coker}(\d_{A1})$, which implies that  the matrix factorization $R\otimes_A R^\ast$ is isomorphic to the identity defect $A$ in $\cM_{d'}/\bZ_{d'}$. Taking the left adjoint of this equation immediately yields $R\otimes_A {}^\ast R\cong A$ as well.
	
	\subsection{\texorpdfstring{The projection defect $P$}{The projection defect P}}
	\label{app:LG P}
	
	Having shown $R\otimes_A R^\ast\cong A$ in the previous appendix, we are now in a position to determine the projection defect $P = R^\ast \otimes_A R$. The projection $P'={}^\ast R\otimes_A R$ based on the left adjoint ${}^\ast R$ can then easily be obtained by left adjunction $P'={}^\ast P$.
	
	The calculation of $P$ follows the same route as the calculation of $R\otimes_A R^\ast$ in appendix~\ref{sec: R otimes R dagger equals A} above. 
First, we fix the chiral fields on all three parts of the worldsheet to be $Y$, $Z$ and $X$:
	\uaeq{
	\tikz[baseline=\offset]{
		\fill[WScolor] (-3,-1) rectangle (-1.5,1);
		\fill[WScolor light] (-1.5,-1) rectangle (1.5,1);
		\fill[WScolor] (1.5,-1) rectangle (3,1);
		\draw[defect, opp arrow position=.5] (-1.5,-1) -- (-1.5,1);
		\draw[defect, arrow position=.5] (1.5,-1) -- (1.5,1);
		\node[DefectColor] at (-1.2,-.8) {$R^\ast$};
		\node[DefectColor] at (1.2,-.8) {$R$};
		\node at (-2.25,.5) {UV};
		\node at (0,.5) {IR};
		\node at (2.25,.5) {UV};
		\node at (-2.25,-.5) {$Y^{d}$};
		\node at (0,-.5) {$Z^{d'}$};
		\node at (2.25,-.5) {$X^{d}$};
	}}
	The matrix factorizations $R$ and $R^\ast$ are described in appendix~\ref{sec: R otimes R dagger equals A}. As in the derivation of $R\otimes_A R^\ast\cong I$, the matrix factorization $R^\ast\otimes R$ is given by the two-periodic part of the free resolution of the $\bZ_{d'}$-invariant part of the $\bC[X,Y]/(Y^d-X^d)\bC[X,Y]$-module 
	\uaeq{{M}':=\text{coker}(d_{R^\ast 1}\otimes\id_{R_0},\id_{R^*0}\otimes d_{R1}).
	}
The latter is generated by
	\uaeq{
		{f}^l_{[k],[i]} := f^\ast_{[k]} Z^l \otimes f_{[i]}
	}
	subject to the relations
	\uaeq{
		Y^{n_{k}} f^\ast_{[k-1]} =Z f^\ast_{[k]}\quad\text{and}\quad Z f_{[i]} = X^{n_{i+1}} f_{[i+1]}.
	}
	These relations allow to reduce the generators to the ones with $l=0$. The remaining generators ${f}^0_{[k],[i]}$ carry $\bZ_d\times\bZ_{d'}\times\bZ_d$-charges
	\uaeq{
		\left[m+\sum_{j=1}^{[k]} n_{j} + 1, -k-1+i,-m-\sum_{j=1}^{[i]} n_{j} \right].
	}
	The $\bZ_{d'}$-invariant part of ${M}'$ is generated by the $\bZ_{d'}$-invariant generators, i.e. those ${f}^0_{[k],[i]}$, for which $[-k-1+i]=0$. These are the 
	$\hat{f}_{[i]}:={f}^0_{[i-1],[i]}$, which are subject to the relations
		\uaeq{
		Y^{n_{i}} \hat{f}_{[i]} = X^{n_{i+1}} \hat{f}_{[i+1]}.
	}
	They carry $\bZ_d\times\bZ_d$-charges
	\uaeq{
		\left[m+\sum_{j=1}^{[i-1]} n_{j} + 1, -m-\sum_{j=1}^{[i]} n_{j}\right].
	}
	Comparing with the matrix factorization $P$ given in equation~\eqref{eqp1}, one finds that $(\widetilde{M}')^{\bZ_{d'}}\cong\text{coker}(p_1)$ (where $Z$ has to be replaced by $Y$ in $p_1$).
Hence,  $R^\ast\otimes R$ is isomorphic to the matrix factorization $P$ given in section~\ref{sec:rg defects in LG orbs}.

	\subsection{\texorpdfstring{Boundary conditions satisfying $B\otimes_A P \cong B$}{Boundary conditions satisfying B otimes P = B}}
	\label{app:B otimes P cong B for LG}
	
	We now determine the boundary conditions, which are invariant under fusion with $P$. Elementary left boundary conditions in $\cM_d/\bZ_d$ are given by the $\bZ_d$-equivariant matrix factorizations
	\uaeq{
		B_\text{UV}: \bC[Z] \begin{pmatrix}[N + k]\end{pmatrix}\,
		\tikz[anchor=base, baseline=0]{
			\draw[->] (0,.2) -- (2,.2);
			\draw[->] (2,0) -- (0,0);
			\node at (1,.3) {$Z^k$};
			\node at (1,-.4) {$-Z^{d-k}$};
		}\, \bC[Z] \begin{pmatrix}[N]\end{pmatrix}.
	}
	of $-Z^d$, where  $k,N\in\bZ_d$, $k\neq 0$, c.f.~section~\ref{sec: LG realization}.
	The aim is to identify those boundary conditions, for which $B_\text{UV}\otimes_A P \cong B_\text{UV}$. To do so, we just calculate the fusion as is done in the previous appendices.
	We denote the generators of $B_0$ and $B_1$ by $b_0$ and $b_1$, respectively. They have $\bZ_d$-charge $[N]$, respectively $[N+k]$. The generators $\hat f_{[i]}$ of $P_0$ have $\bZ_d\times\bZ_d$-charge
	$[m+1+\sum_{l=1}^{[i-1]}n_{[l]},-m-\sum_{l=1}^{[i]} n_{[l]}]$, c.f. appendix~\ref{app:LG P}.
	
	To determine the fusion $B_\text{UV}\otimes P$, we again employ the method 
	described in appendix~\ref{sec: R otimes R dagger equals A}. For this, we determine generators and relations of the $\bZ_d$-invariant part of the $\bC[X]/X^d\bC[X]$-module $M':=\text{coker}(d_{B1}\otimes\id_{P0},\id_{B0}\otimes d_{P1})$:
	\aeq{\label{eq:conditions on generators}
		b_0 Z^k = 0 \\
		Z^{n_{i}} f_{[i]} = X^{n_{[i+1]}} f_{[i+1]}
	}
	For $B_\text{UV}\otimes_A P \cong B_\text{UV}$ to hold, out of all the generators $b_0 Z^q \otimes f_{[i]}$ of the fusion product exactly one generator may survive in $(M')^{\bZ_d}$. It must
	\begin{itemize}
		\item be invariant under the left $\bZ_d$-action, i.e.
			\uaeq{
				\left[ N + q + m +1+ \sum_{l=1}^{[i-1]} n_{l}  \right] = 0
			}
		\item carry right $\bZ_d$-charge $[N]$, i.e.
			\uaeq{
				\left[ -m - \sum_{l=1}^{[i]} n_{l} \right] = \left[N\right]
			}
		\item has to be a generator with respect to $\bC[X]$ and in particular cannot be eliminated by \eqref{eq:conditions on generators}, i.e.
			\uaeq{
				q<n_{i}\quad\text{and}\quad q<k
			}
			and
		\item it has to satisfy $b_0 Z^q \otimes f_{[i]} X^k = 0$.
	\end{itemize}
	The first two conditions fix $N=-m-\sum_{j=1}^{[i]}$ and imply $q=n_{i}-1$ which is consistent with $q<n_{i}$. The last condition becomes
	\uaeq{
		k\in\left\{n_{i},n_{i}+n_{i-1}, ..., n_{i}+...+n_{i-d'-2}\right\}.
	}
	These conditions are equivalent to $B_\text{UV}\otimes_A P\cong B_\text{UV}$ and imply that $B_\text{UV}$ must be of the form
	\uaeq{
		B_\text{UV}: \bC[Z]\,
		\tikz[anchor=base, baseline=0]{
			\draw[->] (0,.2) -- (2,.2);
			\draw[->] (2,0) -- (0,0);
			\node at (1,.3) {$Z^{n_{i} + ... + n_{i-I}}$};
			\node at (1,-.4){$-Z^{d-n_{i} - ... - n_{i-I}}$};
		}\, \bC[Z] \begin{pmatrix}-m-\sum_{l=1}^i n_{l}\end{pmatrix}
	}
	for arbitrary $i\in\bZ_{d'}$ and $I\in\left\{0, ..., d'-2\right\}$.

	\subsection{IR symmetry defects in the UV}
	\label{app:LG IR symmetries}
	
	Following section~\ref{sec: LG realization}, the IR $\bZ_{d'}$-symmetry is realized in the UV by means of the defects
	\uaeq{
		R^\ast\otimes_{A} {}_aI_{d'}\otimes_{A} R=:{}_aP.
	}
	As ${}_aI_{d'}\otimes_{A} R$ is described by the same matrix factorization as $R$ but with all left charges shifted by $+a$, we can employ the same set-up as in appendix \ref{app:LG P} and only shift charges by $+a$ where necessary. The corresponding module $M'$ is generated by
	\uaeq{
		f^l_{[k],[i]} := f^\ast_{[k]} Z^l \otimes f_{[i]}
	}
	with $\bZ_d\times\bZ_{d'}\times\bZ_d$-charges
	\uaeq{
		\left[m+\sum_{j=1}^{[k]} n_{j} + 1, -k-1+l+i+a,-m-\sum_{j=1}^{[i]} n_{j} \right]
	} 
	subject to the relations
	\uaeq{
		Y^{n_{k}} f^\ast_{[k-1]} =Z f^\ast_{[k]}\quad\text{and}\quad Z f_{[i]} = X^{n_{i+1}} f_{[i+1]}.
	}
	While the relations can be used to reduce generators to those with $l=0$, $\bZ_{d'}$-invariance gives the condition $[i+a-k-1]=0$. The remaining generators $\hat{f}_{[i]}:=f^0_{[i-1],[i-a]}$ of $(M')^{\bZ_{d'}}$ obey
	\uaeq{
		Y^{n_{i}} \hat{f}{[i]} = X^{n_{i-a+1}} f_{[i+1]}
	}
	and carry $\bZ_d\times\bZ_d$-charges
	\uaeq{
		\left[m+\sum_{j=1}^{[i-1]} n_{j} + 1, -m-\sum_{j=1}^{[i-a]} n_{j} \right].
	}
	One now easily reads off that $(M')^{\bZ_{d'}}$ is isomorphic to the cokernel of the matrix $p_1$ of the matrix factorization ${}_aP$ given in section~\ref{sec:rg defects in LG orbs}. Thus, the lifted symmetry defects are isomorphic to these matrix factorizations.

	\subsection{\texorpdfstring{$R_\infty\otimes_{U(1)} R_\infty^\ast\cong I_\text{IR}$}{R infty times R infty dagger = I IRi}}
	\label{app:RG defect fusion infinity}
	
	In this appendix we show that one can insert loops of the $U(1)$-equivariant Landau-Ginzburg theory with a single chiral superfield and zero superpotential into the Landau-Ginzburg orbifold models $\cM_{d'}/\bZ_{d'}$, $d'\geq 3$ without affecting correlators. The respective RG defects are described by the matrix factorizations $R_\infty$ of $Z^{d'}$ presented in section~\ref{sec: the limit}:
	\uaeq{
		R_\infty:S^{d'}\begin{psmallmatrix}[k+1,-m]\\ [k+2,-m-n_1]\\ [k+3,-m-n_1-n_2] \\ \vdots \end{psmallmatrix}
		\tikz[baseline=0]{
			\begin{scope}
				\node at (0,1.6) {$\d_{R1}=\begin{psmallmatrix}Z& 0 & ... & 0 & 0 \\ -X^{n_1} & Z &&&\\0&-X^{n_2}&Z&&\\\vdots&&\ddots&\ddots&\\0&&&-X^{n_{d'-1}}&Z\end{psmallmatrix}$};
				\draw[arrow position = 1] (-2.8,.2) -- (2.8,.2);
				\draw[arrow position = 1] (2.8,0) -- (-2.8,0);
				\node at (0,-.5) {$\d_{R0}$};
			\end{scope}
		}S^{d'}\begin{psmallmatrix}[k,-m]\\ [k+1,-m-n_1]\\ [k+2,-m-n_1-n_2] \\ \vdots \end{psmallmatrix}\,.
	}
	Here $m\in\bZ$, $k\in\bZ_{d'}$ and $n_1, ..., n_{d'-1}\in\mathbb{N}$. Moreover, $S=\bC[Z,X]$ and
	\uaeq{
		\d_{R0}=\begin{pmatrix}
					Z^{d'-1}& 0 & ... & ... & 0 \\
					Z^{d'-2}X^{n_1} & Z^{d'-1} &0&...&0\\
					Z^{d'-3}X^{n_1+n_2}&Z^{d'-2}X^{n_2}&Z^{d'-1}&\ddots& \vdots \\
					\vdots&\vdots&\ddots&\ddots&0\\
					X^{n_1+...+n_{d'-1}}&Z X^{n_2+...+n_{d'-1}}&...&Z^{d'-2}X^{n_{d'-1}}&Z^{d'-1}
				\end{pmatrix}.
	}
	The adjoint
	\uaeq{
		R_\infty^\ast:{\tilde S}^{d'}\left(\begin{smallmatrix}[m+1,-k]\\ [m+1+n_1,-k-1]\\ [m+1+n_1+n_2,-k-2] \\ \vdots \end{smallmatrix}\right)
		\tikz[baseline=0]{
			\begin{scope}
				\node at (0,1.2) {$\left(\begin{smallmatrix}Y& -X^{n_1} & &  & \\  & Y &-X^{n_2}&&\\&&\ddots&\ddots&\\&&&Y&-X^{n_{d'-1}}\\&&&&Y\end{smallmatrix}\right)$};
				\draw[arrow position = 1] (-2.4,.2) -- (2.4,.2);
				\draw[arrow position = 1] (2.4,0) -- (-2.4,0);
				\node at (0,-.5) {$\d_{R^\ast 0}$};
			\end{scope}
		}{\tilde S}^{d'}\left(\begin{smallmatrix}[m+1,-k-1]\\ [m+1+n_1,-k-2]\\ [m+1+n_1+n_2,-k -3] \\ \vdots \end{smallmatrix}\right)
	}
	can be obtained by taking the limit $d\to\infty$ of $R^*$. It is a matrix factorization of $-Y^{d'}$. 
	$\tilde S = \bC[X,Y]$ and $\d_{R^\ast 0}$ is given by $-\d_{R0}^T$ with $Z$ replaced by $Y$.

	According to section~\ref{sec:b type defects in LG}, the fusion product $R_\infty\otimes_{U(1)} R^\ast_\infty$ is given by the $U(1)$-invariant part of the tensor product matrix factorization $R_\infty\otimes R^\ast_\infty$. The $U(1)$-invariant generators of the latter are 
		\uaeq{
		g_{(i,j)}&:=f_i\otimes X^{n_i+...+n_{j+1}-1} f^\ast_j\quad&[k+i,-k-1-j] \\
		g_{(d'+i,j)}&:=f_{d'+i}\otimes X^{n_i+...+n_{j+1}-1} f^\ast_j\quad&[k+1+i,-k-1-j]  \\
		g_{(i,d'+j)}&:=f_{i}\otimes X^{n_i+...+n_{j+1}-1} f^\ast_{d'+j}\quad&[k+i,-k-j]  \\
		g_{(d'+i,d'+j)}&:=f_{d'+i}\otimes X^{n_i+...+n_{j+1}-1} f^\ast_{d'+j}\quad&[k+1+i,-k-j]  
	}
	for $1\leq i\leq d'-1$ and $0\leq j\leq i-1$. The 
	$\bZ_{d'}\times\bZ_{d'}$-charges of the generators are specified in square brackets. Here, $f_i$ and $f^\ast_i$ label the generators of $R_\infty$ and $R^\ast_\infty$, respectively. The generators with $0\leq i<d'$ are $\bZ_2$-even and the ones with $d'\leq i<2d'$ are $\bZ_2$-odd. 
	Setting
	\uaeq{
		l= \frac{i(i+1)}2+j, \qquad 0\leq l\leq M:= \frac{(d'+1)(d'-2)}2,
	}
	one can order the generators as follows
	\uaeq{
		g_l &= g_{(i,j)} \\
		g_{M+l} &= g_{(d'+i,d'+j)} \\
		g_{2M+l} &= g_{(d'+i,j)} \\
		g_{3M+l} &= g_{(i,d'+j)}.
	}
	$\left(R_\infty\otimes_{U(1)}R^\ast_\infty\right)_0$ is then generated by the $g_l$ and $g_{M+l}$ for $0\leq l\leq M$ 
	 and $\left(R_\infty\otimes_{U(1)}R^\ast_\infty\right)_1$ by the $g_{2M+l}$ and $g_{3M+l}$ for $0\leq l\leq M$. 
	 
	 In terms of the generators, 	
	the $U(1)$-invariant tensor product matrix factorization
	\uaeq{
		\d=\d_R\otimes_{U(1)}\mathbb 1 + \mathbb 1 \otimes_{U(1)} \d_{R^\ast}=:\begin{pmatrix}&\d_1\\\d_0&\end{pmatrix}
	}
	takes the form
	\aeq{\label{eq:infty matrix unchanged}
		(\d_1)_{(p,q),(d'+i,j)} &= \delta_{q,j} (Z \delta_{p,i} - \delta_{p,i+1}) \\
		(\d_1)_{(d'+p,d'+q),(d'+i,j)} &= \delta_{p,i} \theta(j-q) Y^{d'-1-(j-q)} \\
		(\d_1)_{(p,q),(i,d'+j)} &= \delta_{p,i} (Y\delta_{q,j}-\delta_{q+1,j}) \\
		(\d_1)_{(d'+p,d'+q),(i,d'+j)} &= \delta_{q,j} \theta(p-i) Z^{d'-1-(p-i)} \\
		(\d_0)_{(d'+i,j),(p,q)} &= \delta_{j,q}\theta(i-p)Z^{d'-1-(i-p)} \\
		(\d_0)_{(i,d'+j),(p,q)} &= -\delta_{i,p} \theta(q-j) Y^{d'-1-(q-j)} \\
		(\d_0)_{(d'+i,j),(d'+p,d'+q)} &= -\delta_{i,p} (Y\delta_{j,q}-\delta_{j+1,q}) \\
		(\d_0)_{(i,d'+j),(d'+p,d'+q)} &= \delta_{j,q} (Z\delta_{i,p}-\delta_{i,p+1})
	}
	where $1\leq i,p\leq d'$, $0\leq j<i$, $0\leq q<p$ and $\theta(x)=\begin{cases}1, x\geq 0\\0, x<0\end{cases}$. 
	For example, for $d'=5$ one obtains
	\setcounter{MaxMatrixCols}{20}
	\uaeq{
		\d_1=\begin{pmatrix}
			Z &  &  &  &  &  & & & &  & Y& &  & &  &  & &  &  &  \\
			-1&Z &  &  &  &  & & & &  &  &Y&-1& &  &  & &  &  &  \\
			  &  &Z &  &  &  & & & &  &  & &Y & &  &  & &  &  &  \\
			  &-1&  &Z &  &  & & & &  &  & &  &Y&-1&  & &  &  &  \\
			  &  &-1&  &Z &  & & & &  &  & &  & &Y &-1& &  &  &  \\
			  &  &  &  &  &Z & & & &  &  & &  & &  &Y & &  &  &  \\
			  &  &  &-1&  &  &Z& & &  &  & &  & &  &  &Y&-1&  &  \\
			  &  &  &  &-1&  & &Z& &  &  & &  & &  &  & &Y &-1&  \\
			  &  &  &  &  &-1& & &Z&  &  & &  & &  &  & &  &Y &-1\\
			  &  &  &  &  &  & & & &Z &  & &  & &  &  & &  &  &Y \\
			Y^4&&&&&&&&&&Z^4&&&&&&&&&\\
			&Y^4&Y^3&&&&&&&&Z^3&Z^4&&&&&&&&\\
			&&Y^4&&&&&&&&&&Z^4&&&&&&&\\
			&&&Y^4&Y^3&Y^2&&&&&Z^2&Z^3&&Z^4&&&&&&\\
			&&&&Y^4&Y^3&&&&&&&Z^3&&Z^4&&&&&\\
			&&&&&Y^4&&&&&&&&&&Z^4&&&&\\
			&&&&&&Y^4&Y^3&Y^2&Y&Z&Z^2&&Z^3&&&Z^4&&&\\
			&&&&&&&Y^4&Y^3&Y^2&&&Z^2&&Z^3&&&Z^4&&\\
			&&&&&&&&Y^4&Y^3&&&&&&Z^3&&&Z^4&\\
			&&&&&&&&&Y^4&&&&&&&&&&Z^4\\
		\end{pmatrix}
	}
	Stripping off trivial summands this matrix factorization reduces to the IR identity matrix factorization ($S'=\bC[Z,Y]$)
	\uaeq{
		I_\text{IR}: {S'}^{d'}\begin{pmatrix}[1,0]\\ [2,-1]\\ [3,-2] \\ \vdots \\ [d',-d'+1] \end{pmatrix}
		\tikz[baseline=0]{
			\begin{scope}
				\node at (0,1.8) {$\begin{pmatrix}Z& 0 & ... & 0 & -Y \\ -Y & Z &&&\\0&-Y&Z&&\\\vdots&&\ddots&\ddots&\\0&&&-Y&Z\end{pmatrix}$};
				\draw[arrow position = 1] (-3,.2) -- (3,.2);
				\draw[arrow position = 1] (3,0) -- (-3,0);
				\node at (0,-.5) {$\d_{I_\text{IR}0}$};
			\end{scope}
		}\;{S'}^{d'}\begin{pmatrix}[0,0]\\ [1,-1]\\ [2,-2] \\ \vdots \\ [d'-1,-d'+1] \end{pmatrix}
	}
	In order to see this, we perform a change of basis on \eqref{eq:infty matrix unchanged}:
	\uaeq{
		{\d}_1 = S \cdot {\tilde\d}_1 \cdot T^{-1},\qquad{\d}_0 = T \cdot {\tilde\d}_0 \cdot S^{-1},
	}
	where $S$ and $T^{-1}$ are defined by
	\uaeq{
		(S)_{(p,q),(i,j)} &= \delta_{q,j} (\delta_{p,i}-Z\delta_{p+1,i}) \\
		(S)_{(p,q),(d'+i,d'+j)} &= 0 \\
		(S)_{(d'+p,d'+q),(i,j)} &= -\delta_{p+1,i} Y^{d'-1-(j-q)} \theta(j-q) \\
		&\quad-\delta_{i,j+1}\left[ Z^{d'-p+i-1} Y^{q-i} \theta(q-i)+ Z^{i-1-p} Y^{d'+q-i} \theta(i-2-p) \right] \\
		(S)_{(d'+p,d'+q),(d'+i,d'+j)} &= \delta_{q,j} (\delta_{p,i}+\theta(i-p-1) Z^{i-p}) + \delta_{i,d'-1}\delta_{j,0}\theta(q-1)Z^{d'-1-p}Y^q
	}
	and
	\uaeq{
		(T^{-1})_{(d'+p,q),(d'+i,j)} &= \delta_{q,j} \delta_{p,i} \\
		&\;+ \delta_{i,d'-1} \left(-\delta_{q,j}Z^{d'-1-p} \theta(d'-1-p) +\delta_{p,d'-1} Y^{q-j}\theta(q-j-1)\right)\\
		(T^{-1})_{(p,d'+q),(d'+i,j)} &= -\delta_{i,d'-1}\delta_{q+1,p}\theta(q-j-1)Z^{d'-1-q}Y^{q-1-j} \\
		(T^{-1})_{(d'+p,q),(i,d'+j)} &= -\delta_{q,j} Y \left( \delta_{p+1,i} + Z^{i-p-1} \theta(i-p-2) \right) \\
		&\qquad\qquad+\delta_{q+1,j} Z^{i-p-1}\theta(i-p-1) \\
		(T^{-1})_{(p,d'+q),(i,d'+j)} &= \delta_{p,i}\delta_{q,j} + \delta_{q,j}\delta_{q+1,p} \theta(i-p-1) Z^{i-p}.
	}
	Here again $1\leq i,p\leq d'$, $0\leq j<i$, $0\leq q<p$.
Then 
	\uaeq{
		({\tilde\d}_1)_{(p,q),(d'+i,j)} &= -\delta_{q,j}\delta_{p,i+1} + Z \delta_{i,d'-1}\delta_{p,d'-1}\delta_{q,d'-2}\delta_{j,d'-2} \\
		({\tilde\d}_1)_{(d'+p,d'+q),(d'+i,j)} &= \delta_{p,d'-1}\delta_{i,d'-1} (W\delta_{q,j+1}-Y\delta_{q,0}\delta_{j,d'-2}) \\
		({\tilde\d}_1)_{(p,q),(i,d'+j)} &= -Y \delta_{p,i}\delta_{q,j}\delta_{i,j+1} + Z \delta_{p+1,i}\delta_{p,j}\delta_{p,q+1} \\
		({\tilde\d}_1)_{(d'+p,d'+q),(i,d'+j)} &= \delta_{i,p+1}\delta_{q,j}W + \delta_{i,1}\delta_{j,0}\delta_{p,d'-1}\delta_{q,0}Z 
	}
	and
	\uaeq{
		({\tilde \d}_0)_{(d'+i,j),(p,q)} &= -\delta_{i+1,p}\delta_{j,q} W + \delta_{i,d'-1}\delta_{j,d'-2}\delta_{p,q+1} Z^p Y^{d'-1-p} \\
		({\tilde\d}_0)_{(i,d'+j),(p,q)} &= -\delta_{i,j+1}\delta_{p,q+1}\left( \theta(p-i) Y^{d'-1-p+i} Z^{p-i} \right.\\
			 &\qquad\qquad\qquad\qquad\left.+\theta(i-p-1) Z^{d'-i+p} Y^{i-p-1} \right) \\
		({\tilde\d}_0)_{(d'+i,j),(d'+p,d'+q)} &= \delta_{p,d'-1}\delta_{i,d'-1} \left( \delta_{j+1,q}-Y^{d'-1} \delta_{j,d'-2}\delta_{q,0} \right) \\
		({\tilde\d}_0)_{(i,d'+j),(d'+p,d'+q)} &= -\delta_{i,p+1}\delta_{j,q}+\delta_{p,d'-1}\delta_{q,0}\delta_{i,j+1}Z^{d'-i}Y^j
	} are matrix factorization of $W=Z^{d'}-Y^{d'}$ which reduce to the indentity matrix factorization.

In the example $d'=5$  $\d_1$ above turns into 
	\uaeq{{\tilde\d}_1=\begin{pmatrix}
			  &  &  &  &  &  & & & &  & -Y& &Z& &  &  & &  &  &  \\
			-1&  &  &  &  &  & & & &  &  & &  & &  &  & &  &  &  \\
			  &  &  &  &  &  & & & &  &  & &-Y & &  &Z& &  &  &  \\
			  &-1&  &  &  &  & & & &  &  & &  & &  &  & &  &  &  \\
			  &  &-1&  &  &  & & & &  &  & &  & &  &  & &  &  &  \\
			  &  &  &  &  &  & & & &  &  & &  & &  &-Y & &  &  &Z\\
			  &  &  &-1&  &  & & & &  &  & &  & &  &  & &  &  &  \\
			  &  &  &  &-1&  & & & &  &  & &  & &  &  & &  &  &  \\
			  &  &  &  &  &-1& & & &  &  & &  & &  &  & &  &  &  \\
			  &  &  &  &  &  & & & &Z &  & &  & &  &  & &  &  &-Y \\
			&&&&&&&&&0&0&W&&&&&&&&\\
			&&&&&&&&&&&0&0&W&&&&&&\\
			&&&&&&&&&&&&&&W&&&&&\\
			&&&&&&&&&&&&&&0&0&W&&&\\
			&&&&&&&&&&&&&&&&&W&&\\
			&&&&&&&&&&&&&&&&&&W&\\
			&&&&&&&&&-Y&Z&&&&&&&&&\\
			&&&&&&W&&&&&&&&&&&&&\\
			&&&&&&&W&&&&&&&&&&&&\\
			&&&&&&&&W&&&&&&&&&&&\\
		\end{pmatrix}
	}
	which is easily recognized as the matrix associated to a sum of the identity matrix factorization $I_\text{IR}$ with a number of trivial rank-one matrix factorizations. 
	
	In the general case, 
 the generators not belonging to trivial summands are the ones labelled by the restricted index sets
\uaeq{
	\left\{(i,j)\left\vert i=j+1 \right.\right\} &\subset\left\{ (i,j)\left\vert i=1, ...,d'-1;\;j=0, ..., i-1 \right.\right\} \\
	\left\{ (d'+i,d'+j) \left\vert i=d'-1,\;j=0 \right.\right\} &\subset \left\{ (d'+i,d'+j)\left\vert i=1, ...,d'-1;\;j=0, ..., i-1 \right.\right\} \\
	\left\{ (d'+i,j) \left\vert i=d'-1,\;j=d'-2 \right.\right\} &\subset \left\{ (d'+i,j)\left\vert i=1, ...,d'-1;\;j=0, ..., i-1 \right.\right\} \\
	\left\{ (i,d'+j) \left\vert i=j+1 \right.\right\} &\subset \left\{ (i,d'+j)\left\vert i=1, ...,d'-1;\;j=0, ..., i-1 \right.\right\}
}
Restricting to these generators yields the IR identity defect $I_\text{IR}$.

	\bibliographystyle{JHEP}
	\bibliography{references}
	
\end{document}